\documentclass{jfm}

\usepackage{graphicx}
\usepackage{dcolumn}
\usepackage{epstopdf, epsfig, amsmath, bm}
\usepackage{etoolbox}

\usepackage[utf8]{inputenc}
\usepackage[T1]{fontenc}

\usepackage{newtxtext}
\usepackage{newtxmath}
\usepackage{natbib}
\usepackage{hyperref}
\hypersetup{
    colorlinks = true,
    urlcolor   = blue,
    citecolor  = black,
}

\newcommand{\RomanNumeralCaps}[1]
\linenumbers

\newcommand{\be}{\begin{equation}}
\newcommand{\ee}{\end{equation}}
\newcommand{\bes}{\begin{equation*}}
\newcommand{\ees}{\end{equation*}}
\newcommand{\bea}{\begin{eqnarray}}
\newcommand{\eea}{\end{eqnarray}}
\newcommand{\bi}{\begin {itemize}}
\newcommand{\ei}{\end {itemize}}
\newcommand{\benm}{\begin{enumerate}}
\newcommand{\eenm}{\end{enumerate}}
\newcommand{\bmn}{\begin{minipage}}
\newcommand{\emn}{\end{minipage}}
\newcommand{\bfig}{\begin{figure}}
\newcommand{\efig}{\end{figure}}
\newcommand{\ig}{\includegraphics}
\newcommand{\lnw}{\linewidth}

\newcommand{\bcls}{\begin{columns}}
\newcommand{\ecls}{\end{columns}}
\newcommand{\bcl}{\begin{column}}
\newcommand{\ecl}{\end{column}}

\newcommand{\bmat}{\begin{matrix}}
\newcommand{\emat}{\end{matrix}}
\newcommand{\bpmat}{\begin{pmatrix}}
\newcommand{\epmat}{\end{pmatrix}}
\newcommand{\bvmat}{\begin{vmatrix}}
\newcommand{\evmat}{\end{vmatrix}}
\newcommand{\bbmat}{\begin{bmatrix}}
\newcommand{\ebmat}{\end{bmatrix}}

\newcommand{\ol}{\overline}

\newcommand{\ptl}{\partial}

\newcommand{\lla}{\left\langle}
\newcommand{\rra}{\right\rangle}
\newcommand{\lal}{\langle}
\newcommand{\ral}{\rangle}

\newcommand{\ep}{\epsilon}
\newcommand{\al}{\alpha}

\newcommand{\om}{\omega}

\renewcommand{\k}{{\bm k}}

\newcommand{\e}{{\bm e}}
\newcommand{\f}{{\bm f}}

\newcommand{\x}{{\bm x}}

\newcommand{\bO}{{\bm\Omega}}

\newcommand{\bo}{{\bm \omega}}

\newcommand{\bk}{{\bm k}}

\newcommand{\bu}{{\bm u}}
\newcommand{\bv}{{\bm v}}
\newcommand{\bx}{{\bm x}}

\newcommand{\DD}{{\mathcal D}}

\newcommand{\FF}{{\mathcal F}}

\newcommand{\PP}{{\mathcal P}}

\graphicspath{{./jfm_fig/}}

\title{The conditional Lyapunov exponents and synchronisation of rotating turbulent flows}

\author{Jian Li\aff{1}, Mengdan Tian\aff{1}, Yi
Li\aff{2}\corresp{\email{yili@sheffield.ac.uk}},
Wenwen Si\aff{1}, Huda Khaleel Mohammed\aff{3}
} 
\affiliation{
  \aff{1}School of Naval Architecture and Maritime, Zhejiang Ocean
University, Zhoushan, 316022, China
  \aff{2}School of Mathematics and Statistics, University of
  Sheffield, Sheffield, S3 7RH, UK 
  \aff{3}Department of System and Control Engineering, College of Electronics
  Engineering, Ninevah University, Iraq. 
}

\begin{document}
\maketitle

\begin{abstract}
    The synchronisation between rotating turbulent flows in periodic boxes is
  investigated numerically. The flows
  are coupled via a master-slave coupling, taking the Fourier modes with wavenumber below a given
  value $k_m$ as the master modes. It is found that
  synchronisation happens when $k_m$ exceeds a threshold value $k_c$, 
  and $k_c$ depends strongly on the
  forcing scheme. 
  In rotating Kolmogorov flows, $k_c\eta$ does not change with
  rotation in the range of rotation rates considered, $\eta$ being the Kolmogorov length scale. 
Even though the energy spectrum has a steeper slope, the value
of $k_c\eta$ is the same as that found in isotropic turbulence. 
  In flows driven by a forcing term maintaining
  constant energy injection rate, synchronisation becomes easier when rotation
  is stronger. $k_c\eta$ decreases with rotation, and it is reduced significantly for strong
rotations when the slope of the energy spectrum approaches $-3$.
It is shown that the conditional Lyapunov exponent for a given $k_m$ is reduced by rotation in the 
flows driven by the second type of forcing, but it increases mildly with rotation for the Kolmogorov flows. 
The local conditional Lyapunov exponents fluctuate more strongly as rotation
is increased, 
although synchronisation occurs as long as the average conditional Lyapunov exponents are
  negative. 
  We also look for the relationship between $k_c$ and the energy spectra of the
Lyapunov vectors. We find that the spectra
always seem to peak around $k_c$, and synchronisation fails when the energy spectra of the conditional Lyapunov vectors
have a local maximum in the slaved modes.

\end{abstract}

\section{Introduction }\label{sect:intro}

For some chaotic systems, one may couple two realisations of the system in
specific ways to synchronise the states of the two realisations, in the
sense that the two realisations remain chaotic, but the 
difference between them 
decays over time and approaches zero asymptotically. 
This phenomenon is called
(complete) chaos
synchronisation, which was first discussed in \citet{FujisakaYamada83} 
and attracted wide attention by \citet{PecoraCarroll90} (see, e.g.,
\citet{PecoraCarroll15} for a historical account). 
The phenomenon has applications in, e.g., secure communication, parameter
estimation, and is used as a paradigm to understand a wide range of phenomena. 
The research into these applications as well as the principles behind the phenomenon and other
forms of chaos synchronisation are 
reviewed in \citet{PecoraCarroll15, Erogluetal17, Boccalettietal02}. 

In turbulent simulations, chaos synchronisation is closely linked to 
data assimilation, a practice where observational or measurement data are
synthesised with simulation to produce more accurate predictions of turbulent flows.  
If the aim of data assimilation is to recover the chaotic instantaneous 
turbulent fields, 
it becomes
a problem of chaos synchronization. 
For isotropic turbulence,
typically two flows can be synchronised completely by replacing Fourier modes
with wavenumbers less than $k_m$ 
from one
flow 
with those in the other, 
and synchronisation is achieved only if $k_m$ is larger than a threshold value
$k_c$. 
To the best of our knowledge, \citet{Henshawetal03} are  
the first to investigate the synchronisation of turbulent flows, where 
a theoretical estimate of $k_c$ is derived but numerical
experiments are conducted to show that synchronisation can be achieved with
far fewer Fourier modes. 
Another early work is \citet{Yoshidaetal05}, where it was numerically established
that $k_c \eta\approx 0.2 $ with $\eta$ being the Kolmogorov length scale.
\citet{Lalescuetal13} investigate a similar problem 
with a different forcing scheme as well as anisotropic grids, 
and $k_c  \eta\approx 0.15$ is found. 

When $k_m$ is smaller than $k_c$,  
\citet{VelaMartin21} shows that  
partial synchronisation can be obtained
and that the velocity fields in domains with 
strong vorticity are better synchronized than those with weaker vorticity.
This result suggests that the synchronisation of turbulent flows may have its own
specific features pertinent to the physics of turbulence. 
In Couette flows, \citet{NikolaidisIoannou22} shows that synchronization 
occurs when streamwise
Fourier modes with wavenumber exceeding a threshold value are replicated in
the two systems. They also show that synchronization happens if 
the conditional Lyapunov exponent is negative, inline with result
known from the synchronisation of low-dimensional chaotic systems \citep{Boccalettietal02}.
Channel flows are investigated 
by \citet{WangZaki21}, where data from layers in the flow domain 
with different orientations are used to couple two systems. By doing so,
scaling of the thickness of the layers needed for synchronization is
established, through numerical experiments as well as analyses of the
conditional Lyapunov exponents. 

In the aforementioned research, the coupling of the two flows is always
achieved by replacing part of the velocity field in one flow by the
corresponding part of velocity in the other flow.
This type of coupling is termed master-slave coupling. Another common way to
couple the two systems is through nudging,
where a linear forcing term is introduced in either one or
both of the flow fields. 
The forcing term 
nudges one flow from the other, hence the name `nudging'. 
Nudging is used in \citet{Leonietal18,Leonietal20} to synchronize 
isotropic turbulence with or without
rotation. The efficacy of different nudging schemes is compared. 
In rotating turbulence, they find that synchronisation becomes more effective
due to the presence of large scale coherent vortices, and 
inverse cascade can be reconstructed when nudging is applied to small scales. 

Going beyond the synchronisation between two 
simulations with identical system parameters, 
\citet{BuzzicottiLeoni20} consider the synchronisation 
between large eddy simulations (LES) and direct numerical simulations (DNS). 
using the nudging method. Because the two systems are different in this case, complete
synchronisation is unachievable. However, the authors show that 
the error between the nudged LES
velocity and DNS velocity can be minimised by tuning the parameters in the
subgrid-scale (SGS) models. Chaos synchronisation thus is
used to optimise model parameters. 
\citet{Lietal22} investigate the synchronisation between LES and DNS 
using the master-slave coupling, with a focus on the threshold wavenumber and
the synchronisation error for
different SGS models. They find that the standard
Smagorinsky model under certain circumstances produce smaller synchronisation
error than the dynamic Smagorinsky model and the dynamic mixed model. 

Rotating turbulence, i.e., turbulent flows in a rotating frame of reference,
is ubiquitous in atmospheric, oceanic as well as industrial flows. 
Rotating turbulence possesses features distinct from non-rotating
turbulence, including, for example, the emergence of coherent vortices,
steepened energy spectrum, and quasi-two-dimensionalization of the flow. For
detailed reviews on these phenomena, see, e.g., \citet{GodeferdMoisy15} and
\citet{SagautCambon08}. 
More recently it is also noted that some features strongly 
depend on the forcing scheme \citep{DallasTobias16}.
The synchronisation of rotating turbulence is investigated 
in \citet{Leonietal18,Leonietal20}, 
as is mentioned above. 
These investigations leave some interesting questions unanswered. 
The most important one is how
synchronisation depends on the rate of rotation. 
For example, how does the threshold wavenumber $k_c$ change with the
rotation rate? 
Also, given the strong effects of the forcing term on the
small scales of 
rotating turbulence \citep{DallasTobias16},
how the forcing term affects synchronisation in rotating turbulence remains unclear. 
We intend to address 
these questions in present investigation. 

We use master-slave coupling instead of nudging. The former does not require 
specifying the coupling strength hence reducing the number of control
parameters by one. 
To characterise the synchronised state,
we calculate the conditional Lyapunov exponents of
the slave system and quantify their  
dependence on rotation. Two different forcing mechanisms are considered to
illustrate the effects of the forcing term. 
As we will show later, rotation has significant impacts on the synchronisation
behaviours and the impacts strongly depend on the forcing term. We believe these results
are useful addition to our understanding on 
rotating turbulence, especially on how to enhance its predictability via
simulations equipped with data assimilation functionalities. The impact of the findings may be found in
fields such as numerical weather prediction. 

The manuscript is organised as follows. We introduce the governing equations, 
the controlling parameters, and the definition of conditional Lyapunov exponents in Section
\ref{sect:eqn}. The numerical methods and a summary of the numerical
experiments are presented in Section
\ref{sect:results}, which is followed by the results and discussions. 
Section \ref{sect:conclusions} concludes the article with the main
observations we make from the numerical experiments. 

\section{Governing equations \label{sect:eqn}}

We consider rotating turbulent flows in a $[0,2\pi]^3$ box 
with $\x = (x_1,
x_2, x_3) = (x,y,z)$ representing the spatial coordinates.
The flow 
satisfies the periodic boundary condition in all three directions. 
Let $\bO \equiv \Omega \hat{\bf k}$ be the rotation rate of a rotating frame
of reference, where $\hat{\bf k}$ is the unit vector in the $z$ direction.  
Let $\bu(\x,t)$ be the velocity field. 
For an observer in the rotating frame, the Navier-Stokes equation (NSE) reads
(see, e.g., \citet{greenspan69})
\be \label{eq:nse}
D_t \bu + 2 \bO \times \bu = - \nabla p + \nu \nabla^2 \bu + \f,
\ee
where 
\be
D_t \equiv \ptl_t + (\bu \cdot \nabla) 
\ee
is the material derivative with
$\bu$ as the advection velocity; 
$p = p(\bx,t)$ is the pressure; 
$\nu$ is the viscosity, and $\f = \f(\x,t)$ is the forcing term.  
The density of the fluid has been assumed to be unity. 
The velocity is
assumed to be incompressible so that 
\be \nabla \cdot \bu = 0. 
\ee
Two different forcing terms are considered in this investigation. In the first case,
\be
\f \equiv (a_f\cos k_f x_2,0,0) 
\ee
with $a_f = 0.15$ and $k_f =1$.  
Customarily, the flow driven by forcing terms of this type is called the
Kolmogorov flow \citep{BorueOrszag96}, therefore we call this forcing term the
Kolmogorov forcing. 
Kolmogorov flow in general is inhomogeneous due to the sinusoidal
form of the force, although we do not investigate the effects of the
inhomogeneity in what follows. 
Kolmogorov forcing does not directly inject energy into turbulent velocity
fluctuations. 
Rather, its role is to
maintain the unstable mean velocity profile which generates 
turbulent fluctuations when it loses its stability
\citep{BorueOrszag96}.
The parameter $k_f$ introduces a length scale, which will be at the
order of the integral scale of the flow. A velocity scale can be defined from
$k_f$ and $a_f$, which determines the order of magnitude of the turbulent
kinetic energy of the flow.

In the second case, the forcing term is confined in a range of small
wavenumbers in the Fourier space. Specifically, let $\hat\bu(\bk,t)$ be the
Fourier transform of $\bu$ and $\hat{\f}(\bk,t)$ be that of $\f$, with $\k$
being the wavenumber. The
force is defined by
\be
\hat{\f}(\k,t) = 
\begin{cases}
  A(t) \hat\bu(\k, t),  & \vert \k\vert \le k_{f, \text{max}} \\
  0, & \vert \k \vert > k_{f, \text{max}}.
\end{cases}
\ee
where $k_{f,\text{max}} = 2$, and $A(t)$ is given by 
\be
A(t) = \frac{ \ep_f}{\sum_{\vert \k \vert \le k_{f,\text{max}}}
\hat{\bu}(\k,t)\hat{\bu}^*(\k,t)}, 
\ee
with $\ep_f = 0.05$ and $^*$ representing complex conjugate. 
This forcing term injects kinetic energy into the flow field at a constant rate
equal to $\ep_f$, 
via Fourier
modes with $\vert \k \vert \le k_{f,\text{max}}$. In the stationary stage, the
mean energy dissipation rate of the flow would be the same as $\ep_f$. We call
this forcing term `constant power forcing'.

Obviously the two forcing terms are different in many ways, although 
both are commonly used in
turbulent simulations. As will be shown below, the flow fields driven by the
two forces are different in many ways. 
To put this observation in context, we note that 
\citet{DallasTobias16} investigate the effects of the forcing term on the
evolution of rotating turbulence. They used a Taylor-Green forcing with a
memory time scale $\tau_m$. With different $\tau_m$ one may obtain different
stationary states. Among others, the energy spectrum may display different
slopes in different stationary states. 
In our simulations, the Kolmogorov forcing
term is a constant, therefore has an infinite memory time. The constant power
forcing has
a memory time of the
order of $(\ep_f k_{f, \text{max}}^2)^{-1/3} \approx 2$.  
Therefore, it is not surprising to find significant difference between the 
flows driven by the two different forces. The difference allows us to explore
how the forcing terms affect the synchronizability of the flows. 

The synchronisation of two flows is investigated by
simulating them with same parameters concurrently. 
Let $\bu^{(1)}$ and $\bu^{(2)}$ be the velocity
fields of the two flows, respectively. 
The two velocity fields are initialised 
with different initial conditions, then evolve over time simultaneously according to the NSE. 
To synchronise the two flows,
the Fourier modes of $\bu^{(2)}$
with $\vert\k\vert\le k_m$ are replaced by those of $\bu^{(1)}$ at each time
step. As such,
\be \label{eq:coupling}
\hat{\bu}^{(2)}(\k,t) = \hat{\bu}^{(1)}(\k,t), 
\ee
for $\vert \k \vert \le k_m$ 
at all time. 
This way of coupling the two flows together is usually termed 
master-slave coupling \citep{Boccalettietal02}. In this
case, $\bu^{(2)}$ is the slave whereas $\bu^{(1)}$ is
the master. 

It is expected that, under suitable conditions, 
$\bu^{(1)}$ and $\bu^{(2)}$ will
remain turbulent (chaotic) but they will synchronise, i.e., $\bu^{(2)}$ will gradually approach
$\bu^{(1)}$. 
Let the norm of a generic vector field $\bm w$ be
$$
\Vert \bm w \Vert^2 = \frac{1}{(2\pi)^3} \int_{[0,2\pi]^3} \bm w \cdot \bm w dV.
$$
The synchronisation error  
\be \label{eq:error}
\Delta(t)\equiv \Vert \bu^{(1)} -
\bu^{(2)}\Vert
\ee
will decay exponentially towards zero \citep{Henshawetal03, Yoshidaetal05} when the two flows
synchronise. 

The ability to synchronise the two flows crucially depends on $k_m$, which we will call \emph{the coupling
wavenumber}. The Fourier modes in the two velocity fields with $\vert \k \vert
> k_m$ are the slave modes, whereas those with $\vert \k \vert \le
k_m$ are the master modes. 

Synchronisation depends on various statistics of the flow field, which
will be briefly introduced next.
As $\bu^{(1)}$ and $\bu^{(2)}$
are both stationary turbulent flows with identical governing equations 
and control parameters,
these statistics can be calculated from either of them. 
Therefore we will only use $\bu$ to indicate the velocity field. 
Let
$\bu'\equiv \bu - \lal \bu \ral$ be the velocity fluctuations, 
where $\lal ~ \ral$ denotes ensemble average. 
The mean energy dissipation rate $\ep$ is defined as
\be
\ep = 2 \nu \lal s'_{ij} s'_{ij} \ral,
\ee
where $s'_{ij} = (\ptl_j u'_i + \ptl_i u'_j)/2$ is the fluctuating strain rate
tensor. The small scales of
the flow are characterised by the Kolmogorov length
scale $\eta$ and 
the Kolmogorov time scale $\tau_k$, which are defined by
(see, e.g., \citet{Pope00})  
\be
\eta = (\nu^3/\ep)^{1/4} \quad \text{and}\quad
\tau_k =
(\nu/\ep)^{1/2}, 
\ee
respectively. 

When two isotropic turbulent flows are synchronised with the coupling
described above, it has been
found \citep{Yoshidaetal05, Lalescuetal13, Lietal22} that 
\be
\Delta(t) \sim \exp( \al
t/\tau_k), 
\ee
where $\al$ is the decay rate (note that the error decays only when $\alpha
<0$). The decay rate $\al$ is a function of $k_m
\eta$. The value of $k_m$ for which $\alpha =0$ is
the threshold wavenumber and is denoted by $k_c$. 
The normalised threshold wavenumber $k_c\eta$ is found to be $0.15 \sim
0.2$ for isotropic turbulence \citep{Yoshidaetal05, Lalescuetal13}.

For rotating turbulence, it is expected that 
the Rossby number will play a role. The Rossby number can be defined using the
small scale parameters, leading to the micro-scale Rossby number \citep{GodeferdMoisy15}
\be Ro_k = \frac{1}{2\Omega
\tau_k}. 
\ee
The large scale Rossby
number $Ro_\ell$ is defined as 
\be
Ro_\ell = \frac{u_{\text{rms}}}{2\Omega \ell}, 
\ee
where 
$u_{\text{rms}}\equiv  (\lal u'_i u'_i\ral/3)^{1/2}$ is the root-mean-square (RMS) velocity, 
and  
$\ell$ is the integral length scale defined \citep{Yoshidaetal05}
as
\be
\ell = \frac{\pi}{2 u_{\text{rms}}^2}\int_0^\infty k^{-1} E(k) dk, 
\ee
with $E(k)$ being the energy spectrum given by
\be
E(k) = \frac{1}{2}\sum_{\vert \k \vert = k} \lal
\hat{\bu}(\k,t)
\cdot \hat{\bu}^*(\k,t)\ral. 
\ee

Synchronisation of chaotic systems is related to the
conditional Lyapunov exponent (CLE) of the slave system. To introduce the concept,
let $\bu$ be the master velocity field, and $\bu^\delta$ be an
infinitesimal perturbation to \emph{the slaved modes} of $\bu$. Thus, by definition,
\be
\hat{\bu}^\delta(\k,t) = 0 \quad \text{for}\quad \vert \k \vert \le k_m. 
\ee
In the meantime,
$\bu^\delta$ obeys the linearised NSE equation
\be \label{eq:udel}
D_t \bu^\delta  + (\bu^\delta \cdot
\nabla)\bu + 2 {\bm \Omega} \times \bu^\delta = -
\nabla p^\delta + \nu \nabla^2 \bu^\delta + \f^\delta , 
\ee
and the continuity equation $\nabla \cdot \bu^\delta = 0$, where $p^\delta$
and $\f^\delta$ are 
the pressure perturbation and the perturbation in the forcing term, respectively.

The CLE, denoted by $\lambda(k_m)$, is defined as \citep{Boccalettietal02, NikolaidisIoannou22} 
\be \label{eq:lyadef}
\lambda(k_m) = \lim_{t\to \infty} 
\frac{1}{t}\log \frac{\Vert \bu^\delta(\x, t+t_0)\Vert}{\Vert \bu^\delta (\x, t_0)\Vert},
\ee
where $t_0$ is the initial time. 
$\lambda(k_m)$ is a function of the coupling wavenumber $k_m$. 
$\lambda(k_m=0)$ is the 
traditional (unconditional) Lyapunov exponent. 
As the unconditional Lyapunov exponent measures the average growth rate of a
generic velocity perturbation
over the turbulent attractor, $\lambda(k_m)$ measures the average growth rate
of the \emph{slaved modes} along a generic orbit $\bu(\x,t)$. 
It is known that for canonical chaotic systems
synchronisation occurs only when the CLE is negative \citep{Boccalettietal02}.
The same is confirmed 
for turbulent channel flows \citep{NikolaidisIoannou22}. One of the questions to be
addressed in present investigation is how the CLE $\lambda(k_m)$ depends on 
the Rossby number. 

For sufficiently large $t$, 
the velocity field $\bu^\delta$ 
gives a measure on the most unstable perturbation
to the slaved modes,
thus is also of interests. This velocity field is called the
Lyapunov vector \citep{OhkitaniYamada89, Bohretal98}, 
which is another quantity we will look into.  

An equation for $\Vert \bu^\delta \Vert$ can be deduced from Eq.
(\ref{eq:udel}), which reads
\be \label{eq:udelnorm}
\frac{d}{dt} \frac{\Vert \bu^\delta\Vert^2}{2} = \PP - \DD + \FF, 
\ee
where
\be \label{eq:pdf}
\PP \equiv \ol{- u_i^\delta u_j^\delta s_{ij}}, ~~
\DD \equiv \nu\ol{\ptl_j u_i^\delta \ptl_j u_i^\delta}, ~~
\FF \equiv \ol{f_i^\delta u^\delta_i}, 
\ee
are the production term, the dissipation term, and the forcing term,
respectively, and $s_{ij} = (\ptl_j u_i + \ptl_i u_j)/2$ is the strain rate
tensor.  
In the above expressions, the overline represents spatial average. The periodic
boundary condition has been used when deriving Eq. (\ref{eq:udelnorm}). 

By virtue of Eq. (\ref{eq:udelnorm}), we obtain
 \be \label{eq:gamma}
 \gamma(k_m,t) \equiv \frac{d}{dt} \log \Vert \bu^\delta \Vert= \frac{\PP - \DD + \FF}{\Vert
\bu^\delta\Vert^2} , 
\ee
where $\gamma(k_m,t)$ 
is called the local CLE. Using Eq.
(\ref{eq:gamma}), we can write 
\begin{align} 
  \lambda(k_m) &=\lim_{t\to\infty}\frac{1}{t} \int_{t_0}^{t+t_0} \gamma(k_m,t)
  dt \label{eq:lyadef1} 
  \\
  &=\lim_{t\to\infty}\frac{1}{t} \int_{t_0}^{t+t_0} \frac{\PP - \DD + \FF}{\Vert
  \bu^\delta\Vert^2} dt. \label{eq:lyadef2}
\end{align}
Therefore, 
the CLE 
$\lambda(k_m)$ is the long time average of $\gamma(k_m,t)$.
Whilst $\lambda(k_m)$ is a time-averaged quantity, $\gamma(k_m,t)$ 
fluctuates over time. Its variance contains information related to the
stability of the synchronised state, and as such is also of some interests.  

The rotation rate $\bm \Omega$ does not  
appear in Eq. (\ref{eq:lyadef2}). Therefore the rotation affects $\Vert
\bu^\delta\Vert$ only indirectly through its effects on the
production and dissipation terms. 
Insights into the effects of rotation on $\lambda(k_m)$, hence the
synchronisation process, can be obtained from analyses of $\PP$, $\DD$ as well
as $\FF$. For example, the production term $\PP$ crucially depends on the alignment
between $\bu^\delta $ and the eigenvectors of the strain rate
tensor $s_{ij}$, as well as the eigenvalues of $s_{ij}$.
These aspects will be looked into in our analyses.  

The CLEs can be calculated according to Eq.
(\ref{eq:lyadef}) once $\bu^\delta$ and $\bu$ are available.  
To find $\bu^\delta$, one might seek to integrate Eq. (\ref{eq:udel})
numerically. However,
this method suffers from the fact that $\bu^\delta$ normally grows
exponentially, so the numerics would fail before a sufficiently long time sequence of
$\bu^\delta$ could be obtained (which is needed to calculate
$\lambda(k_m)$).  
We thus use a common alternative method \citep{Wolfetal85,
BoffettaMusacchio17}, 
where we simulate two coupled flows
$\bu^{(1)}$ and $\bu^{(2)}$ concurrently in the same way as described
previously, except for two differences. Firstly, 
$\bu^{(2)}$ is initialised in such a way that the error
$\Delta(0)$ [c.f. Eq.(\ref{eq:error})] is a small quantity. Secondly, $\bu^{(2)}$ is re-initialised 
repeatedly after each short time interval $\Delta t$, by rescaling $\bu^{(2)}-\bu^{(1)}$
to restore $\Vert
\bu^{(2)}-\bu^{(1)} \Vert$ back to its initial (small) value. 
The interval
$\Delta t$ is chosen to be
short enough such that the evolution of $\bu^{(2)}-\bu^{(1)}$ can be
accurately approximated by the linearised NSE. 
As a result, $\bu^\delta\approx \bu^{(2)}-\bu^{(1)}$. Therefore, we have
\be \label{eq:gamma1}
\gamma \approx \frac{1}{\Delta t}\log \frac{\Vert \bu^{(2)}(\x, t+\Delta t) -
\bu^{(1)}(\x, t+\Delta t) \Vert}{\Vert \bu^{(2)}(\x, t) -
\bu^{(1)}(\x, t) \Vert}, 
\ee
from which we then can calculate $\lambda$ according to Eq. (\ref{eq:lyadef1}). 
For more
details on the algorithm, see, e.g., \citet{BoffettaMusacchio17}. 

We remark that Eq. (\ref{eq:lyadef2}) 
gives us a way to calculate the CLEs via $\PP$,
$\DD$ and $\FF$, once $\bu^\delta$ has been obtained in the way described
above. We used both methods to
cross check the numerics and found no difference in the results. 

Finally, we note that 
$\Delta(t)$ is the same as $\Vert \bu^\delta\Vert$ when the two flows
are synchronised. 
However, they are not interchangeable, because 
they would be significantly different 
when the two flows do not synchronise. 

\section{Numerical simulations and results \label{sect:results}}


  \begin{table} 
    \begin{center}
      \def~{\hphantom{0}}

\begin{tabular}{ccccccccccccccc}
  Case & Force & $N$ & $\Omega$ &  
   $\nu$ & $\delta t$ &
  $u_\text{rms}$ & $\ep$ & $\lambda$ & $\tau_k$ & $\eta$ & $Ro_k$ & $Re_\lambda$
  & $\ell$ & $Re_\ell$ \\  
  \hline
  F1N128$\Omega$01 & 1 & 128 & 0.1 & 0.0060 & 0.0025 & 0.44 & 0.05 & 0.59& 0.36
  & 0.046 & 14.43 & 43 & 1.74 & 128 \\
  F1N128$\Omega$05 & 1 & 128 & 0.5 & 0.0060 & 0.0025 & 0.54 & 0.10 & 0.51 & 0.25
  & 0.038 & 4.08 & 46 & 2.04 & 183 \\
  F1N128$\Omega$1\phantom{0} & 1 & 128 & 1.0 & 0.0060 & 0.0025 & 0.55 & 0.16 & 0.41 & 0.20
  & 0.034& 2.58 & 38 & 2.18 & 200 \\
  F1N128$\Omega$5\phantom{0} & 1 & 128 & 5.0 & 0.0060 & 0.0006 & 0.58 & 1.07 &
  0.17 & 0.08 & 0.021 & 1.34 & 16 & 2.32 & 224 \\
  F1N192$\Omega$01 & 1 & 192 & 0.1 & 0.0044 & 0.0015 & 0.47 & 0.05 & 0.54 &
  0.29 & 0.036 & 16.85 & 58 & 1.69 & 181 \\
  F1N192$\Omega$05 & 1 & 192 & 0.5 & 0.0044 & 0.0015 & 0.53 & 0.10 & 0.43 &
  0.22 & 0.030 & 4.77 & 52 & 2.00 & 240 \\
  F1N192$\Omega$1\phantom{0} & 1 & 192 & 1.0 & 0.0044 & 0.0015 & 0.53 & 0.16 & 0.34 &
  0.17 & 0.027 & 3.02 & 41 & 2.16 & 260\\
  F1N192$\Omega$5\phantom{0} & 1 & 192 & 5.0 & 0.0044 & 0.0004 & 0.66 & 1.04 &
  0.17 & 0.07 & 0.017 & 1.54 & 26 & 2.31 & 347 \\
  F1N256$\Omega$01 & 1 & 256 & 0.1 & 0.0030 & 0.0013 & 0.46 & 0.05 & 0.44 &
  0.24 & 0.027 & 20.41 & 68 & 1.63 & 250\\
  F1N256$\Omega$05 & 1 & 256 & 0.5 & 0.0030 & 0.0013 & 0.49 & 0.10 & 0.33 & 0.17
  & 0.023 & 5.77 & 54 & 1.98 & 323\\
  F1N256$\Omega$1\phantom{0} & 1 & 256 & 1.0 & 0.0030 & 0.0013 & 0.50 & 0.16 & 0.27 & 0.15
  & 0.020 & 3.65 & 45 & 2.15 & 358 \\
  F2N128$\Omega$01 & 2 & 128 & 0.1 & 0.0060 & 0.0025 & 0.50 & 0.05 & 0.67 &
  0.35 & 0.046 & 14.43 & 56 & 1.66 & 138 \\
  F2N128$\Omega$05 & 2 & 128 & 0.5 & 0.0060 & 0.0025 & 0.38 & 0.05 & 0.51 &
  0.35 & 0.046 & 2.89 & 32 & 2.15 & 136 \\
  F2N128$\Omega$1\phantom{0} & 2 & 128 & 1.0 & 0.0060 & 0.0025 & 0.38 & 0.05 &
  0.51 & 0.35 & 0.046 & 1.44 & 32 & 2.29 & 145 \\
  F2N192$\Omega$01 & 2 & 192 & 0.1 & 0.0044 & 0.0015 & 0.50 & 0.05 & 0.57 &
  0.30 & 0.036 & 16.85 & 65 & 1.57 & 178 \\
  F2N192$\Omega$05 & 2 & 192 & 0.5 & 0.0044 & 0.0015 & 0.40 & 0.05 & 0.46 &
  0.30 & 0.036 & 3.37 & 42 & 2.04 & 186 \\
  F2N192$\Omega$1\phantom{0} & 2 & 192 & 1.0 & 0.0044 & 0.0015 & 0.40 & 0.05 &
  0.46 & 0.30 & 0.036 & 1.69 & 42 & 2.25 & 204 \\
\end{tabular}
  \caption{\label{tab:cases} Parameters for the cases.
  $N^3$: the number of grid points.
  $\Omega$: the rotation rate. 
  $\nu$: viscosity. 
  $\delta t$: time step size. 
  $u_\text{rms}$: root-mean-square velocity. 
  $\epsilon$: mean energy dissipation rate. 
  $\eta$: Kolmogorov length scale. 
  $\lambda$: Taylor length scale. 
  $\tau_k$: Kolmogorov time scale.
  $\eta$: Kolmogorov length scale.  
  $Ro_k$: micro-scale Rossby number. 
      $Re_\lambda\equiv u_\text{rms}\lambda/\nu$: the Taylor micro-scale Reynolds number.  
    $\ell$: the integral length scale. 
      $Re_\ell \equiv u_\text{rms}\ell/\nu$: the integral scale Reynolds number. 
  }
\end{center}
\end{table}

Eq. (\ref{eq:nse}) is numerically integrated in the Fourier space 
with the pseudo-spectral method. As is common for the simulation of rotating turbulence, 
the Fourier component $\hat{\bu}$ is decomposed into 
helical modes $a_+(\k,t)$ and $a_-(\k,t)$ 
and the equations for $a_+$ and $a_-$ are integrated. $\hat{\bu}$ are
reconstructed from $a_\pm$ using the helical decomposition. 
With this approach, 
the different components of the Coriolis force are 
decoupled in the equations for $a_\pm$, so that they (as well as the viscous
diffusion term) can be treated with
an integration factor which increases the stability of the algorithm.

The advection term is de-aliased according to the two-thirds rule 
so that the maximum
effective wavenumber is $4\pi/3N$ where $N^3$ 
is the number of grid
points in the simulations. 
Time
stepping is conducted with an explicit second order Euler scheme with a first-order
predictor and a corrector based on the trapezoid rule \citep{Lietal20}. 

Simulations with $N^3=128^3$, $192^3$ and $256^3$ grid points are conducted.
The majority of the analyses focuses on rotation rates $\Omega = 0.1$,
$0.5$ or $1$. For the flows driven by Kolmogorov forcing, 
test cases with $\Omega = 5$ are also simulated to demonstrate that
two-dimensionalization has happened at this rotation rate.  
Table \ref{tab:cases} summarizes the parameters for all the cases. We label
the cases with a code of the form `F$a$N$b$$\Omega$$cd$' or 
`F$a$N$b$$\Omega$$c$', where letters $a$ to $d$ are
numbers. The code records the type of
forcing (with $1$ for Kolmogorov forcing and $2$ for constant power
forcing), the number of grid points, and the rotation rate of the
case. For each case in Table \ref{tab:cases}, 
sometimes multiple simulations are conducted with different $k_m$. 
To differentiate these simulations, we append `K' and the value of
$k_m$ to the end of
the code. Thus, for example, case F1N128$\Omega$01K5 is a $128^3$ simulation driven by 
Kolmogorov forcing with rotation rate being $0.1$ and the coupling wavenumber
$k_m$ being $5$, whereas case F2N256$\Omega$1K7 is a $256^3$ simulation driven by
constant power forcing with rotation rate being 1 and $k_m = 7$. 

Multiple realisations of a case are simulated in some cases 
to obtain convergent statistics for some quantities (e.g., for the variance of the
CLEs shown in Fig. \ref{fig:lya_rot_var}).

Since the main focus of this investigation is on the effects of
rotation, the simulations have only moderate Reynolds
numbers. On the other hand, 
Table \ref{tab:cases} shows that the micro-scale Rossby number in some cases are as small as
$1.34$ and $1.44$. Therefore the range of cases does cover flows where 
rotation will have significant impacts on 
the small scales. 

The CLEs are calculated according to the method explained in Section
\ref{sect:eqn}. $\bu^{(1)}$ is initialized with a fully
developed velocity field.  $\bu^{(2)}$ is initialised with $\bu^{(1)} + \delta \bu$ 
where $\delta \bu$ is composed of random numbers uniformly distributed in the interval $[0, 10^{-6}
u_\text{rms}]$. When we calculate the CLEs with a threshold wavenumber $k_m$, 
$\bu^{(2)}$ is coupled with $\bu^{(1)}$ such that Eq. (\ref{eq:coupling}) is
true at all time. 
The time interval $\Delta t$ between rescaling the magnitude
of 
$\bu^{(2)}-\bu^{(1)}$ is $\Delta t
\approx 0.1 \tau_k$. These values are approximately the same as the ones used
in \citet{BoffettaMusacchio17}. 

\subsection{Basic features of the flow fields}

\bfig
\centering
\ig[width=0.48\lnw]{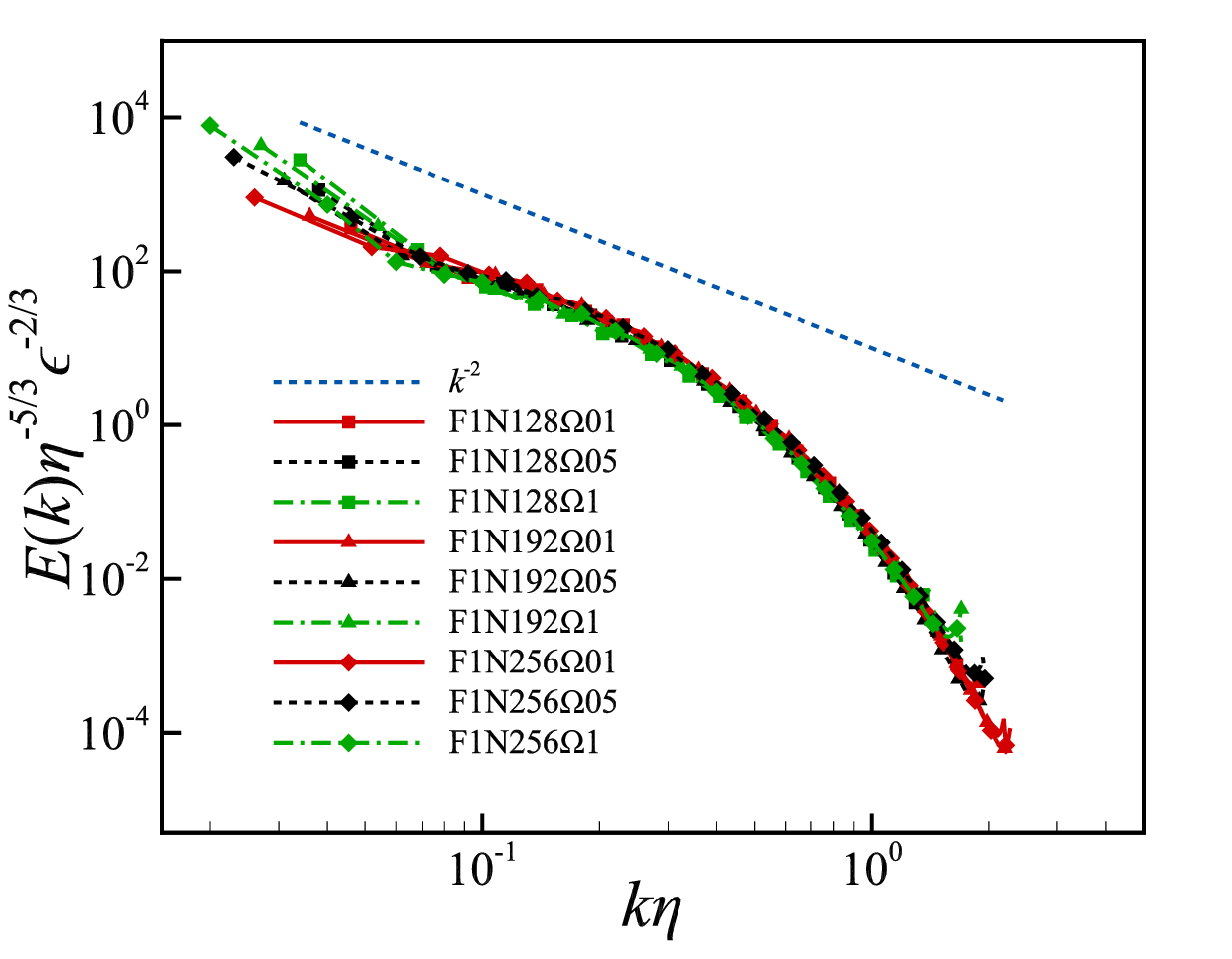} %
\ig[width=0.48\lnw]{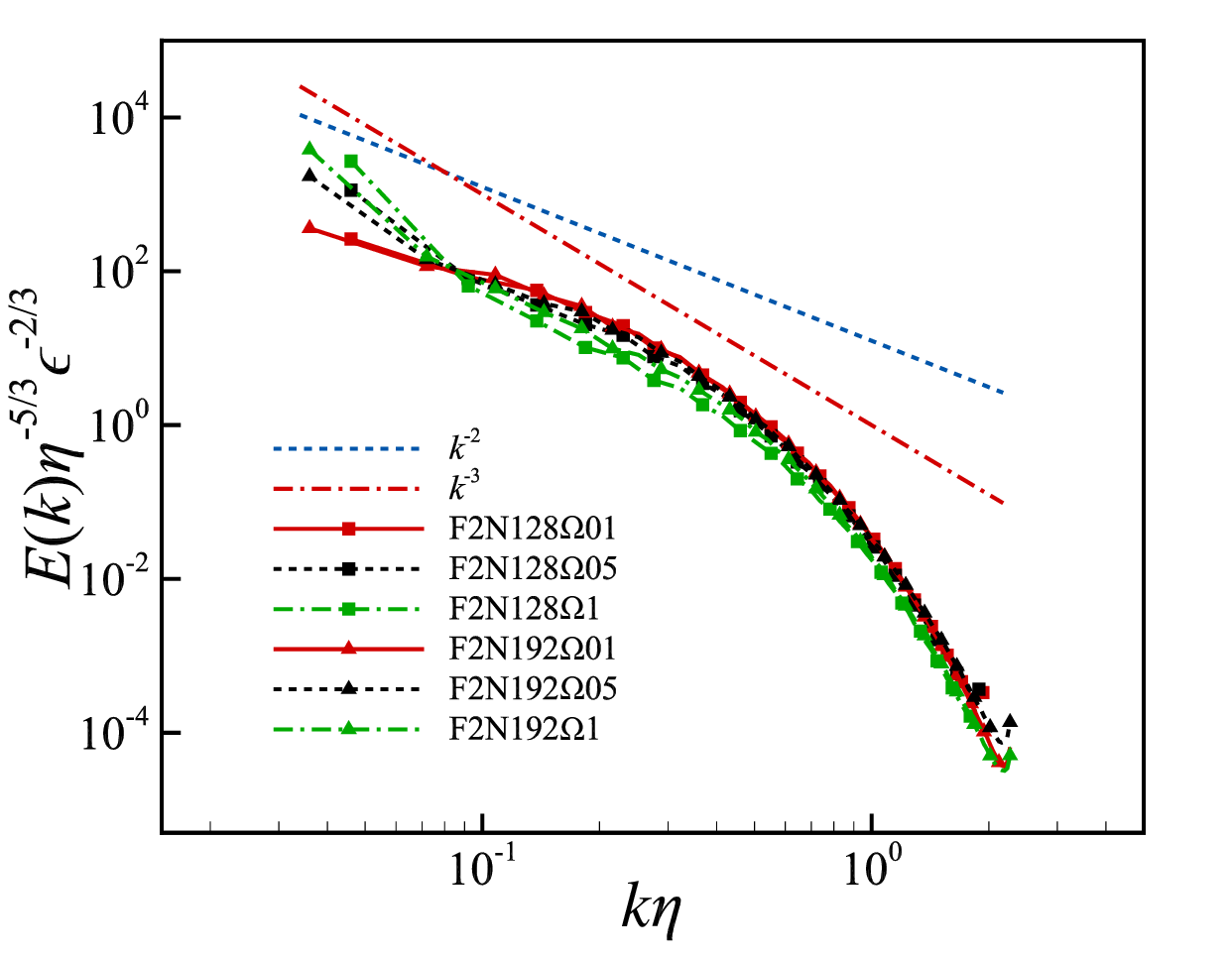}
\caption{\label{fig:Ek} The energy spectra. Left: cases with Kolmogorov forcing. 
Right: cases with constant power forcing. Dashed line without symbols: 
the $k^{-2}$ power law. Dash-dotted line without symbols: the $k^{-3}$ power
law.}
\efig

\bfig
\centering
\ig[width=0.48\lnw]{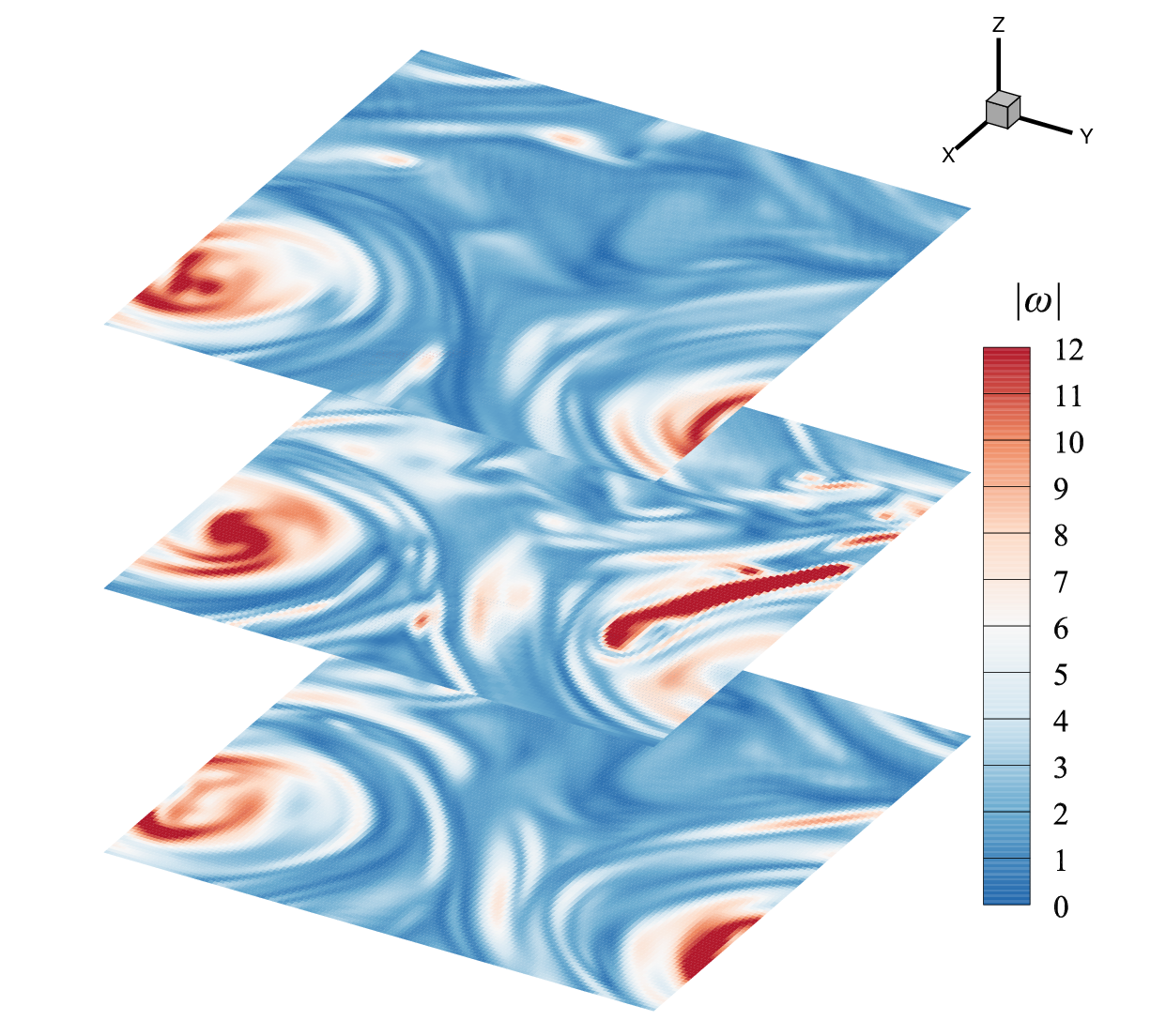} %
\ig[width=0.48\lnw]{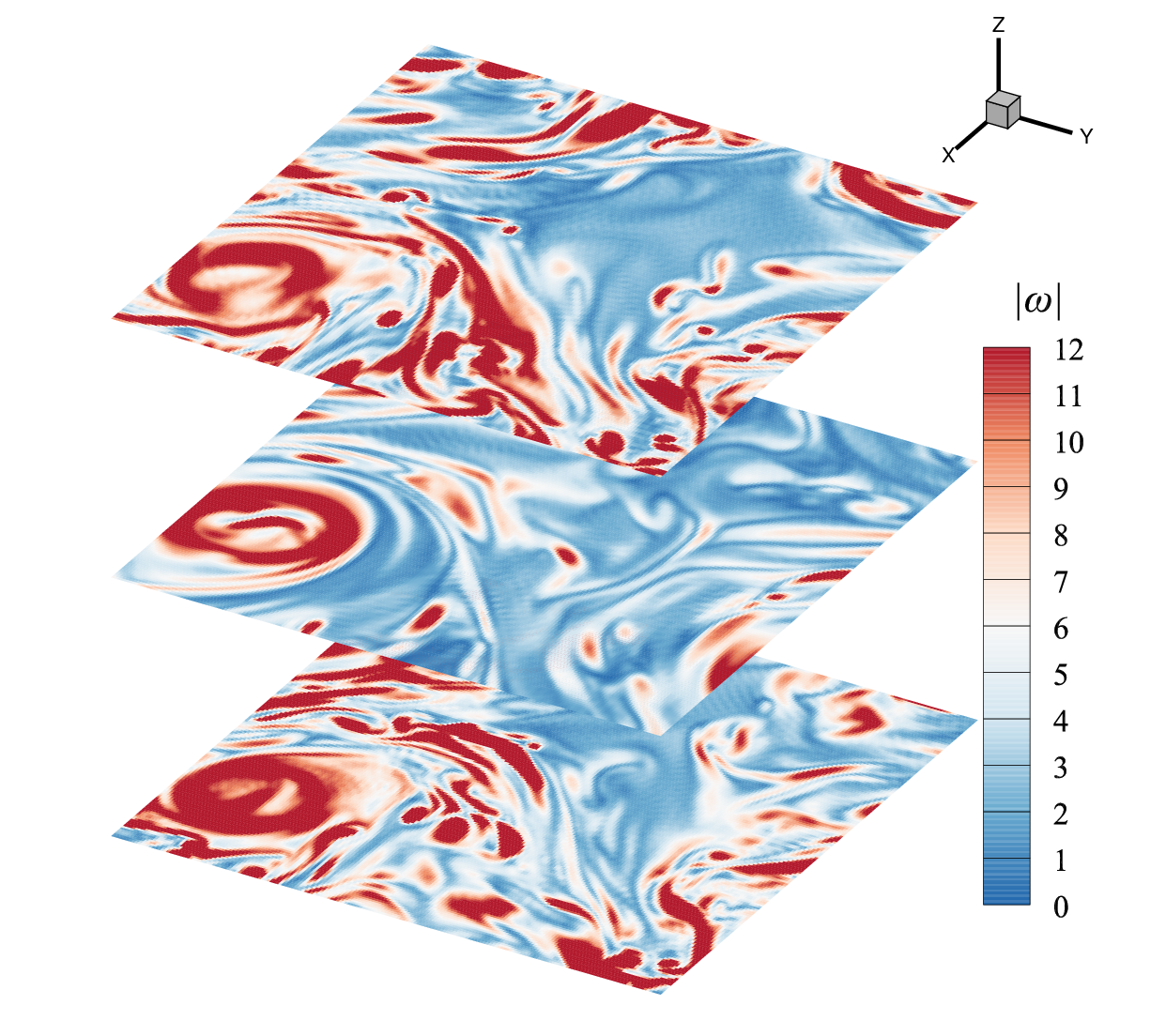}
\caption{\label{fig:vor} Snapshots of $\vert
\bo\vert$ distribution taken at three horizontal layers at the same time $t$ for $\Omega = 1$
with Kolmogorov forcing. Left:
from a case with $N=128$. Right: from a case with $N=192$. }
\efig


We present some results in this subsection to illustrate the basic features of
the flow fields. The energy spectra normalised by Kolmogorov parameters
are shown in Fig. \ref{fig:Ek}. For the flows driven by Kolmogorov forcing
shown in the left panel, the normalised spectra collapse onto a single curve except for the
few lowest
wavenumbers. 
At the lowest wavenumbers, the spectra increase with the rotation rate, which
shows increased energetics for the large scales, consistent with our
understanding of rotating turbulence. 

The Reynolds number for the flows is relatively small so
no clear inertial range can be identified. Nevertheless,
the spectra appear to be consistent with the $k^{-2}$
scaling law which has been reported in previous research \citep{YeungZhou98,
DallasTobias16}.  

For the flows driven by constant power forcing, similar behaviours
are observed for lower rotation rates, as shown in the right panel. 
However, for $\Omega = 1$,
the spectra have
steeper slopes in the mid-wavenumber range, and they appear to be more
consistent with the $k^{-3}$ power law. 
The spectra in the 
dissipation range also appear to drop off
at a faster rate.  
The
contrast between the left and right panels shows that the forcing
terms can
lead to significant quantitative differences in the flows. 

In both flows, energy pile-up is observed at the
lowest wavenumber end of the spectra, and the 
pile-up increases slightly with the rotation rate.  
The pile-up is an indication of 
the emergence of large scale columnar vortices, which is a common feature of rotating
turbulence. Columnar vortices are indeed visually observable in our simulations with the larger
rotation rates, which are illustrated 
in Fig. \ref{fig:vor} for two simulations with $\Omega = 1$. 
The figure shows a snapshot of the distribution of
$\vert\bo\vert$ on three horizontal cross sections of the flow domain, where
$\bo \equiv \nabla \times \bu$ is the vorticity. 
A columnar vortex is visible at the left corner in both flows shown in the two
panels. 
The left
panel shows a simulation with a smaller Reynolds number. In this case, the
diameter of the columnar vortex is roughly half of the size of the domain. 
For the
flow with a larger Reynolds number (right panel), 
the background vorticity is stronger,
and the columnar vortex appears to
be slightly smaller in size but it is still clearly visible. 
We will not show the results for other rotation rates, 
but we can confirm that columnar vortices are also quite
prevalent for $\Omega = 0.5$, while they are rare for $\Omega = 0.1$.

\bfig
\centering
\ig[width=0.48\lnw]{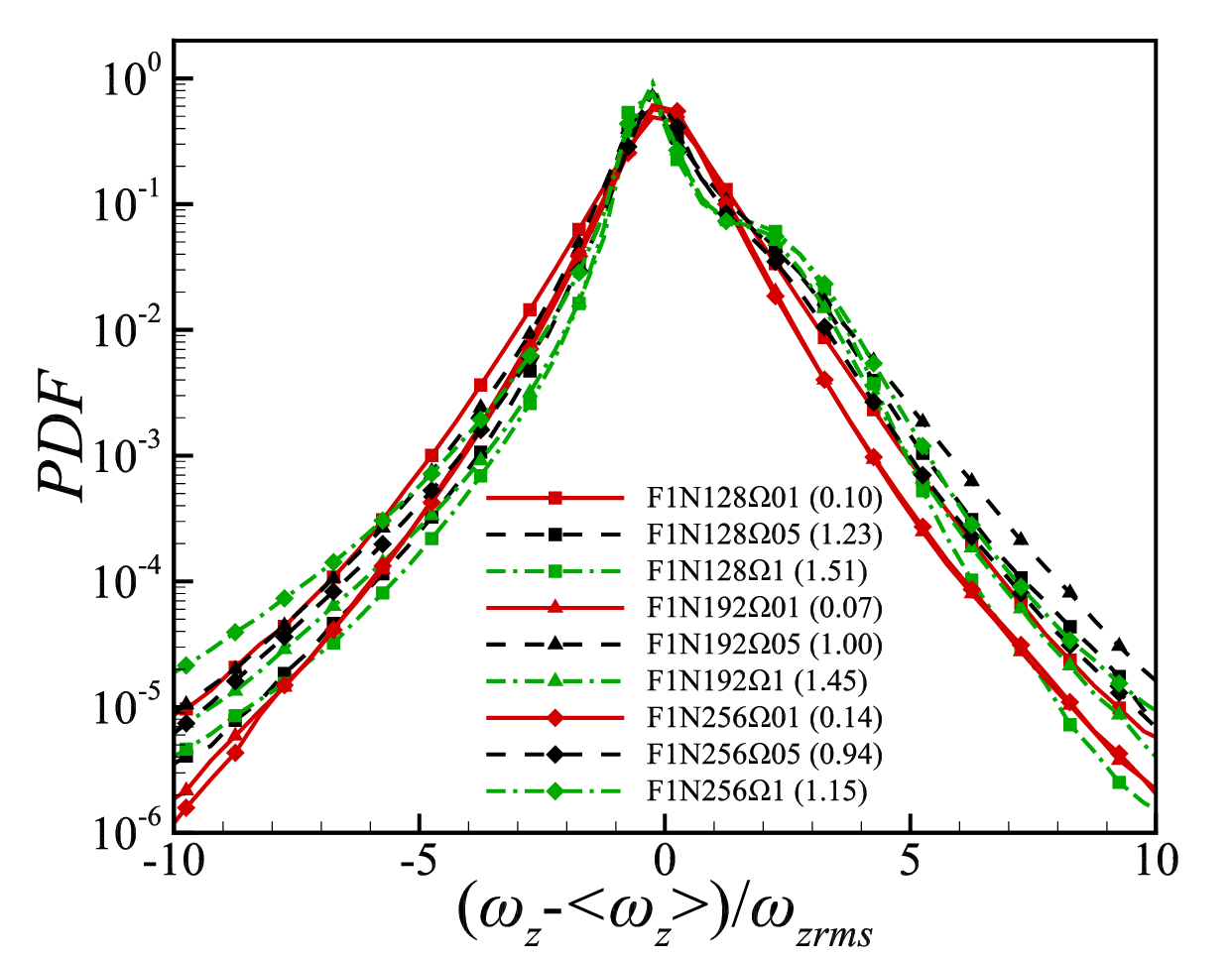}
\ig[width=0.48\lnw]{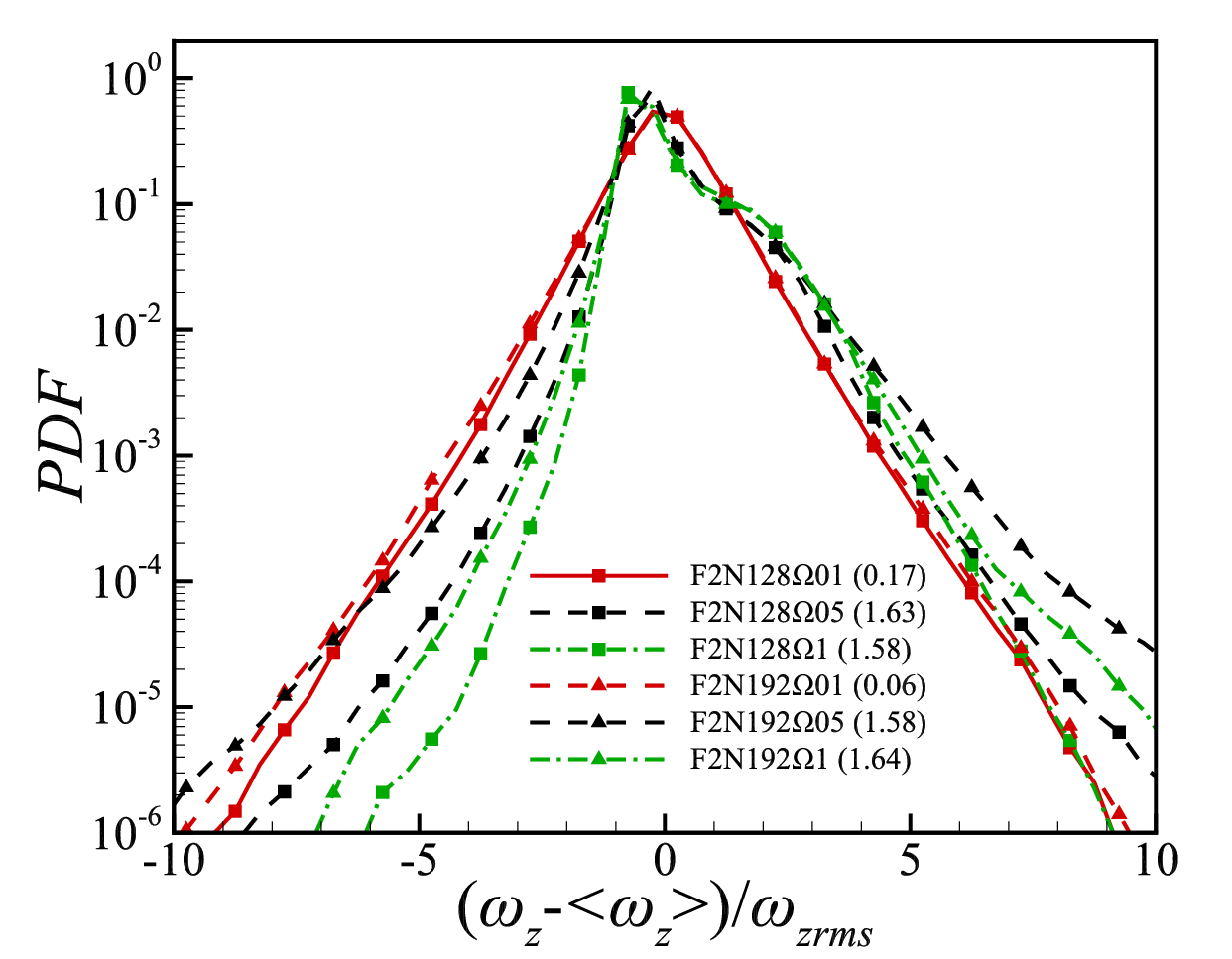}
\caption{\label{fig:vor_skewness} The PDF of the vorticity component along the
rotation axis $\om_z$.
Left: cases with Kolmogorov forcing. Right: cases with constant power forcing.
}
\efig

The probability density function (PDF) of the vorticity component
along the rotation axis is 
also of interest because it is well known that the PDF displays a positive
skewness \citep{Bartelloetal94, Morizeetal05} in rotating turbulence, 
due to the prevalence of cyclonic vortices over the
anti-cyclonic ones. 
The skewness emerges as rotation is introduced, 
peaks at an intermediate rotation rate, and then decreases when the
rotation rate further increases as
the flow is two-dimensionalized under strong rotation. 
The PDFs for our simulations 
are plotted in Fig. \ref{fig:vor_skewness}. The PDFs are indeed skewed towards the
positive values with the corresponding skewness given in the parentheses. 
For flows driven by constant power forcing with $N=128$, the skewness for
$\Omega = 1$ is slightly smaller than that for $\Omega = 0.5$. In other cases,
the skewness increases with the rotation rate. These PDFs show, from another
angle, that the
effects of rotation are clearly significant. 

\bfig
\centering
\ig[width=0.32\lnw]{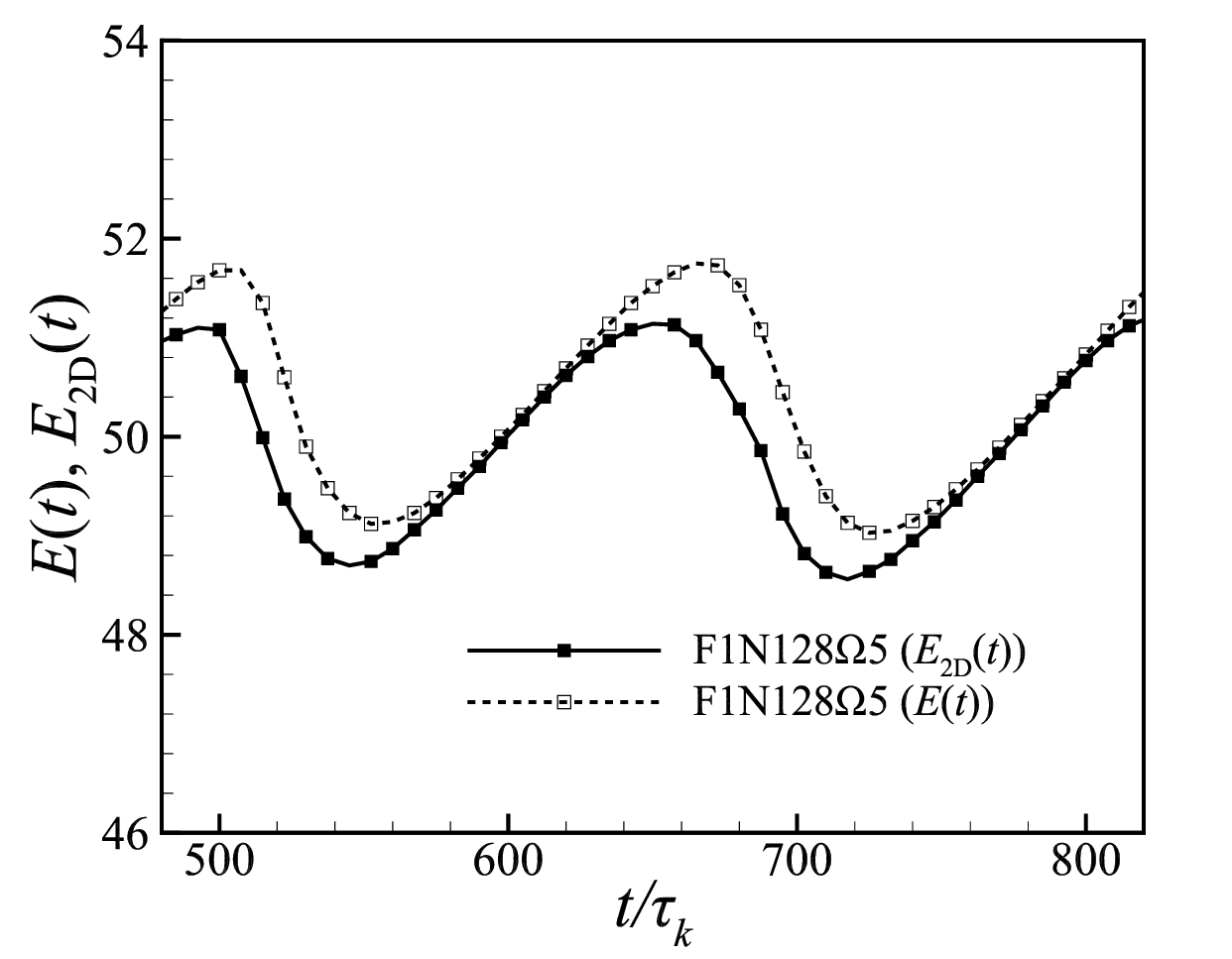}
\ig[width=0.32\lnw]{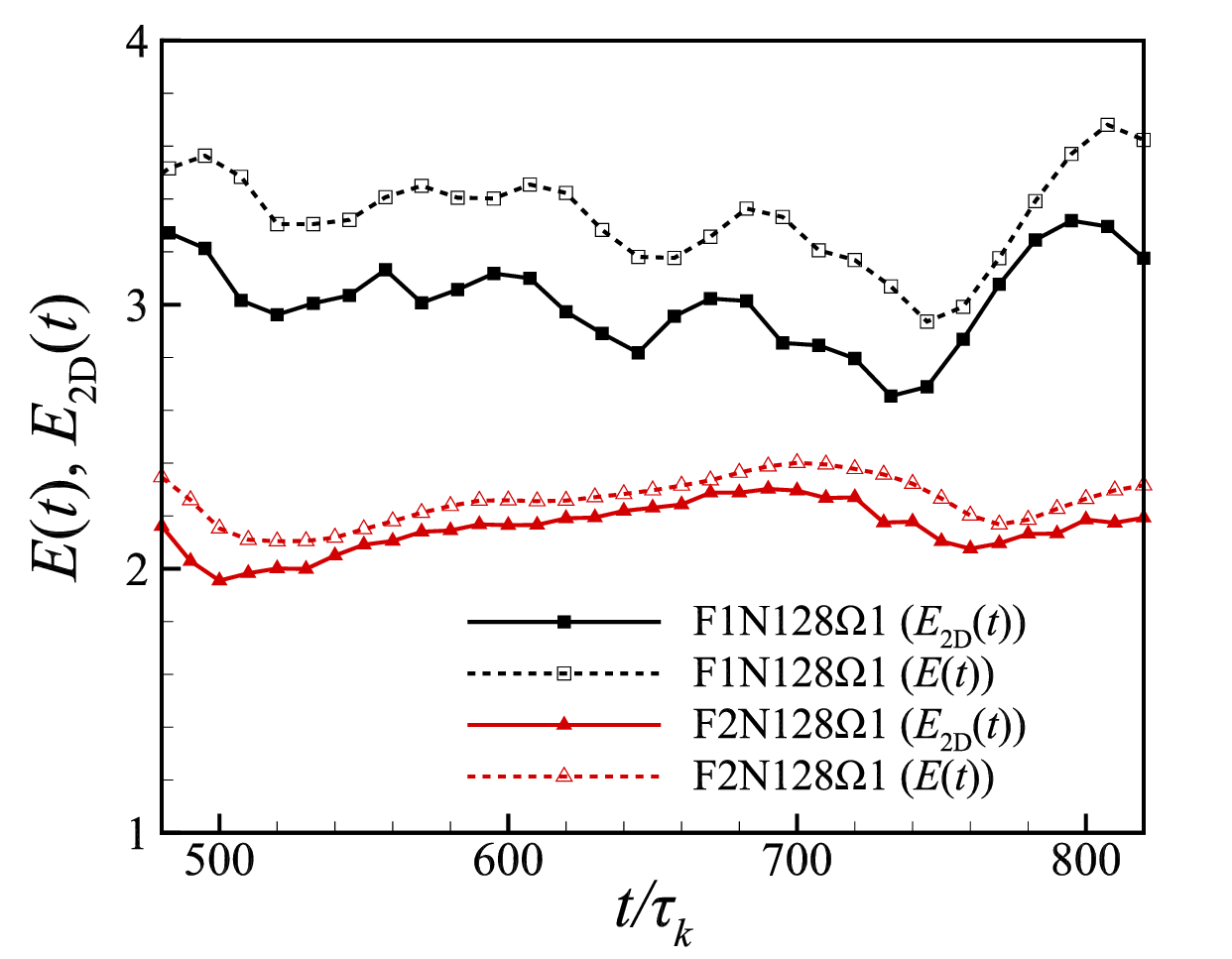}
\ig[width=0.32\lnw]{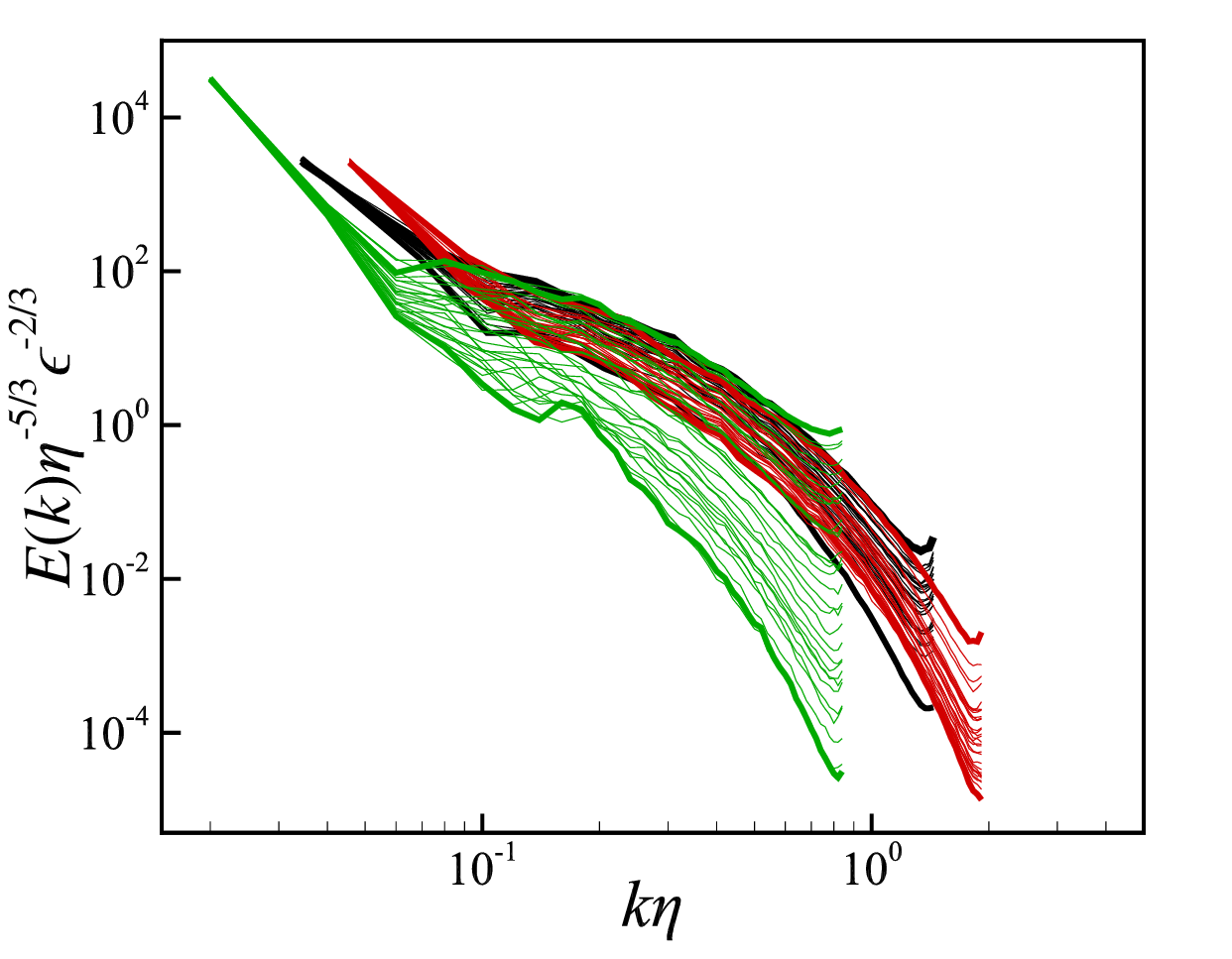}
\caption{\label{fig:rot5} Left: kinetic energy of the flow field and that in
the two dimensional modes for $\Omega = 5$. Middle: those for cases with
$\Omega = 1$ and $N=128$. Right: instantaneous energy spectra for cases with
$N=128$, plotted every $10\tau_k$
for time spanning $300\tau_k$. Green lines:
$\Omega = 5$; red lines: $\Omega = 1$ with constant power forcing; black
lines: $\Omega = 1$ with Kolmogorov forcing.}
\efig
Table \ref{tab:cases} shows that, compared with the flows driven by constant
power forcing, those 
driven by Kolmogorov forcing tend to have larger micro-scale Rossby
numbers $Ro_k$
for a given rotation rate $\Omega$. In order to obtain even smaller $Ro_k$ for
the latter flows, we computed a few test cases with $\Omega = 5$, and
found that the flows are strongly two-dimensionalised at this rotation rate.
Let $E_{2D}(t)$ be the kinetic energy in the two dimensional Fourier modes with
$k_z = 0$, and $E(t)$ be the total kinetic energy, i.e.,
\be
E_{2D}(t) = \frac{1}{2}\sum_{\{\k:k_z = 0\}} \hat{\bu}(\k,t) \cdot
\hat{\bu}^*(\k,t), \quad 
E(t) = \frac{1}{2}\sum_{\k} \hat{\bu}(\k,t) \cdot
\hat{\bu}^*(\k,t).
\ee
The results for $E_{2D}(t)$ and $E(t)$ for the flows with $\Omega = 5$ (i.e.,
cases F1N128$\Omega$5 and F1N192$\Omega$5) are
shown in the left panel of Fig. \ref{fig:rot5}. As a comparison, the results for $\Omega = 1$ are shown
in the middle panel. It can be observed that, for $\Omega = 5$, both $E(t)$ and $E_{2D}(t)$ are an
order of magnitude higher than for $\Omega = 1$, and almost all energy is
contained in the two dimensional modes as $E_{2D}(t)$ deviates from $E(t)$
only slightly. There are regular periods of time in which $E_{2D}(t)$ is
indistinguishable from $E(t)$. These behaviours suggest that, at $\Omega = 5$, the flows are
quasi-two-dimensionalised with large-scale, 2D columnar vortices, where instability sets in
periodically which leads to temporary small deviation between $E_{2D}(t)$ and
$E(t)$. A detailed discussion of this process can be found in
\citet{Alexakis15}. 
The energy spectra for the flow with $\Omega=5$ at various times are shown in the right panel of Fig.
\ref{fig:rot5} in green lines, together with those for $\Omega = 1$ with
both forcing terms (shown in black or red). The high wavenumber ends of the spectra swing
violently over time, in a range spanning five orders of magnitude. Though
oscillations are also seen in the spectra for the flows with $\Omega = 1$, the
amplitude is much smaller. 

To summarize, the results in this subsection show that
for $\Omega = 0.1$, $0.5$ and $1$, the flows are
still predominantly turbulent while displaying strong effects of rotation. 
The
flows where $\Omega = 5$, on the other hand, appear to be mostly
two-dimensionalised and only display weakly turbulent behaviours. 
We will limit our interests in the synchronisation of flows where turbulence dominates.
Therefore we will focus on the first three rotation
rates, 
and the cases with
$\Omega = 5$ will not be discussed further in what follows. 

\subsection{Synchronisation error}

\bfig
\centering
\ig[width=0.48\lnw]{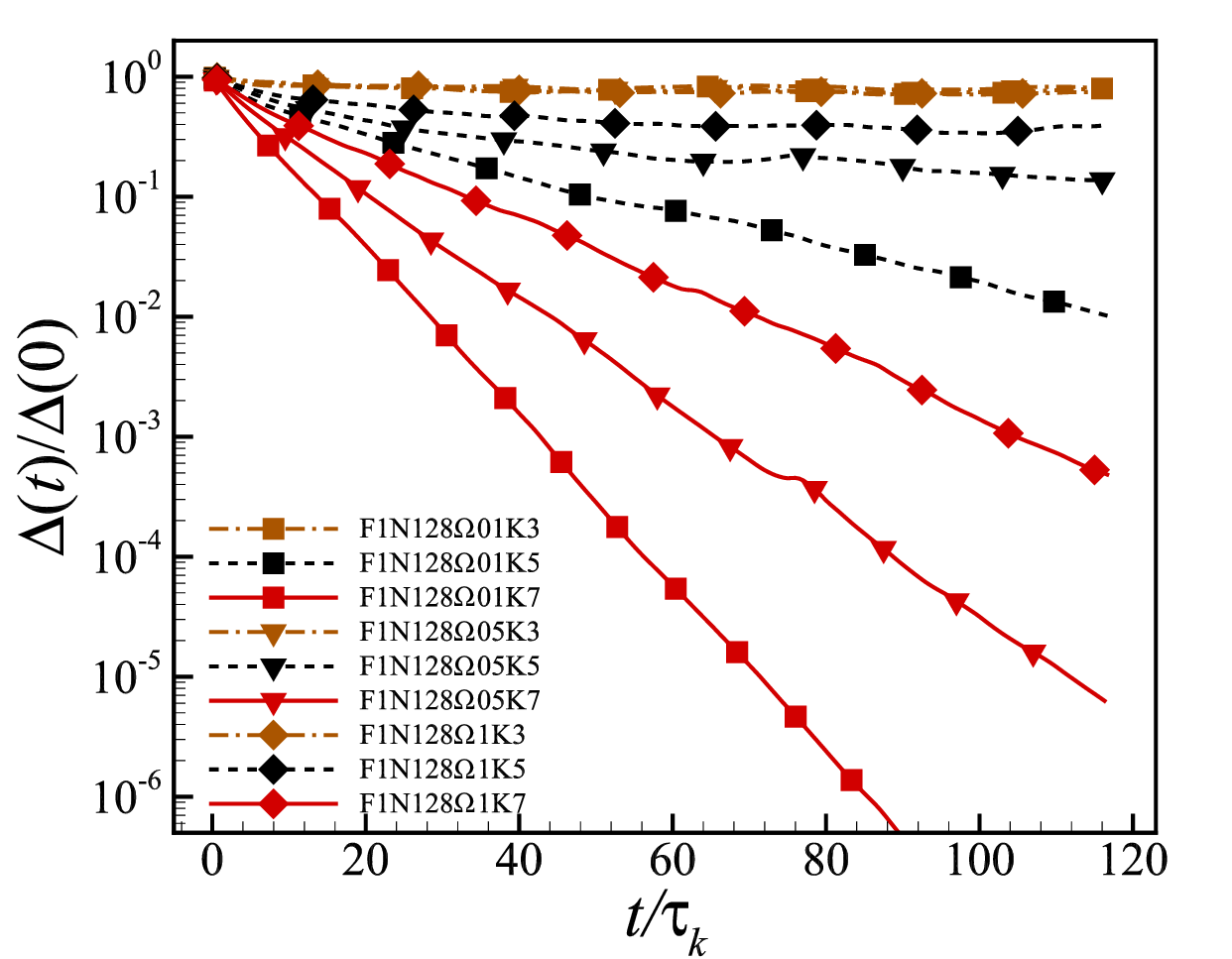}
\ig[width=0.48\lnw]{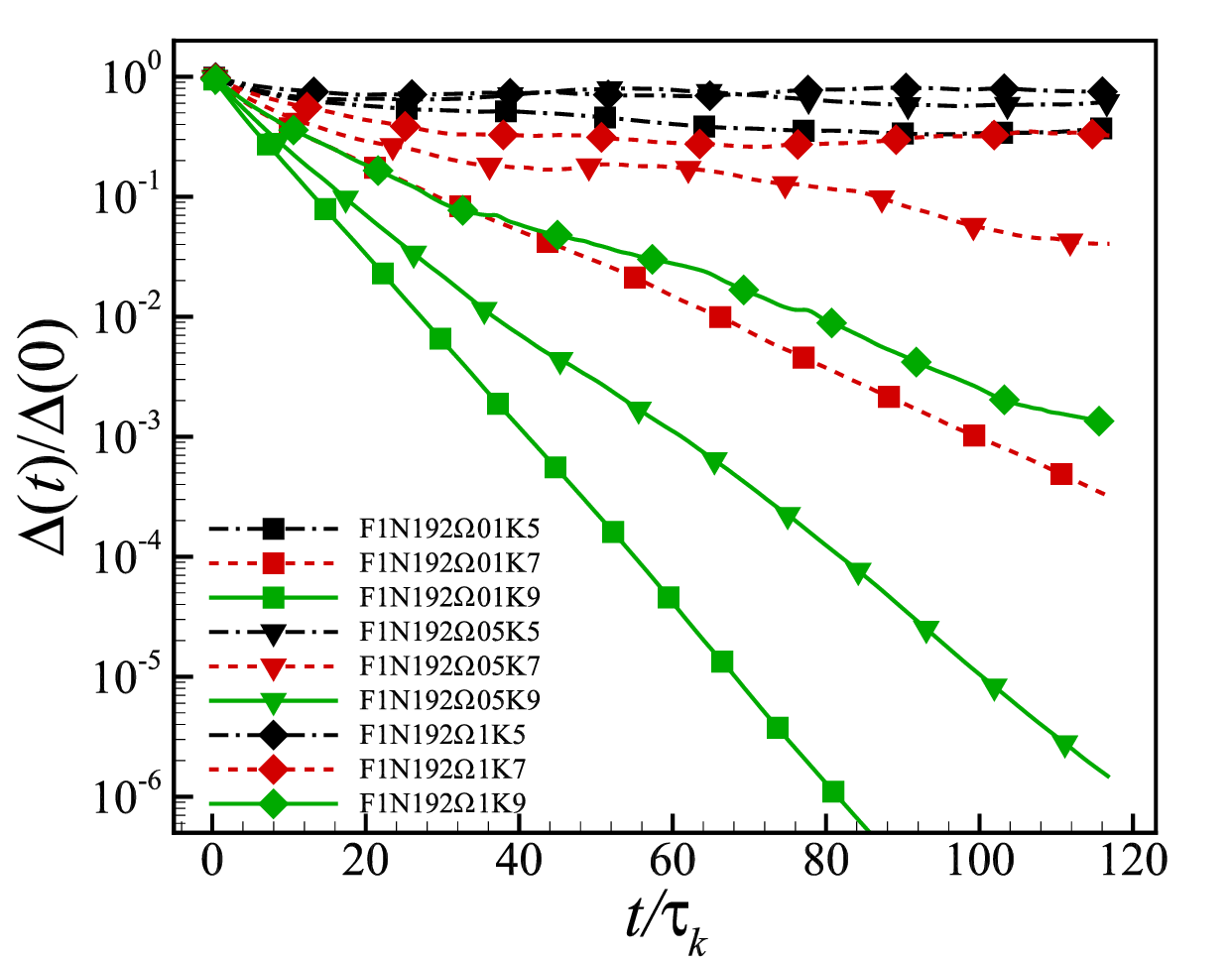}

\ig[width=0.48\lnw]{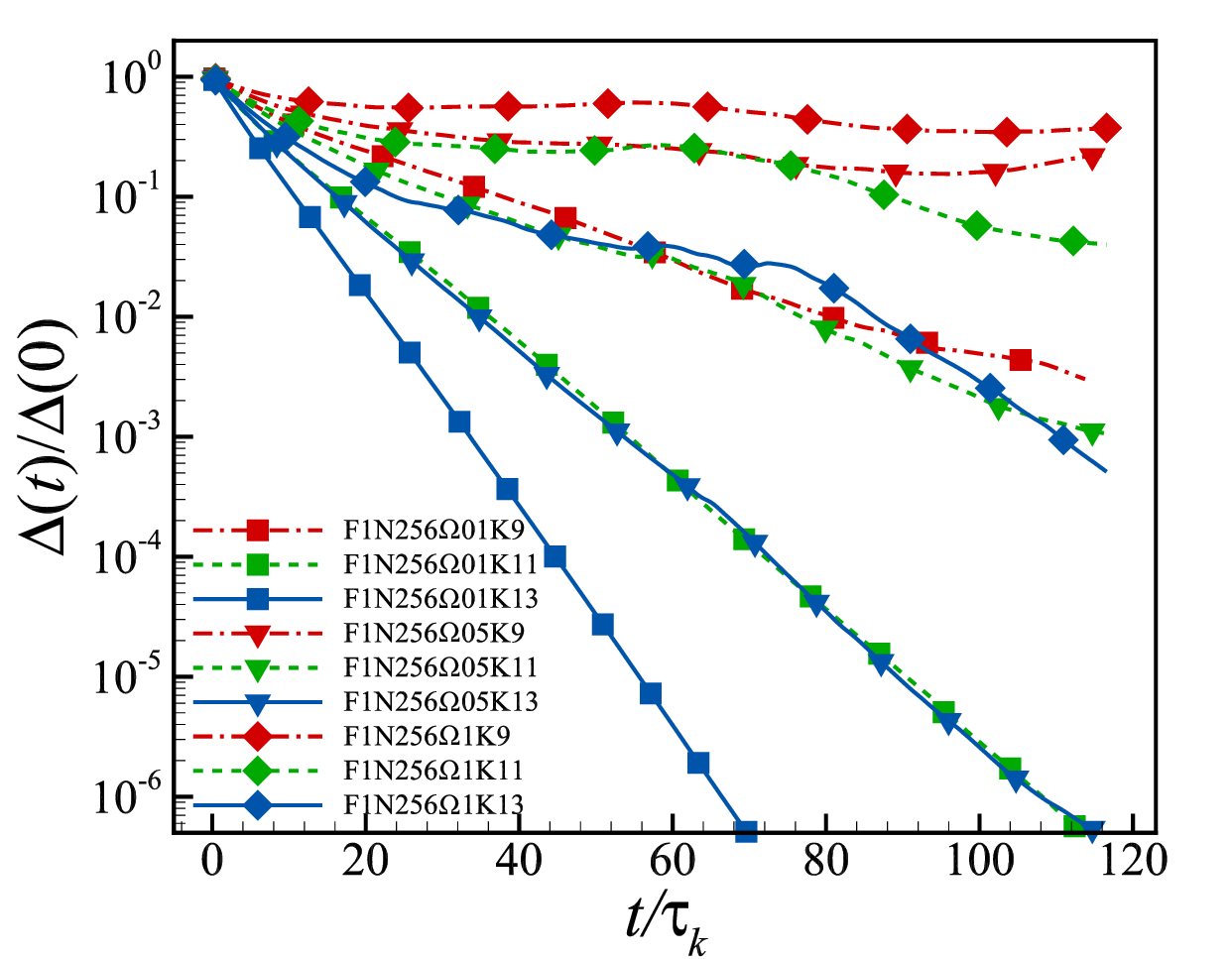}
\ig[width=0.48\lnw]{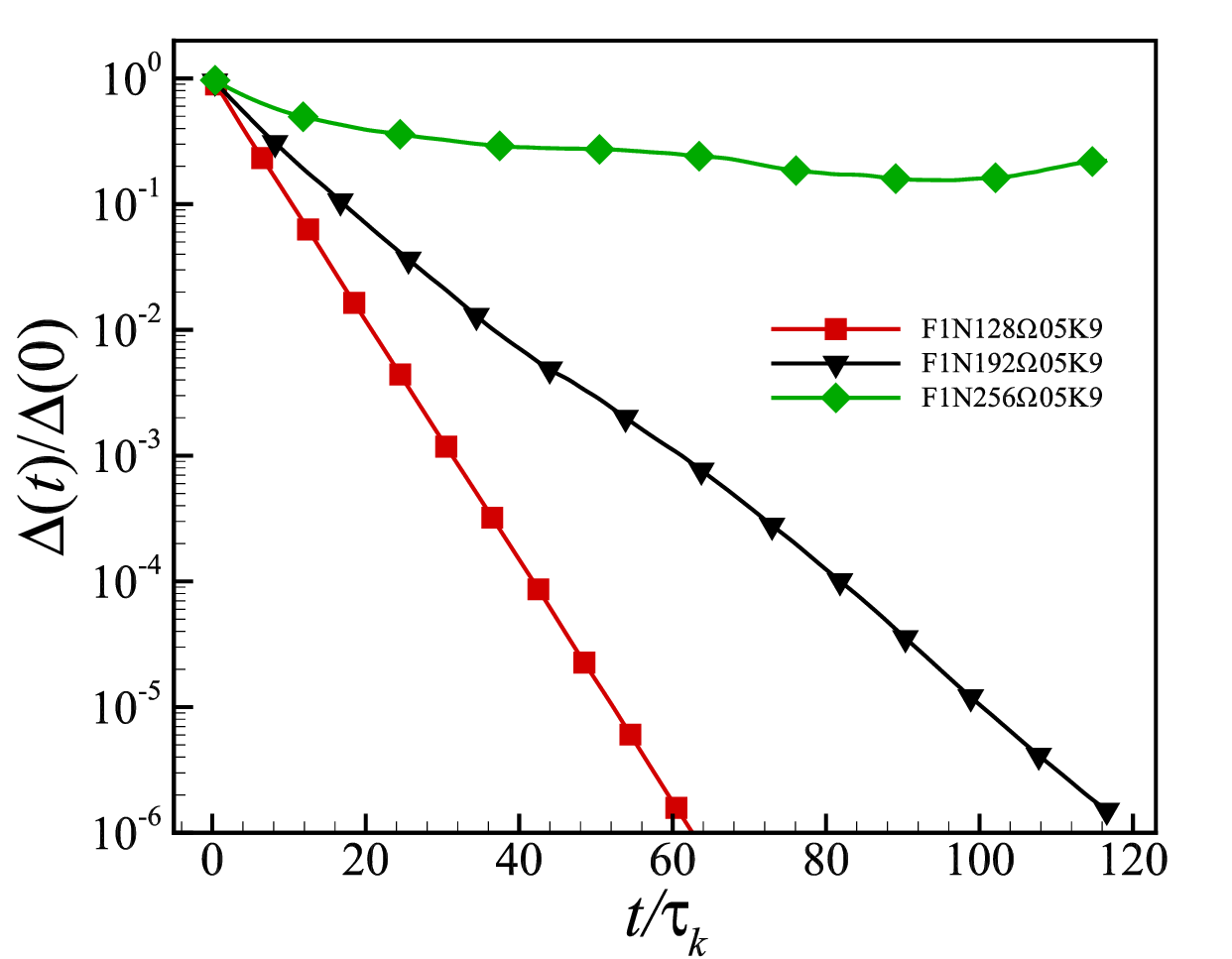}
\caption{\label{fig:decay_f2_tauk} The normalised synchronisation error
$\Delta(t)/\Delta(0)$ for the cases
with Kolmogorov forcing. Top-left: $N=128$. Top-right: $N=192$. Low-left:
$N=256$. Low-right: comparison between cases with different Reynolds numbers. }
\efig

\bfig
\centering
\ig[width=0.48\lnw]{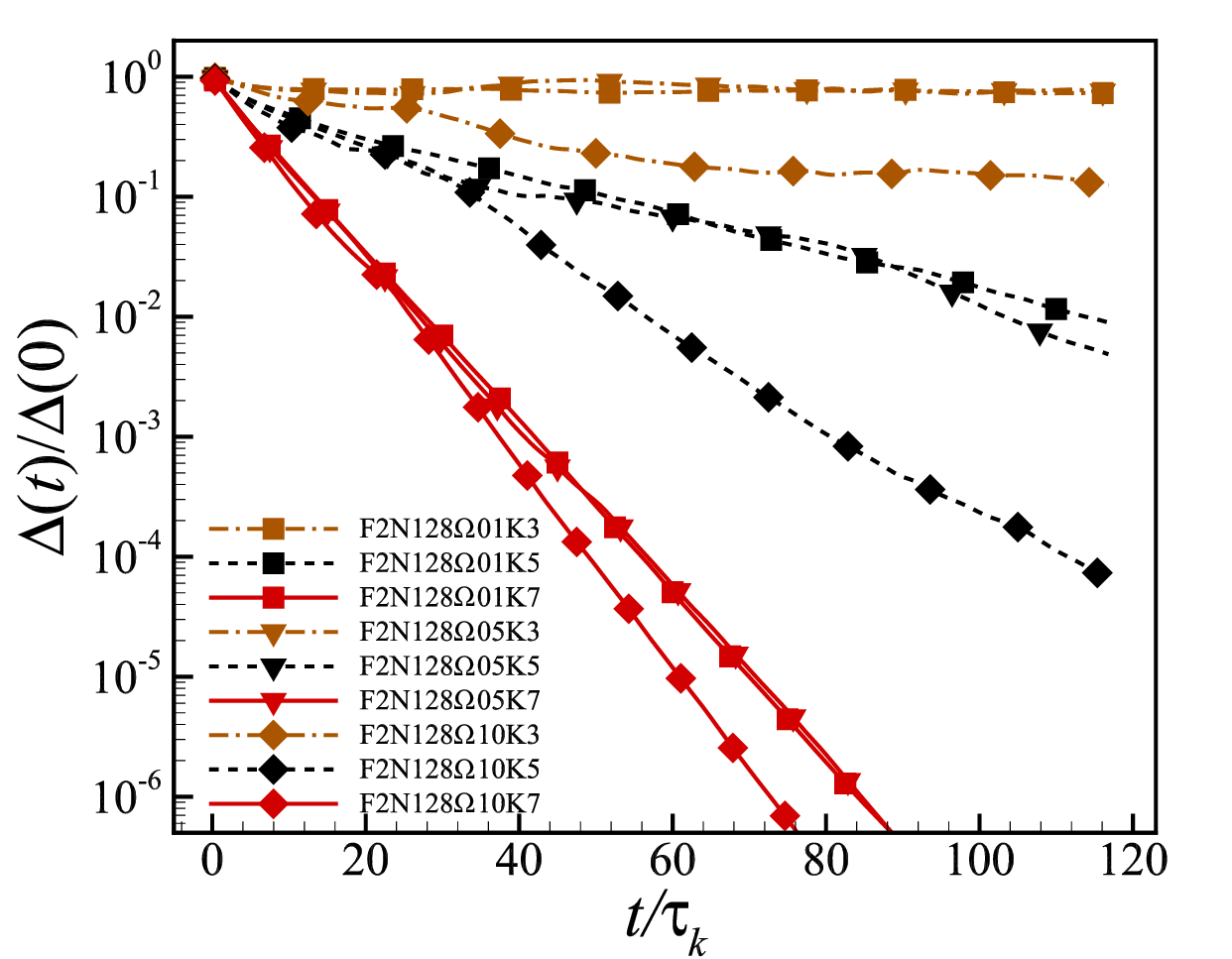}
\ig[width=0.48\lnw]{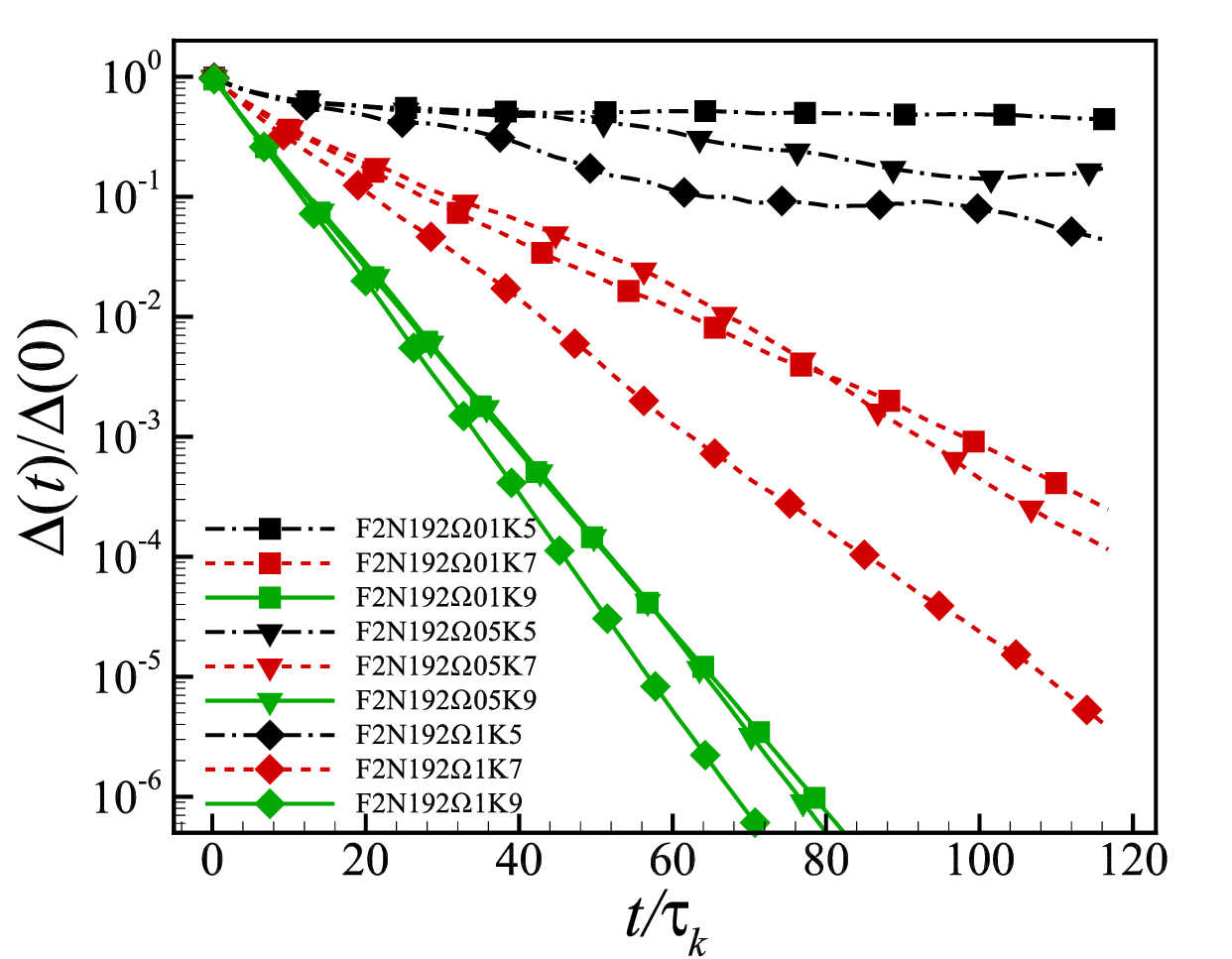}
\caption{\label{fig:decay_f3_tauk} The synchronisation error $\Delta(t)$
for the cases with constant power forcing. Left: $N=128$. Right: $N=192$. }
\efig


We now look into the synchronisation of the flows. To obtain smoother
results, the data shown in this subsection are
the averages of five realisations. 

Fig. \ref{fig:decay_f2_tauk} shows the decay of the synchronisation error 
$\Delta(t)/\Delta(0)$ for different $k_m$ and $\Omega$ with Kolmogorov
forcing.
The top-left, top-right, and bottom-left panels correspond to three different Reynolds numbers.
There are three common trends across all cases included in these three panels. Firstly, 
the error decays exponentially when $k_m$ is
sufficiently large. Secondly, the decay rate increases with $k_m$. 
Thirdly, the error decays only when $k_m$ is greater than some
threshold value, referred to as $k_c$, and $k_c$ clearly is different in different cases. 
For $k_m$ close to but still greater than $k_c$, the error still decays over
time, but the rate of decay fluctuates, so exponential functions do not always
provide a good fit. 

Comparison across the above three panels in Fig.
\ref{fig:decay_f2_tauk} shows 
that 
the decay rate of the error displays the known dependence on the Reynolds number,
namely, everything else being equal, the decay rate 
decreases as the Reynolds number increases. This trend is illustrated in the
bottom-right panel with selected cases with $\Omega = 0.5$ and $k_m = 9$. 
As this effect has been reported
multiple times in previous research, we will not delve too much into it. 
For the same reason, we consider only the cases with $N=128$ and $N=192$ for
flows with
constant power forcing. 

More pertinent to our objectives is the observation that rotation has 
a strong effect on the decay rate. 
Fig. \ref{fig:decay_f2_tauk} shows that, for the same $k_m$, the decay rate
decreases with $\Omega$. 
The same trend is observed
for different Reynolds numbers, as is shown 
in the first three panels of the figure.

The results corresponding to constant power forcing are plotted in Fig.
\ref{fig:decay_f3_tauk}. Not surprisingly, 
$\Delta(t)$ decays exponentially for sufficiently large $k_m$. Moreover, the dependence of
the decay rate on
$k_m$ and $Re_\lambda$ is qualitatively similar to that which is observed in Fig.
\ref{fig:decay_f2_tauk}. 
However, interestingly,
the dependence on rotation is significantly different. The black
lines in the left panel of Fig. \ref{fig:decay_f3_tauk} illustrate the difference clearly.
The three black lines correspond to same $k_m$ but three different rotation rates. While the
decay rates for $\Omega = 0.1$ and $0.5$ show no clear differences, the decay
rate for $\Omega = 1$ is clearly larger. That is, in this case, it appears
that the
decay rate for $\Delta(t)$ increases with rotation. 
The same trend is seen in the right panel of the figure, which is for flows with a
larger Reynolds number. 
This observation is opposite to the
trend we observe in the cases with Kolmogorov forcing (c.f. Fig.
\ref{fig:decay_f2_tauk}), where the decay rate for the same $k_m$ is found to decrease with
rotation. 
The difference in the results for the two forcing terms 
has not been reported before.

\subsection{Conditional Lyapunov exponents and the threshold wavenumbers}


The synchronisability of the slaved flow is related to the conditional Lyapunov
exponents. 
We calculate the CLEs $\lambda(k_m)$ as well as the
local CLEs $\gamma(k_m,t)$ using the
algorithm outlined in Section \ref{sect:eqn}. The results are presented in
terms of the non-dimensionalised CLEs $\Lambda$ and the non-dimensionalised local
CLEs $\Gamma$, which are defined
as 
\be \Lambda
= \lambda \tau_k, ~~ \Gamma = \gamma \tau_k. 
\ee
$\Gamma$ is time dependent and fluctuates over time.
Without showing the time sequences, we note that, 
after a period of transience,
$\Gamma$ stabilizes and fluctuates around a
constant value. 
The magnitude of the fluctuations appears to increase with
rotation, but decreases as $k_m$ increases. 
We will quantify some of these behaviours in what follows, starting
with $\Lambda$, which is the average
of $\Gamma$ in the stationary stage.

\bfig
\centering
\ig[width=0.5\lnw]{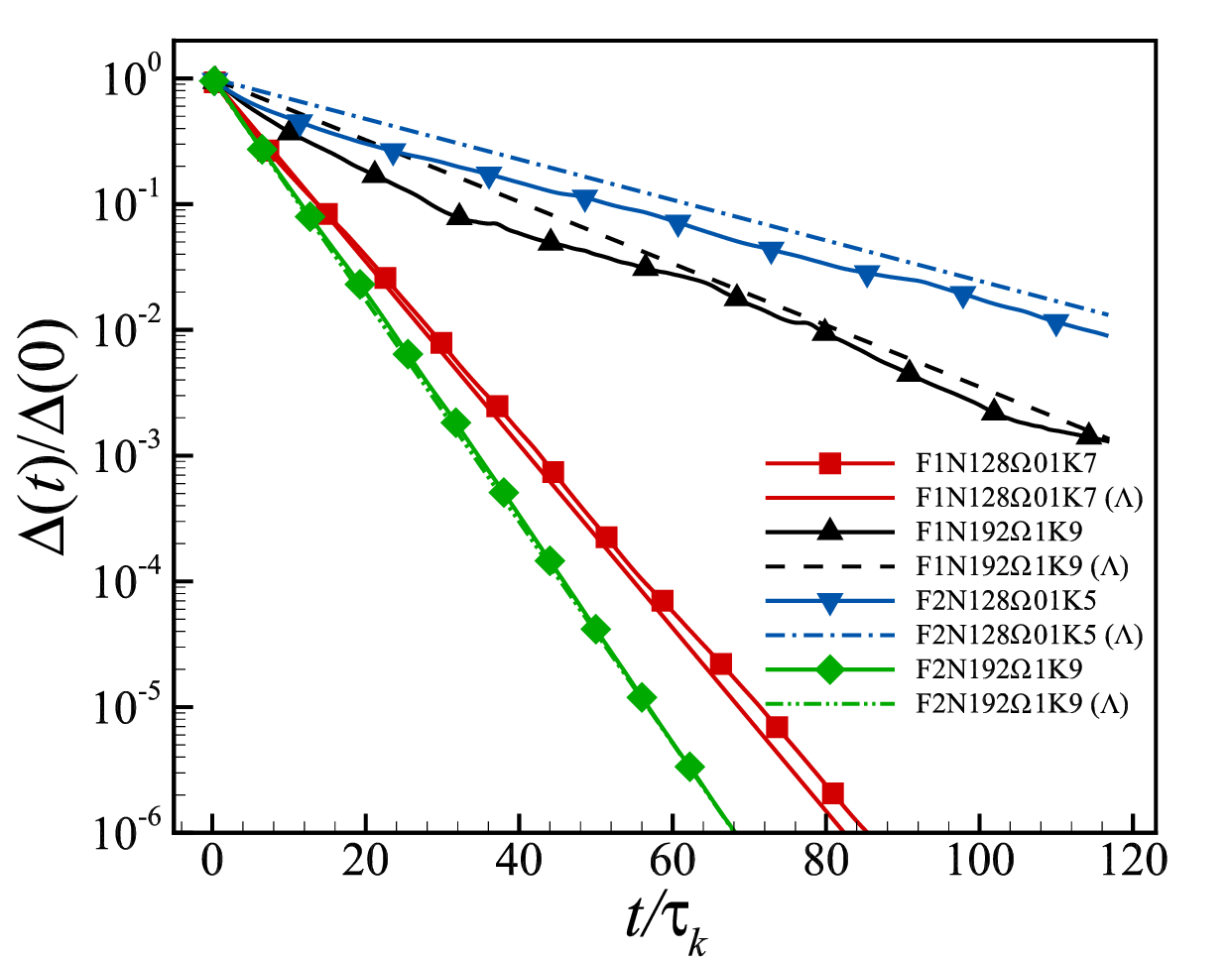}
\caption{\label{fig:lya_decay_check} Comparison between the decay rates of
$\Delta(t)$ and the conditional Lyapunov exponents.}
\efig

Fig. \ref{fig:lya_decay_check} is shown first to 
establish the relationship between the decay rate of $\Delta(t)$ and
the CLE $\Lambda$.
Shown with symbols in the figure are 
$\Delta(t)/\Delta(0)$ for  
a number of cases already discussed in
Figs. \ref{fig:decay_f2_tauk} and \ref{fig:decay_f3_tauk}. 
The lines without symbols represent functions $\exp(\Lambda t/\tau_k)$, where
$\Lambda$ is the CLE for the corresponding flow.  
Some small discrepancies are seen between the two, which we attribute to
statistical uncertainty in the data. We note that the discrepancies are in-line
with those found in previous research (e.g., \citet{NikolaidisIoannou22}).  
The overall agreement between the two 
shows that in most cases the error decays exponentially and 
the decay rate $\alpha$ equals $\Lambda$. For the
case shown with the black line and solid triangles, $\Delta(t)$ does not decay
exponentially. However, it undulates mildly around the
exponential function in such a way that $\Lambda$ appears to
capture the long time mean decay rate. 
Overall, we may conclude that the decay rate of $\Delta(t)$ 
is equal to $\Lambda$, and the synchronisation between two flows can be
fully characterised by $\Lambda$. 

As a side note, we note that larger discrepancy is observed for case F1N128$\Omega$01K7 than for case
F2N192$\Omega$1K9. This observation appears counter intuitive at first sight, since $\Omega$ is
larger in the latter case which should lead to larger fluctuation in $\Gamma$ hence
larger statistical error in $\Lambda$ (or the corresponding decay rate $\al$). 
However, there is another difference between these
two
cases, which is that case F2N192$\Omega$1K9 is computed with
a larger $k_m$. As
the fluctuation in $\Gamma$ is smaller for larger $k_m$, it is possible that 
the statistical discrepancy in case F2N192$\Omega$1K9 is smaller despite the
fact that it is
computed with a larger $\Omega$.

\bfig
\centering
\ig[width=0.48\lnw]{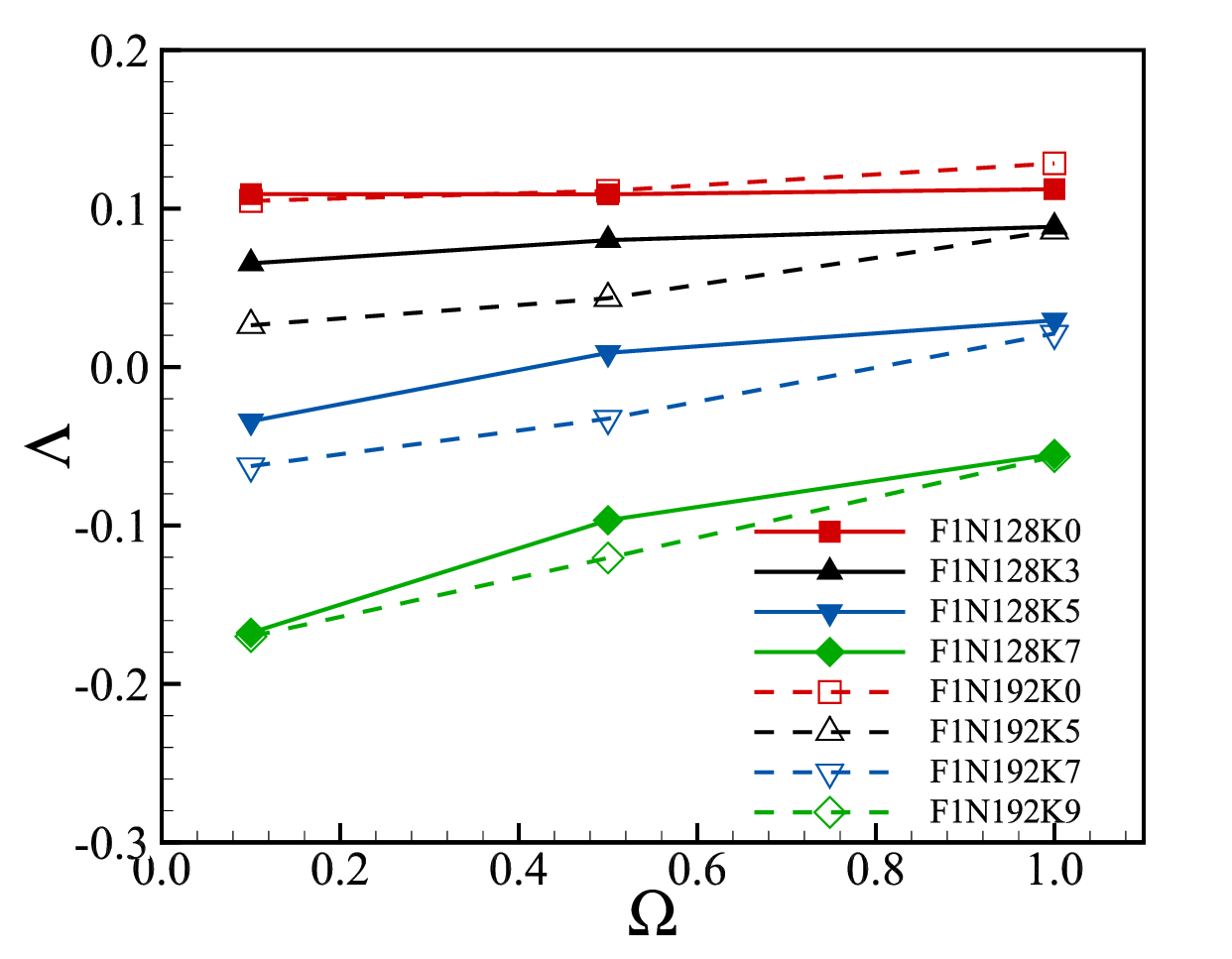}
\ig[width=0.48\lnw]{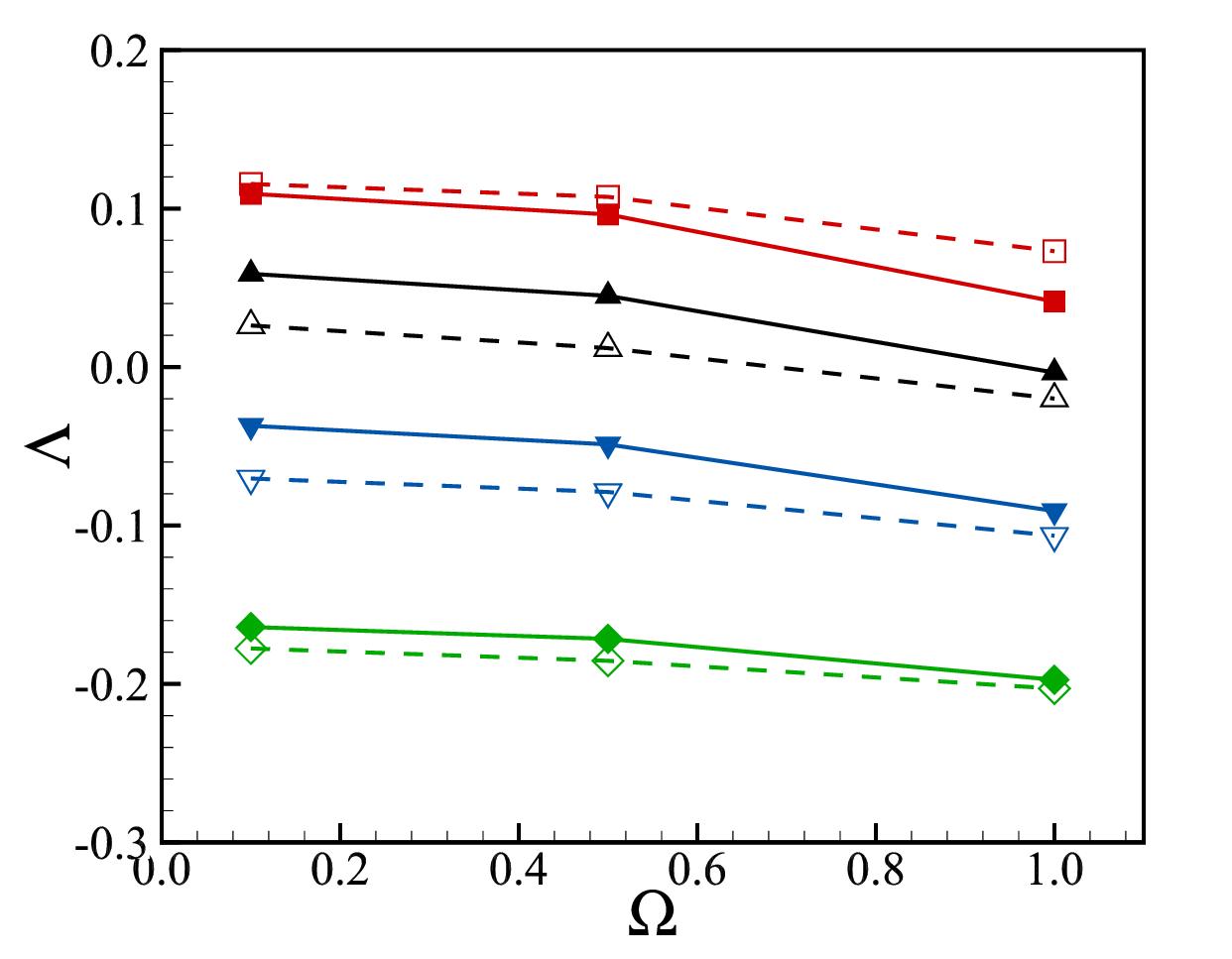}
\caption{\label{fig:lya_rot_f2f3} Normalised conditional Lyapunov exponent
$\Lambda$ as functions of
the rotation rate $\Omega$ for the cases with Kolmogorov forcing (left) and
constant power 
forcing (right). Solid lines:
$N=128$; dashed lines: $N=192$. For $N=128$, $k_m=0$ (squares), $3$
(deltas), $5$ (gradients), and $7$ (diamonds). For $N=192$,
 $k_m=0$ (squares), $5$
(deltas), $7$ (gradients), and $9$ (diamonds).
}
\efig


\bfig
\centering
\ig[width=0.48\lnw]{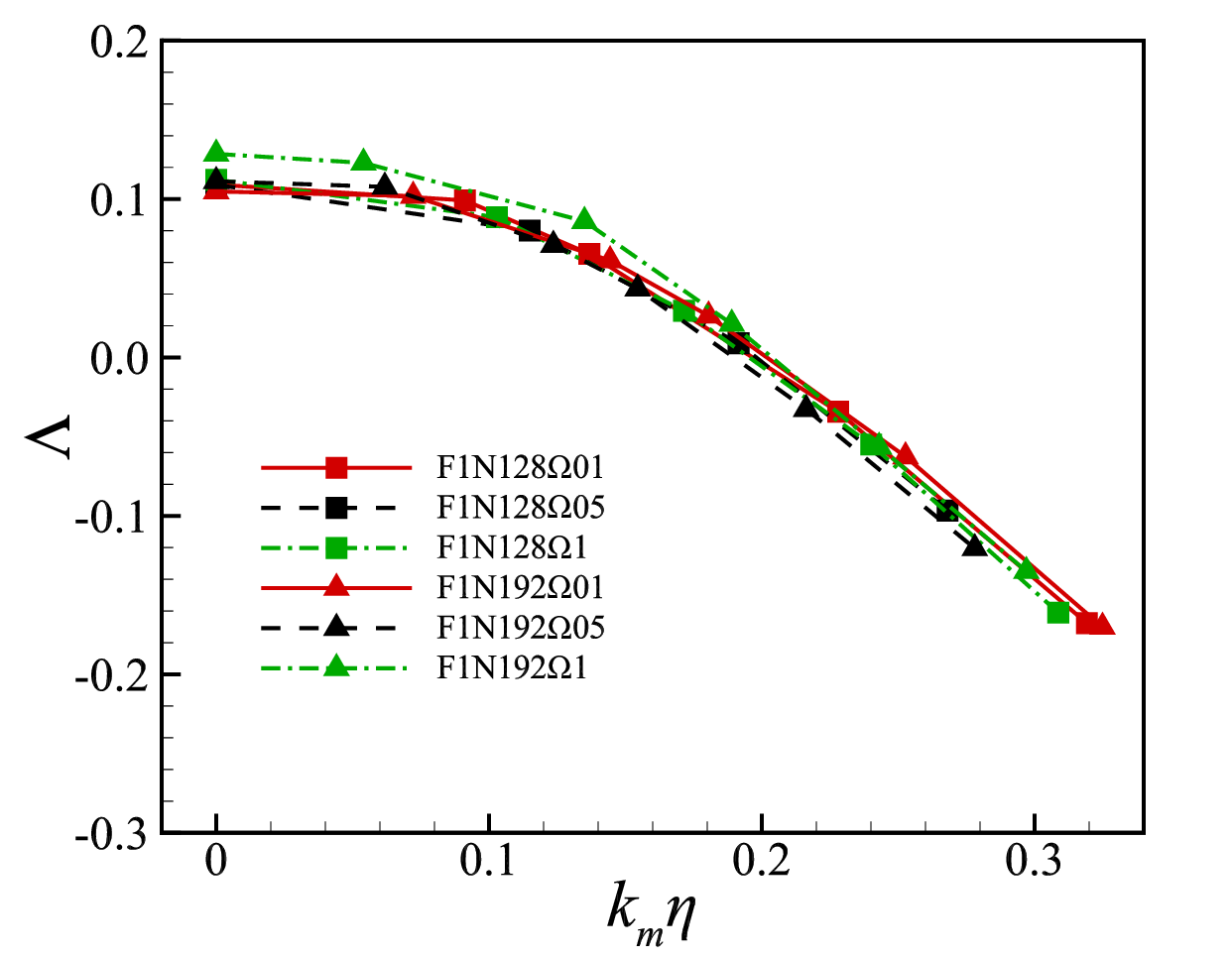}
\ig[width=0.48\lnw]{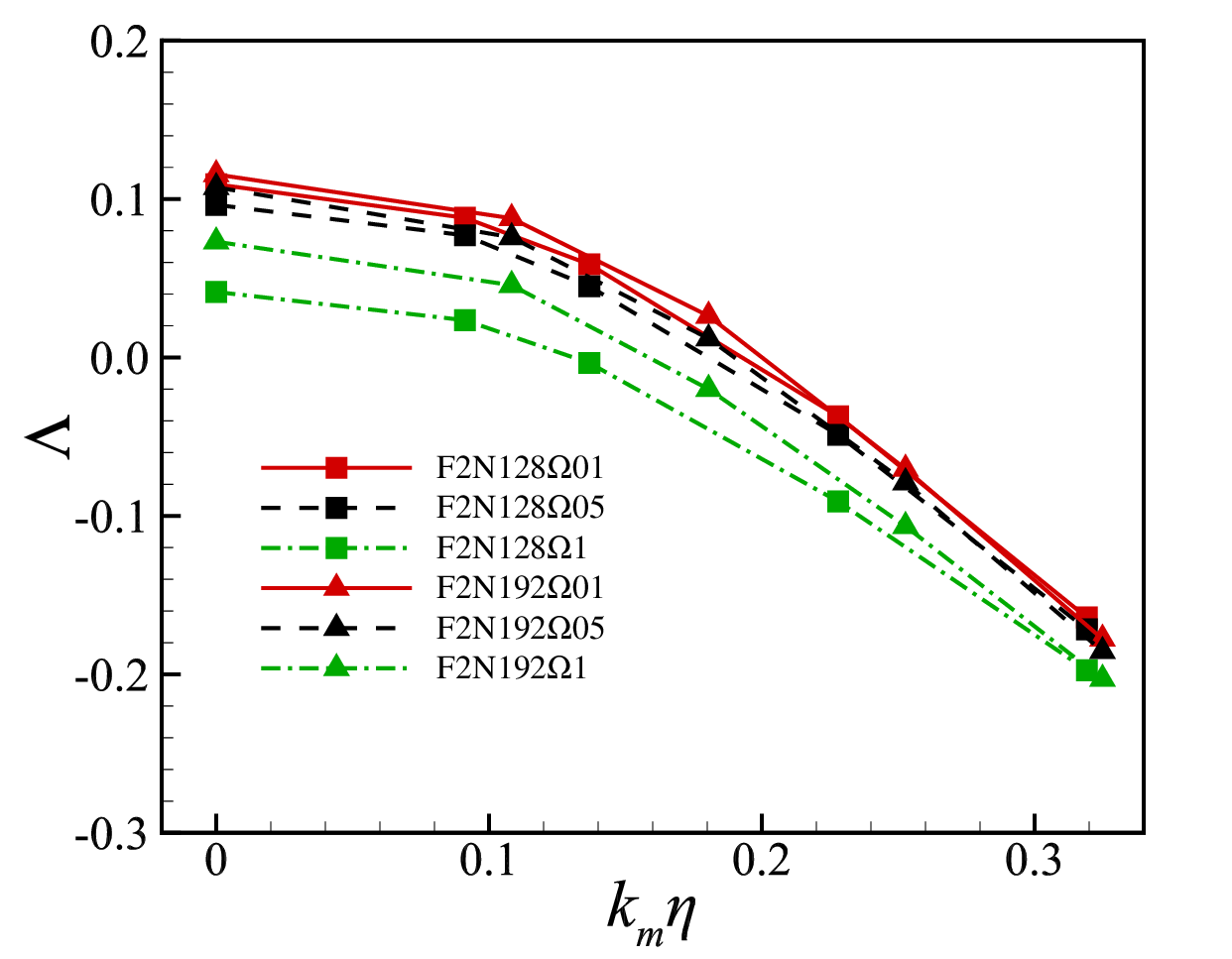}
\caption{\label{fig:lya_keta_f2f3} Normalised conditional Lyapunov exponents
$\Lambda$ as
functions of $k_m\eta$. Left: the cases with Kolmogorov forcing. Right:
the cases with constant power forcing. }
\efig


\bfig
\centering
\ig[width=0.6\lnw]{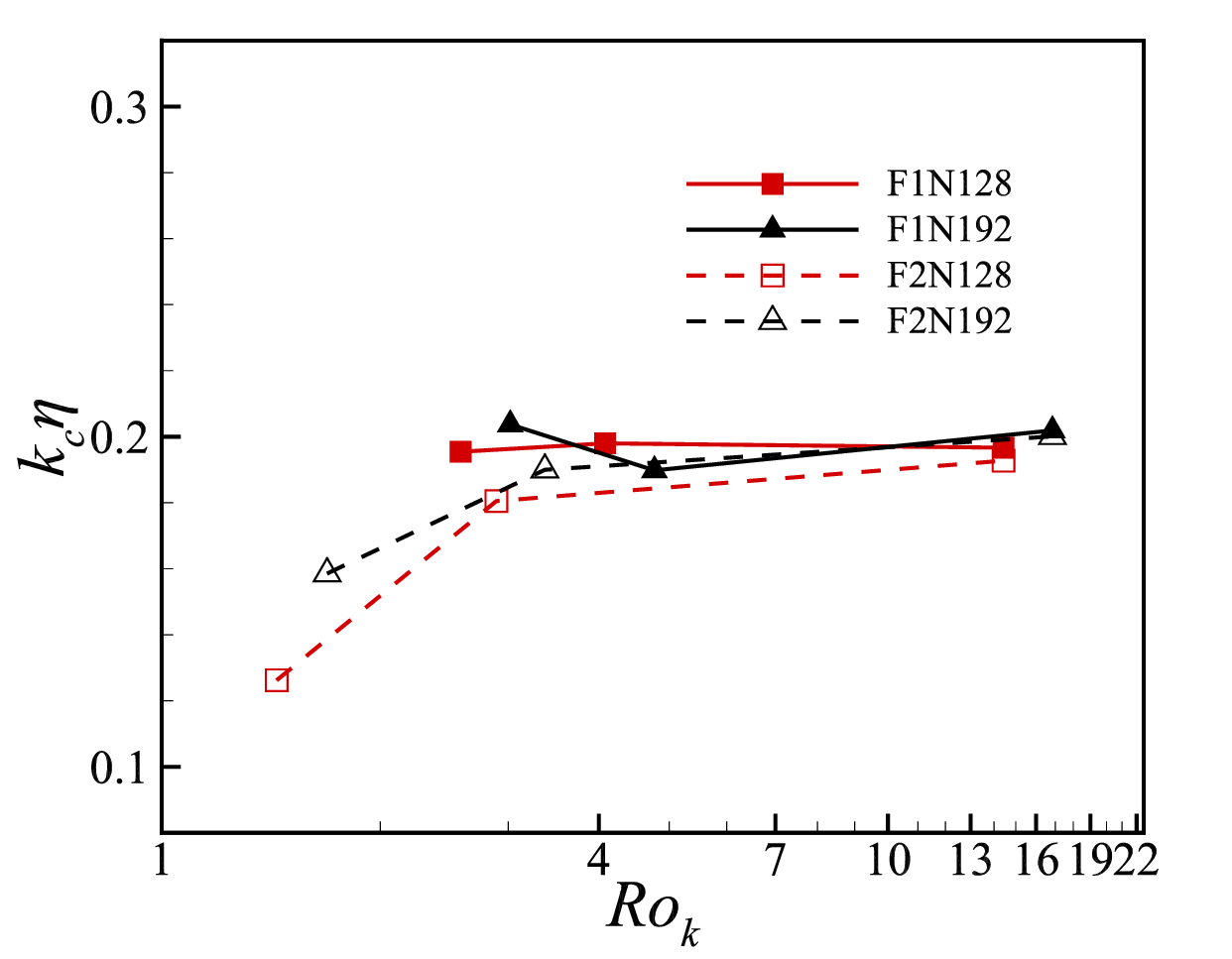}
\caption{\label{fig:kc_R03} Threshold coupling wavenumber $k_c$ as a function
of the micro-scale Rossby number $Ro_k$.}
\efig

We now focus on the results for
$\Lambda$. 
The dependence of 
$\Lambda$ 
on the rotation rate $\Omega$ and the coupling wavenumber $k_m$ is shown 
in Fig. \ref{fig:lya_rot_f2f3}, 
including cases with $k_m=0$ where
$\Lambda$ represents the unconditional Lyapunov exponent. 
The left panel presents the cases with Kolmogorov forcing. 
In these cases, $\Lambda$ always increases
with $\Omega$, and $\Lambda$ increases with $\Omega$ 
quicker for larger $k_m$. The
blue lines, which correspond to $k_m=5$ for $N=128$ and $k_m=7$ for $N=192$,
are particularly instructive. 
In these cases $\Lambda$ increases from a negative value to a positive one as $\Omega$
increases from $0.1$ to $1$. Therefore, the two flows synchronise when $\Omega = 0.1$, but they do
not when $\Omega = 1$, which emphatically shows that rotation makes the flows more difficult to
synchronise when the flow is driven by Kolmogorov forcing. 
However, the observation is different for the flows maintained by constant
power 
forcing, which
is shown in the right panel of Fig. \ref{fig:lya_rot_f2f3}. 
In fact, the trend is reversed in this case: here $\Lambda$ decreases as
$\Omega$ increases, so the flow is easier to synchronise as rotation is
increased. Also, the unconditional Lyapunov exponent appears more 
sensitive to the rotation rate.

Another observation we can make from 
Fig. \ref{fig:lya_rot_f2f3} is that $\Lambda$
decreases with $k_m$, which can be seen by comparing different curves in the
same panel. 
This trend is further investigated 
by plotting $\Lambda$ as a function of $k_m\eta$, which is given in Fig.
\ref{fig:lya_keta_f2f3}. 
We first note the values of $\Lambda$ at $k_m = 0$ for $\Omega = 0.1$. As
$\Omega$ is relatively small, one expects $\Lambda$ to be close to the value
found in non-rotating turbulence. 
Fig. \ref{fig:lya_keta_f2f3} shows that 
$\Lambda$ in this case is around $0.1$, though it weakly depends on the Reynolds number 
as well as the forcing term. This value is indeed close to 
those found previously 
for non-rotating turbulence \citep{BoffettaMusacchio17}.

Fig. \ref{fig:lya_keta_f2f3} shows that, for cases with Kolmogorov forcing,
$\Lambda$ decreases as $k_m\eta$ increases. 
More interestingly, the curves corresponding to different cases
collapse on each other approximately. The one for
$N=192$ and $\Omega = 1.0$ is slightly larger than the rest.
Nevertheless, overall, as a function of
$k_m\eta$, $\Lambda$ depends on rotation only weakly. 
Note that this
observation does not contradict with the results in Fig.
\ref{fig:lya_rot_f2f3}, as the values of
$k_m$ in the latter are not non-dimensionalised by $\eta$ and $\eta$ is
different for different $\Omega$. 

For the cases with
constant power forcing, the right panel in
Fig. \ref{fig:lya_keta_f2f3} shows that $\Lambda$ decreases with
$k_m\eta$ in a similar manner. However, the curves corresponding to
different $\Omega$ do not collapse well. In fact, 
$\Lambda(k_m\eta)$ tends to 
decrease with $\Omega$, in particular for stronger rotations. 

The threshold wavenumber $k_c$ where $\Lambda$ is zero is of particular
interests, as it is the value of $k_m$ for which 
synchronisation fails.
The values of $k_c$ can be found from 
Fig. \ref{fig:lya_keta_f2f3}, as they are 
the values of
$k_m$ where the curves cross the horizontal axis $\Lambda = 0$, which can be
read from the figures directly. The values are plotted as functions of the micro-scale
Rossby number $Ro_k$ in Fig. \ref{fig:kc_R03}. 

Interestingly, the figure shows that $k_c\eta$ essentially does not
depend on rotation when the flows are driven by Kolmogorov forcing, 
within the range of rotation rates we have considered.
To two decimal places, $k_c\eta = 0.20$ or $0.19$ for all rotation rates, which is the value obtained in
\citet{Yoshidaetal05} for non-rotating turbulence. 

This result seems to be contradictory to the observation that the decay rate
for a given $k_m$ decreases with rotation. However, it can be 
explained as follows. The decay rates
of the synchronisation error
are reduced when rotation is
introduced, which leads to increased $k_c$. However, as Table \ref{tab:cases}
shows, $\eta$ is decreased by rotation in this type of flows. 
The end result is that 
$k_c\eta$ remains roughly a constant.

For the flows driven by 
constant power forcing, it appears from Fig. \ref{fig:kc_R03} that there is a consistent trend where
$k_c\eta$ decreases as $Ro_k$ decreases (i.e., as rotation rate increases). For
the smallest $Ro_k$, $k_c\eta$ is reduced to below
$0.15$. Therefore, for the flows driven by constant power forcing,
rotation does increase the synchronizability of the flow, and this is reflected
in both an increased decay rate for the synchronisation error, and a reduced
 $k_c\eta$.  

\citet{Leonietal20} reported that rotating turbulence was easier to
synchronize and they attributed it to the coherent
vortices induced by rotation. Our results for
constant power forcing are consistent with their finding, 
which thus might be explained qualitatively in a similar way. However, the 
results for Kolmogorov forcing show that the physical picture can 
depend on the forcing scheme.


\bfig
\centering
\ig[width=0.6\lnw]{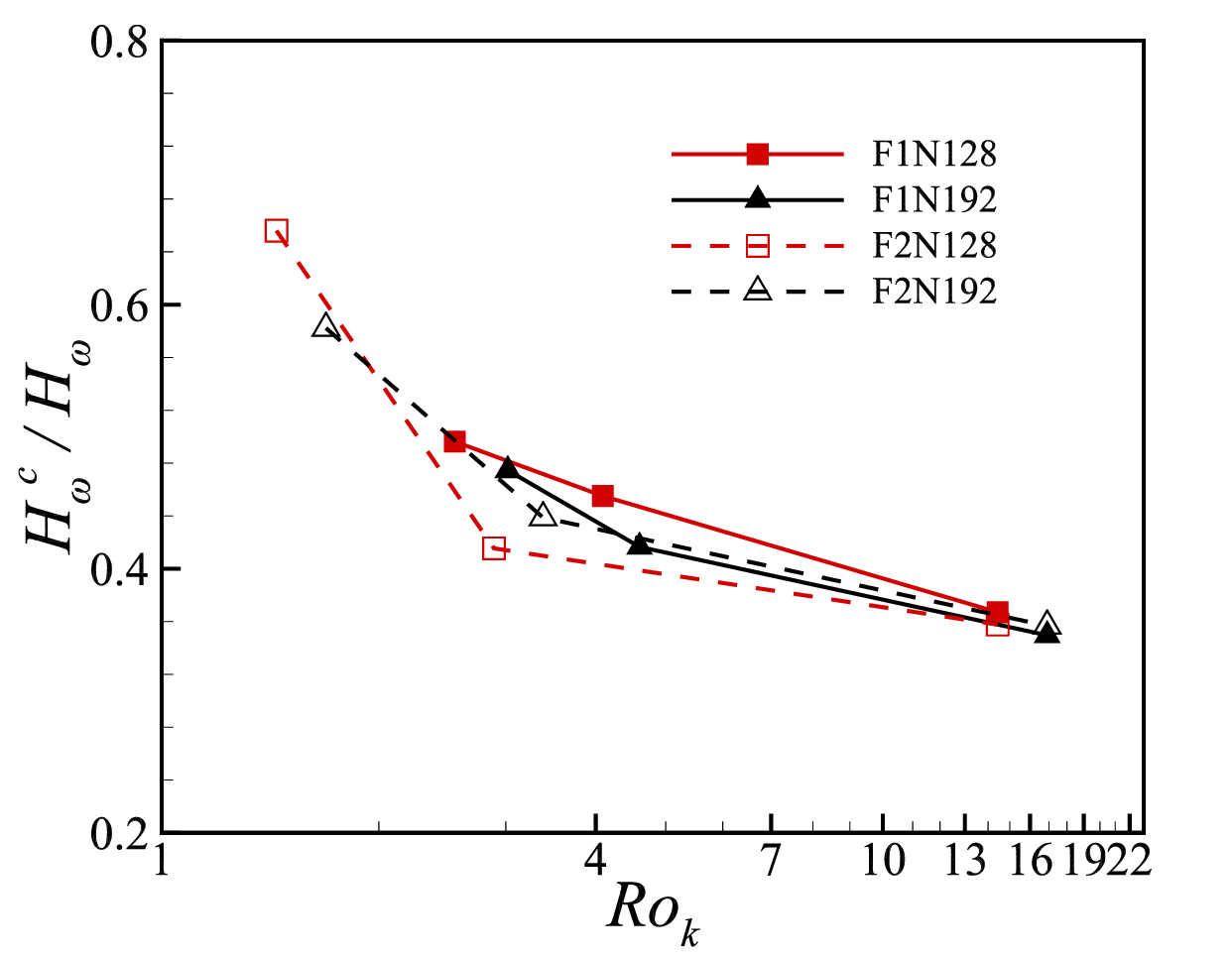}
\caption{\label{fig:kcEk_R03} Enstrophy ratio $H_\omega^c/H_\omega$ 
as a function
of the micro-scale Rossby number $Ro_k$.}
\efig

It is instructive to cross-check the results for $k_c$ with the energy
spectra of the flows. 
Note that in the majority cases the spectra are consistent with a
$k^{-2}$ power law (c.f. Fig. \ref{fig:Ek}). Therefore, the dimensionless threshold wavenumber $k_c\eta$
remains approximately unchanged from the value for isotropic turbulence when 
the spectrum
steepens from $k^{-5/3}$ to $k^{-2}$. 
On the other hand, 
when the slope of the energy spectra is further steepened, reaching approximately that of the
$k^{-3}$ power law,
$k_c\eta$ does become smaller, 
as in the cases with the
constant power forcing when $\Omega = 1$. 

To parametrise the decay rate of the synchronisation error with a physical
quantity, 
\citet{Yoshidaetal05} look into the enstrophy content in the master modes.
Let $H_\om^m = \sum_{k < k_m} k^2 E(k)$ be the enstrophy contained in the master
modes and $H_\om = \sum_k k^2 E(k)$ be the enstrophy of the whole velocity
field.
They find that, in isotropic turbulence, the decay 
rate $\alpha$ is a universal function of the ratio
$H_\om^m/H_\om$, 
and the ratio 
at the threshold
wavenumber $k_c$, denoted by 
$H_\om^c/H_\om$, is approximately $0.35$. 
We plot $H_\om^c/H_\om$ as a function of $Ro_k$ in Fig. \ref{fig:kcEk_R03} for
our simulations. For the largest $Ro_k$, the ratio is approximately $0.36$,
which is close to the value found in \citet{Yoshidaetal05}. The ratio
consistently
increases with rotation for both forcing terms, 
even for larger values of $Ro_k$ where
$k_c\eta$ remains a constant in the flows with Kolmogorov forcing. 
However, the results for different cases do not
collapse on a unique curve. Therefore, it seems that $H^c_\omega/H_\om$ does 
not provide a simple way to characterise $k_c$ in rotating turbulence. 

\bfig
\centering
\ig[width=0.48\lnw]{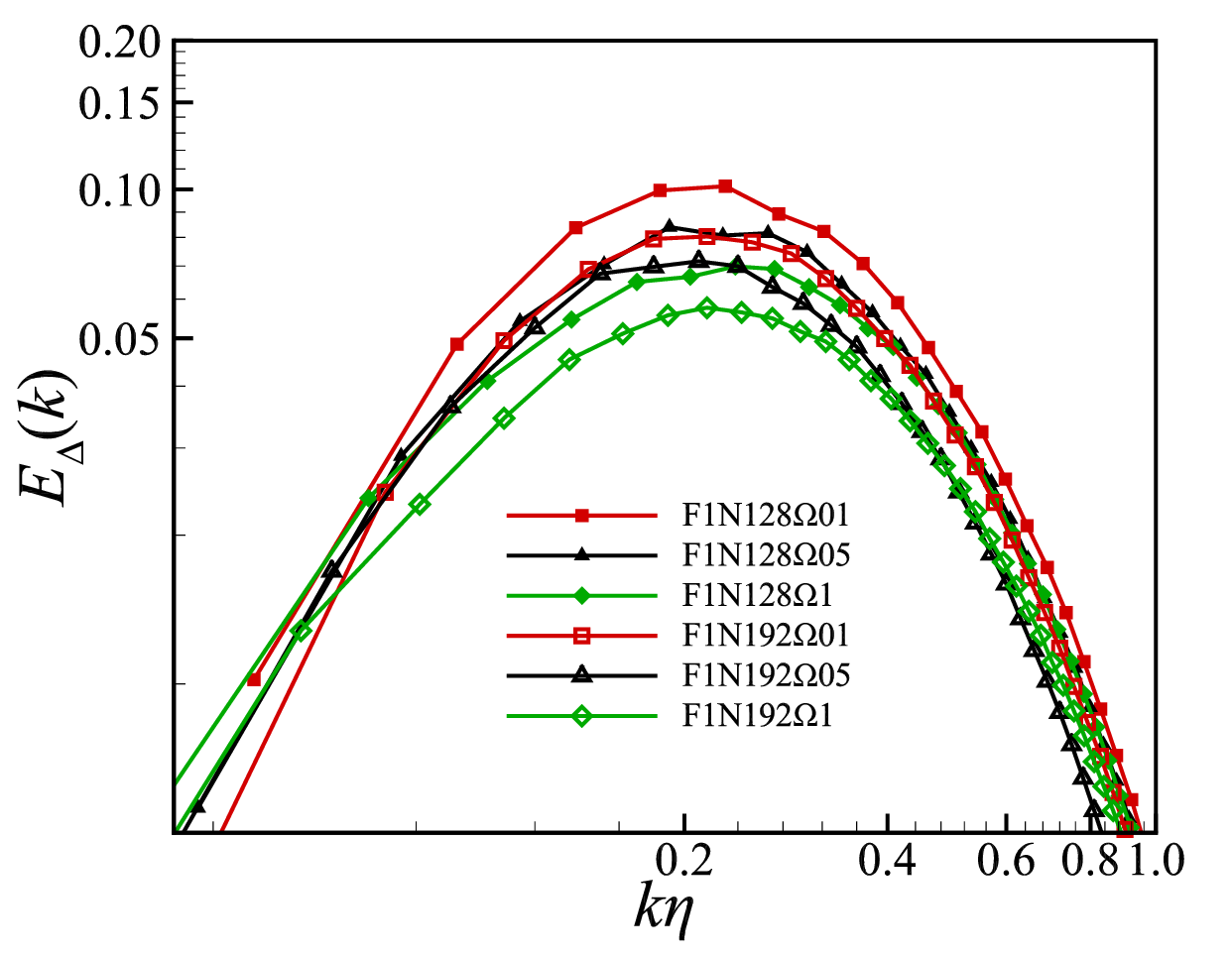} %
\ig[width=0.48\lnw]{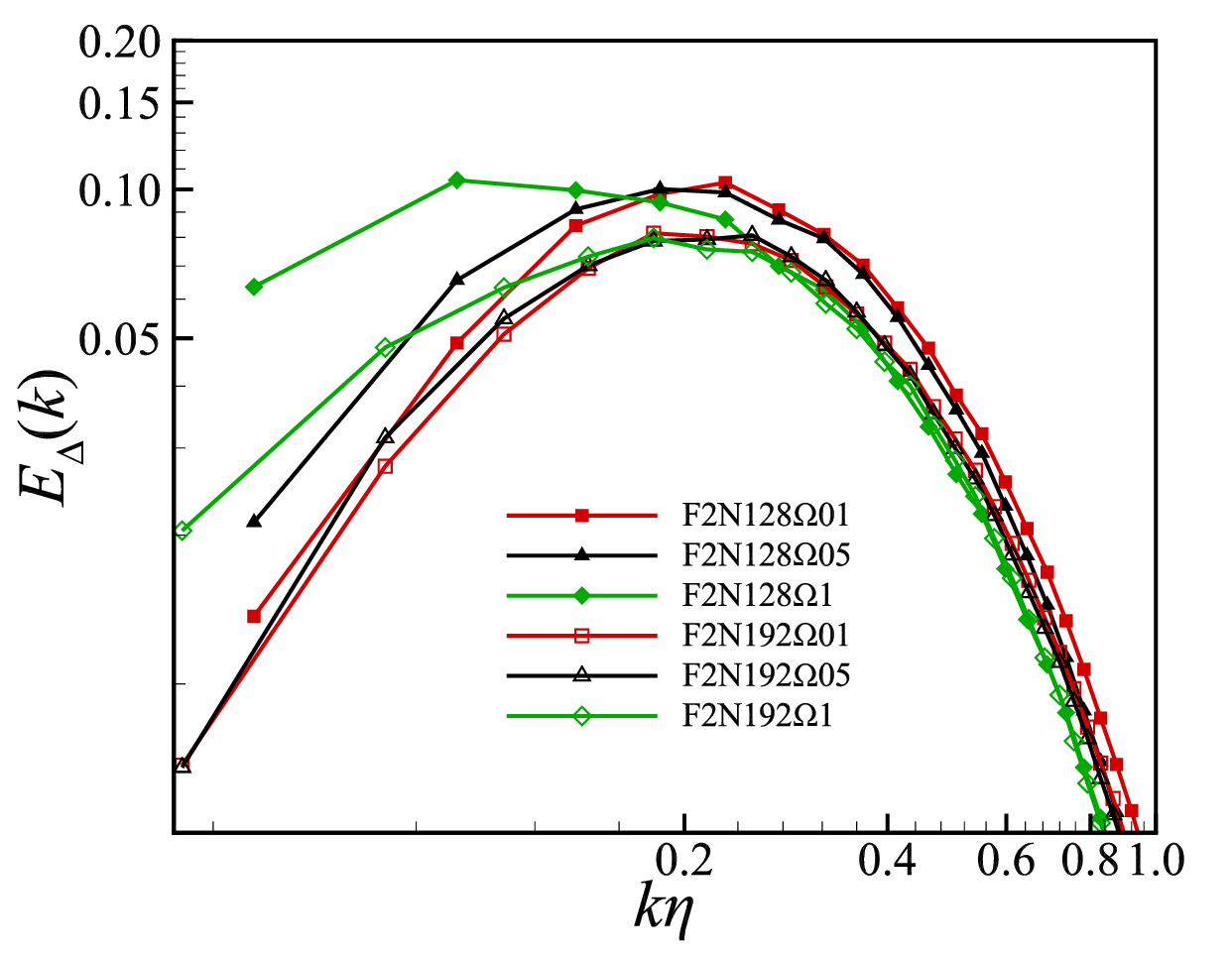}
\caption{\label{fig:lya_vector_errEk_f2f3} The averaged energy spectra of the Lyapunov
vectors $\bu^\delta$ for $k_m=0$. Left: for the cases with Kolmogorov forcing. Right:
for the cases with constant power forcing. The spectra have been normalised
in such a way that the total energy is unity.}
\efig

\bfig
\centering
\ig[width=0.48\lnw]{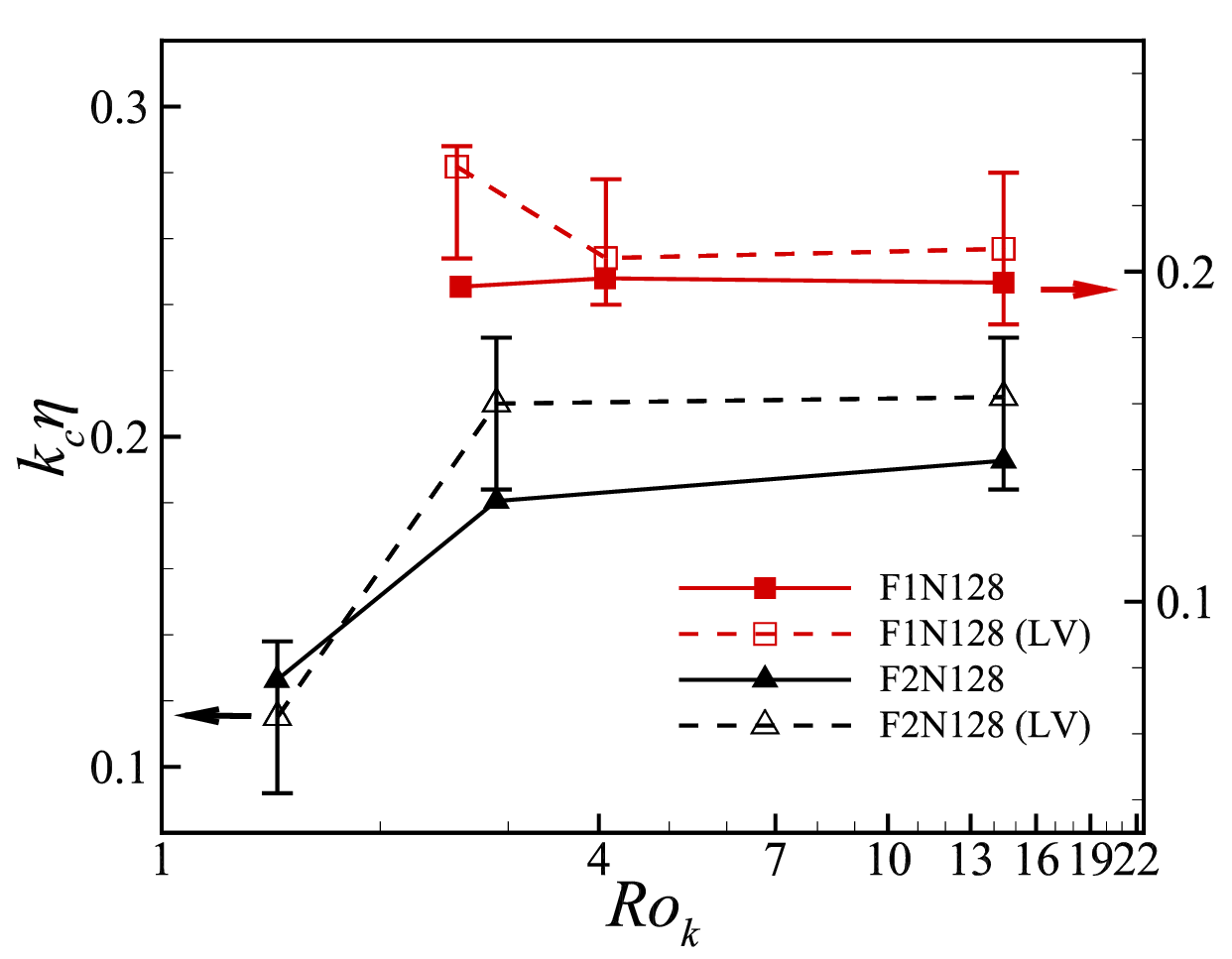} %
\ig[width=0.48\lnw]{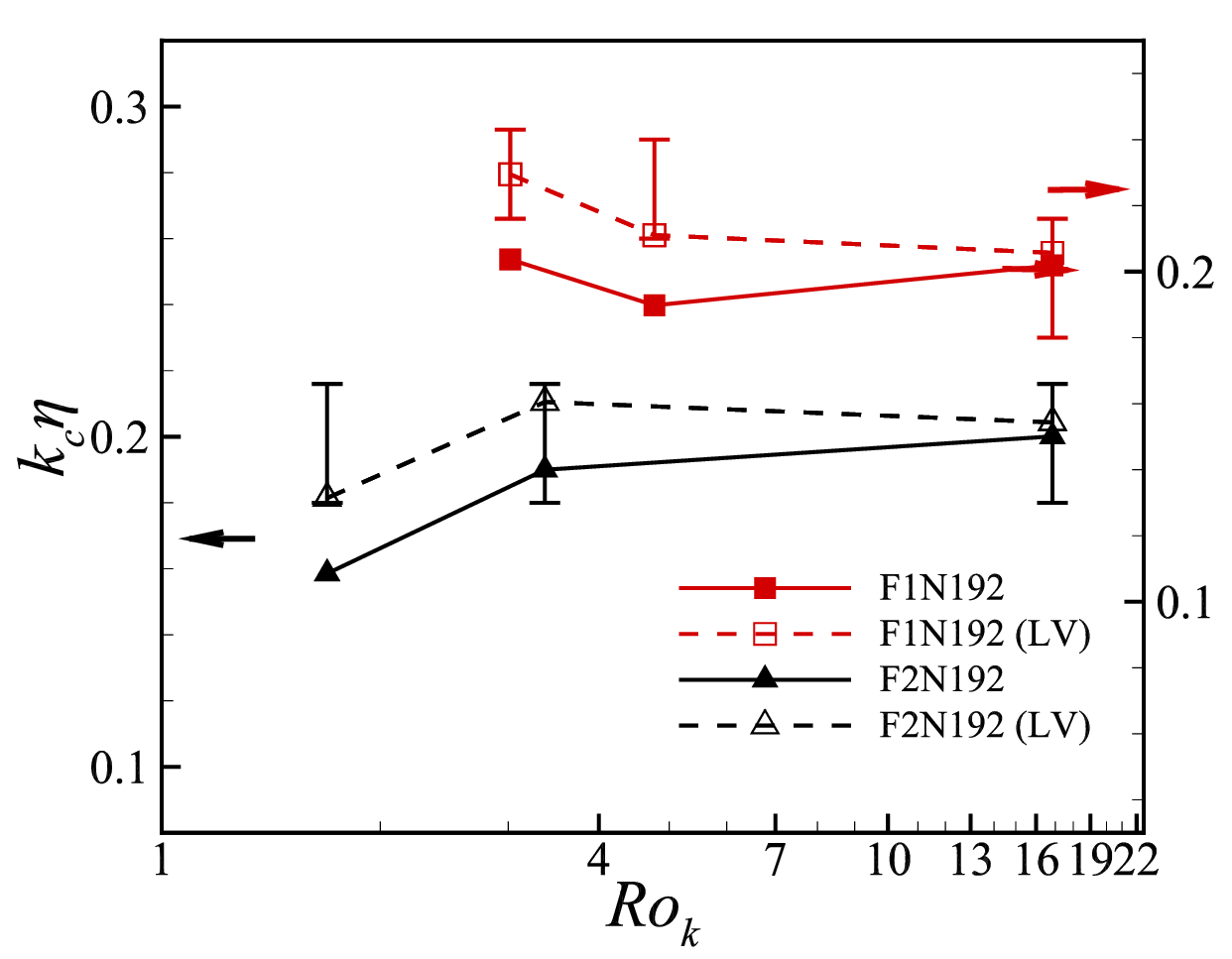}
\caption{\label{fig:kc_R0_errbar} Comparison between the peak wavenumbers for
the spectra of the Lyapunov vectors and $k_c$. Left: cases with $N=128$. Right: cases with
$N=192$. Solid lines and solid symbols: $k_c\eta$ (same as Fig.
\ref{fig:kc_R03}). Dashed lines and empty symbols: normalised peak wavenumbers of the
spectra of the
Lyapunov vectors. Lower groups and left $y$-axes: cases with constant power
forcing. Upper groups and right $y$-axes: cases with Kolmogorov forcing. The
error bars correspond to the two adjacent integer wavenumbers. ) 
}
\efig

Another way to characterize $k_c$ is put forward by \citet{Leonietal20}, where
they observe
that $k_c$ roughly marks the end of the inertial range. The observation is
corroborated in
\citet{NikolaidisIoannou22}. As the Reynolds
numbers for our simulations are relatively small, this observation is not
assessed here even though it is highly desirable to do so. 
Rather, we comment on a potential relationship between $k_c$ and the energy spectrum of 
the Lyapunov vector $\bu^\delta$, 
which provides another perspective into the
threshold wavenumbers.  

Fig. \ref{fig:lya_vector_errEk_f2f3} plots the
energy
spectra of $\bu^\delta(\x,t)$
averaged over $t$ in the stationary stage. As
the magnitude of $\bu^\delta$ is irrelevant, the energy spectra have been
normalised such that the total energy is unity. Also note that included in this figure
are the results with $k_m=0$, i.e., they are the spectra of the unconditional
Lyapunov vectors. 

The left panel of Fig. \ref{fig:lya_vector_errEk_f2f3} is 
for the flows driven by Kolmogorov forcing.
First of all, the energy spectra peak at an intermediate
wavenumber. That is, the perturbations with energy localised on intermediate
wavenumbers are
the most unstable. This observation is consistent with
\citet{OhkitaniYamada89} where
the Lyapunov vector for a shell model is calculated, and they 
find the energy spectrum of the Lyapunov vector is localised in the inertial
range. 
Interestingly, the peaks of the spectra here are all found around $k\eta =
0.2$, 
i.e. around the threshold wavenumber. Similar features are found in the right panel of the
figure, which
is the results for constant power forcing. Again in most cases the peaks are found around
$k_c\eta$. In particular, for the two cases with $\Omega = 1$, the peaks
are found to shift to lower $k\eta$, consistent with Fig.
\ref{fig:kc_R03} which shows $k_c\eta$ is also reduced in these two cases. 

It is desirable to compare the peak wavenumbers with $k_c$ quantitatively. 
There are some challenges in extracting precise peak
wavenumbers due to two factors: firstly, the spectra of $\bu^\delta$ at lower wavenumbers display stronger
statistical fluctuations; secondly, the gap between two data points on the
spectra is $\Delta k =
1$, which is fairly large and potentially introduces error into the reading
of the peak wavenumbers. In order to reduce the uncertainty, we average the
spectra over five realisations.
We then fit a smooth curve to the spectra using cubic splines. The peak
wavenumber of the fitted curve is taken to be the peak wavenumber of the
spectrum.

The cubic spline fitting is conducted using \verb=scipy= function
\verb=UnivariateSpline=,
with smoothing factor $s$ chosen as $0.01\%$ of the maximum of the spectrum, which
implies the 2-norm of the residue of the fitting is smaller than $s$. In other
words, only a very small amount of smoothing is allowed. 

The peak wavenumbers extracted in the above manner are plotted in 
Fig. \ref{fig:kc_R0_errbar} together with $k_c$ which has been shown in
Fig. \ref{fig:kc_R03}. The peak wavenumber obtained this way usually falls between two integer
wavenumbers. These two wavenumbers are used to define the error bars in Fig.
\ref{fig:kc_R0_errbar}. The figure confirms the qualitative comments we made
previously. The peak wavenumbers are slightly larger than $k_c$ in most cases.
However they do display same trends as $k_c$. In particular, for flows driven
by constant power forcing, the peak wavenumber clearly drops off significantly
for the smallest $Ro_k$, despite the uncertainty in the data. 

One plausible explanation of the correlation between $k_c$ and the peak
wavenumber of the energy spectrum of $\bu^\delta$ is as follows. 
Let the coupling wavenumber be $k_m$ in a synchronisation experiment. 
The peak wavenumber 
corresponds to the Fourier modes most susceptible to
infinitesimal perturbations (on average). 
One may hypothesize that, to synchronise two flows, the perturbations to these
most unstable Fourier modes should be suppressed by the coupling in the
synchronisation experiments. This suggests 
that the coupling wavenumber
$k_m$ should be larger than the peak wavenumber. 
However, even though only Fourier modes with wavenumbers up to $k_m$ in the
two flows are
coupled by design (in fact, they are exact copies of each other), the
Fourier modes with wavenumbers slightly larger than $k_m$ are also strongly
coupled, due to the fact that they are linked to the master modes through
nonlinear inter-scale interactions.  
The coupling suppresses
the growth of the synchronisation errors in these modes. Therefore, synchronisation can
still be achieved even if $k_m$ is slightly
smaller than the peak wavenumber. As a result, the threshold wavenumber $k_c$ could be slightly smaller
than the peak wavenumber in the spectrum of $\bu^\delta$.

\bfig
\centering
\ig[width=0.48\lnw]{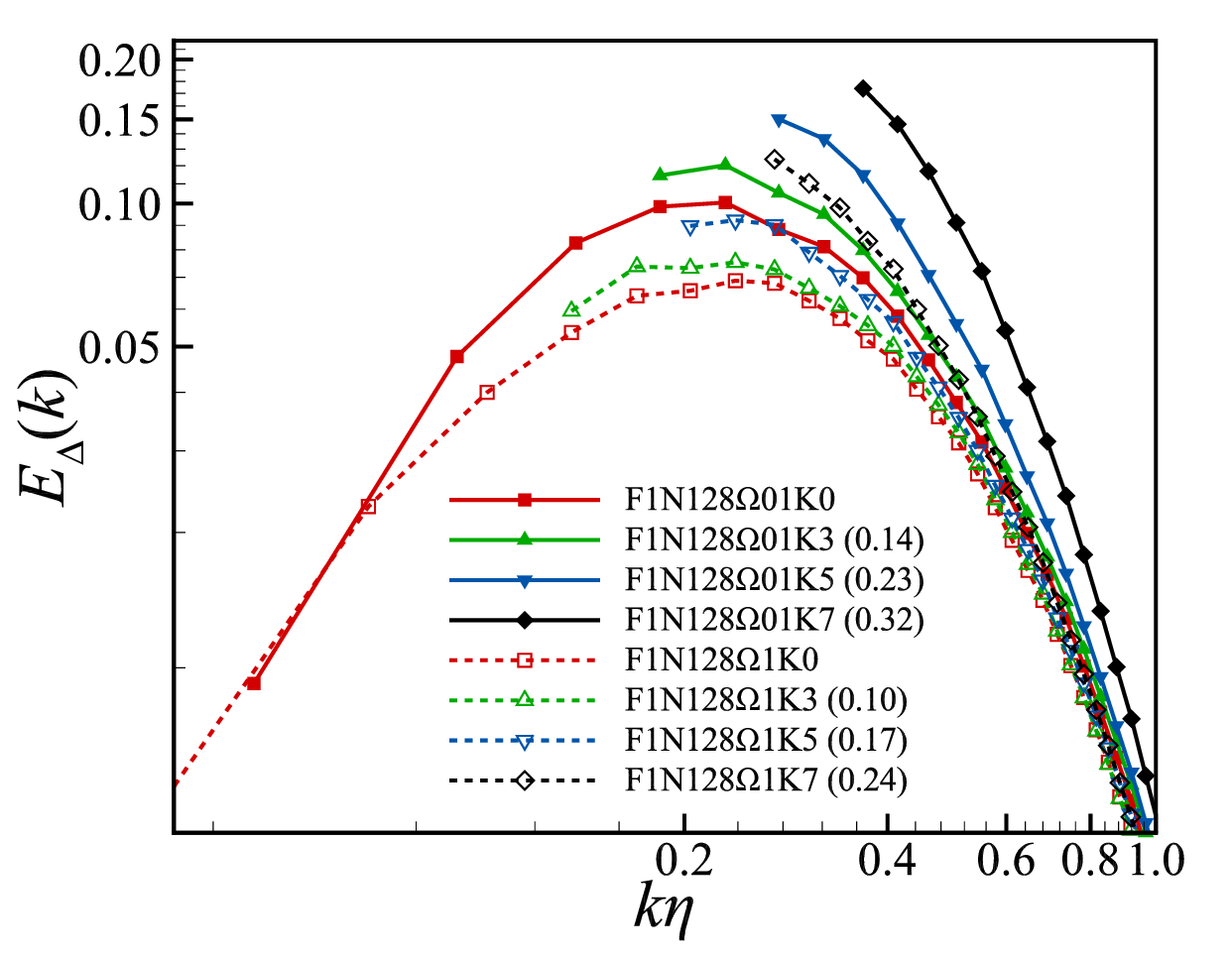} %
\ig[width=0.48\lnw]{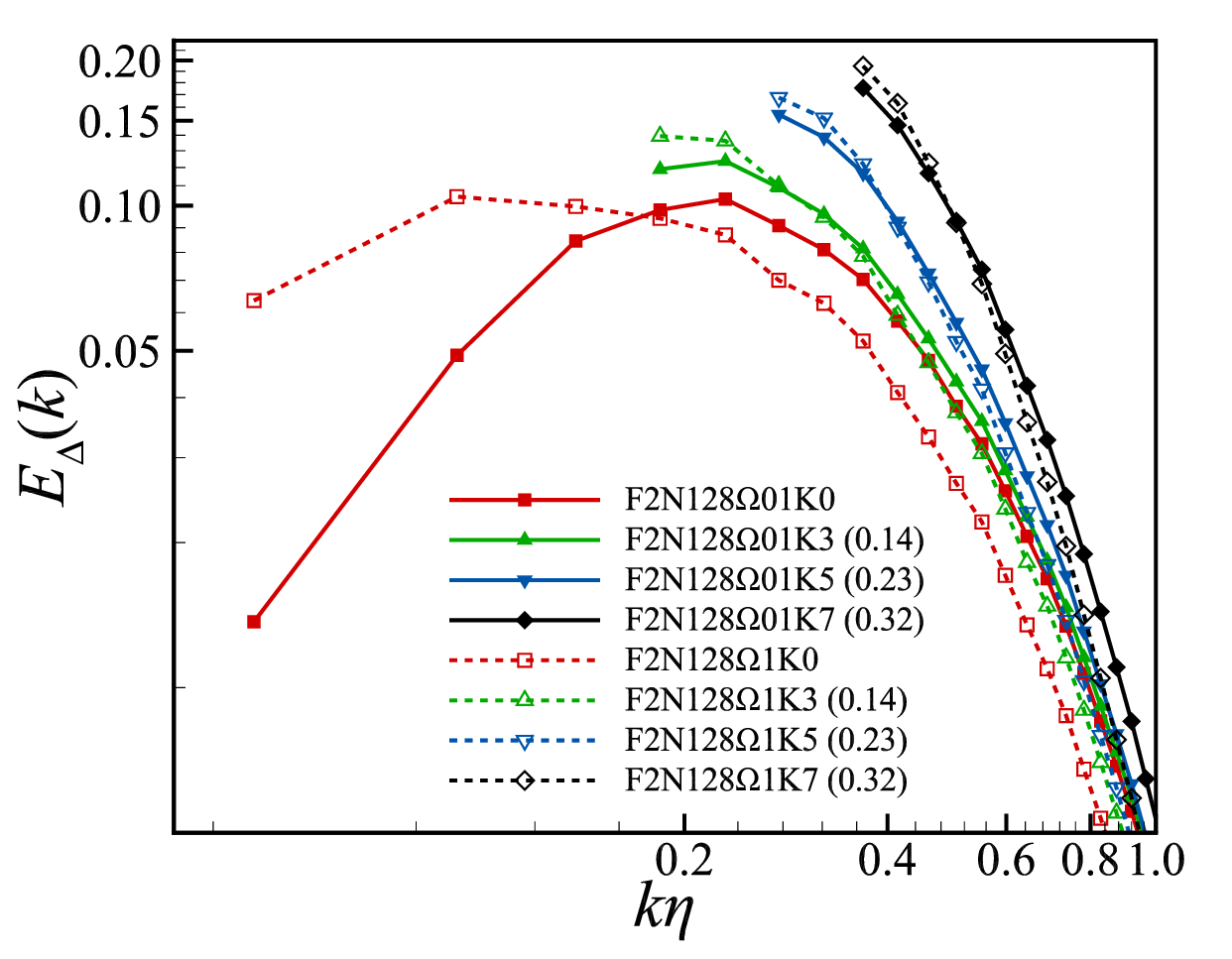}
\caption{\label{fig:err_Ek_lya_f2f3_128} The averaged energy spectra of the
conditional Lyapunov
vectors $\bu^\delta$ for different $k_m$ and $N=128$. The values of $k_m\eta$
are shown in parentheses. Left: cases with Kolmogorov
forcing, where $k_c \eta = 0.20$ for both $\Omega = 0.1$ and $1$. 
Right: cases with constant power forcing, where $k_c \eta = 0.19$ for $\Omega
= 0.1$, and $0.13$ for $\Omega = 1$.}
\efig
\bfig
\centering
\ig[width=0.48\lnw]{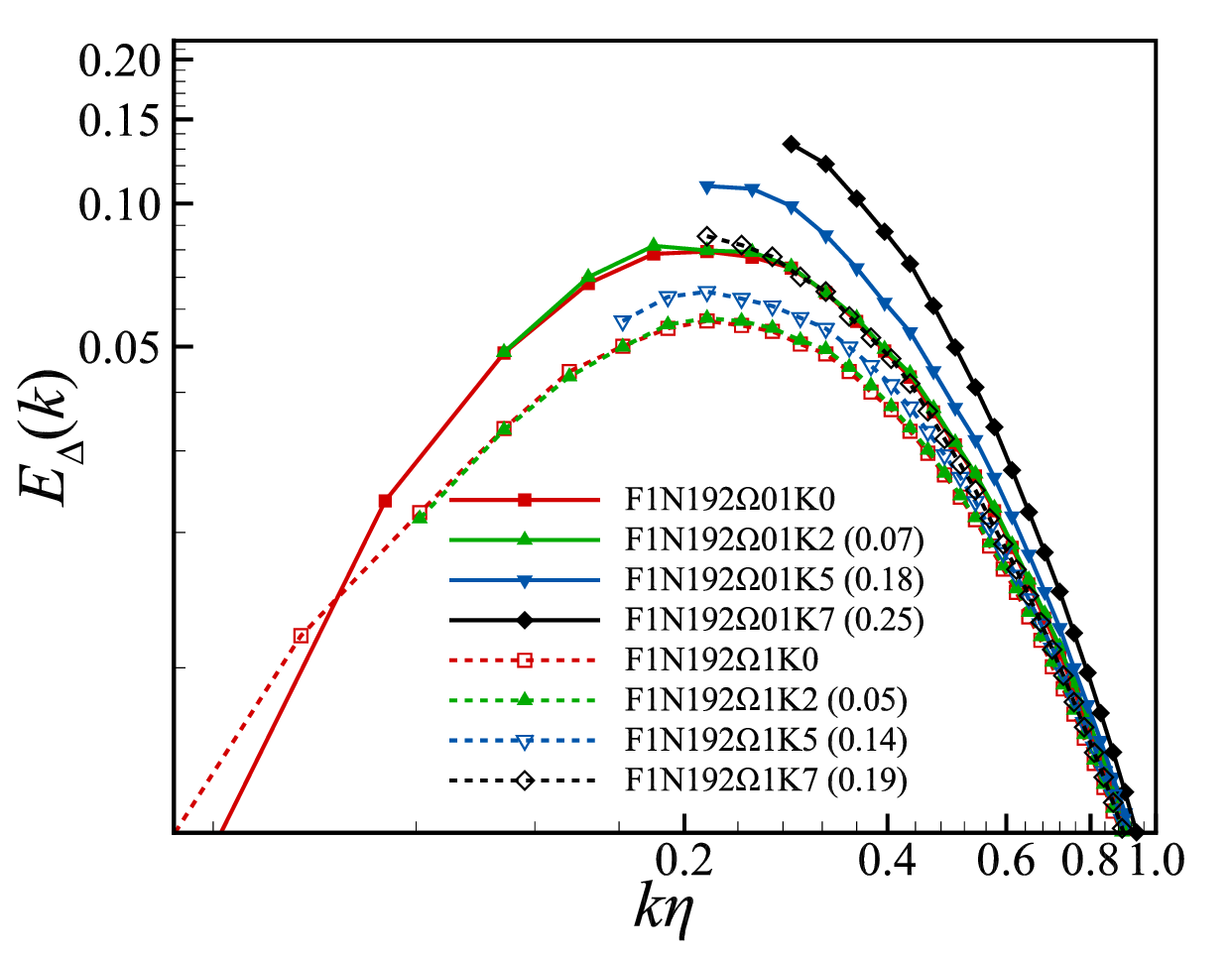} %
\ig[width=0.48\lnw]{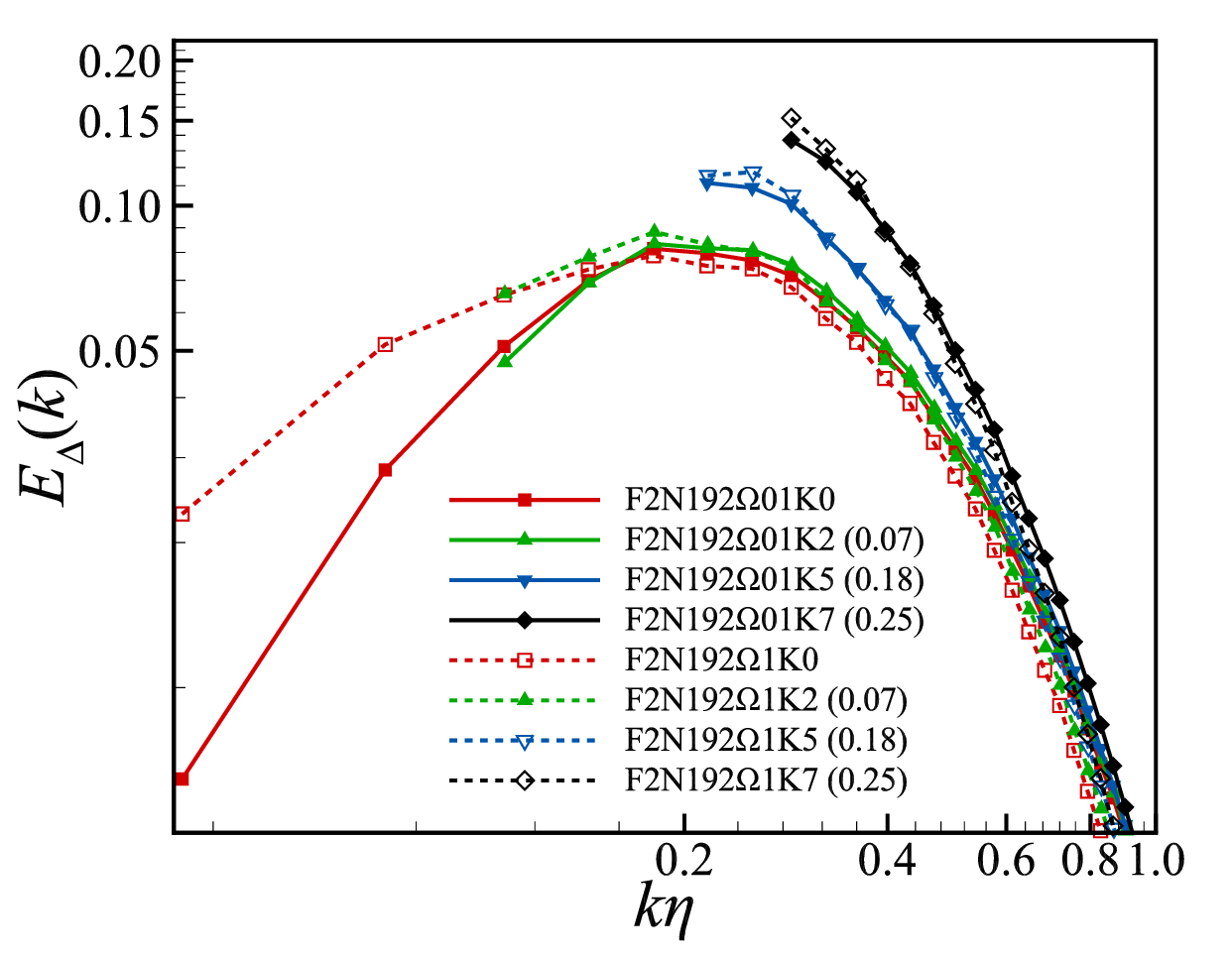}
\caption{\label{fig:err_Ek_lya_f2f3_192} Same as Fig.
\ref{fig:err_Ek_lya_f2f3_128} but for $N=192$. Left: cases with Kolmogorov
forcing, where $k_c \eta = 0.20$ for both $\Omega = 0.1$ and $1$.
Right: cases with constant power forcing, where $k_c \eta = 0.20$ for $\Omega
= 0.1$, and $0.16$ for $\Omega = 1$. 
}
\efig

The spectra of the Lyapunov vectors corresponding to the conditional CLEs,
namely the conditional Lyapunov vectors, are given
in Fig. \ref{fig:err_Ek_lya_f2f3_128} for the cases where $N=128$, and compared 
with the unconditional ones. For readability of the figures, only 
cases where $\Omega = 0.1$ and $1$ with selected $k_m$ are included. 
Note that the spectra for the conditional Lyapunov vectors start from wavenumber $k_m+1$
as $\hat{\bu}^{\delta}(\k,t) = 0$ for $\vert \k \vert \le k_m$. The value of
$k_m\eta$ for each case is shown in the parentheses, which can be compared
with the value of $k_c\eta$ to determine if synchronisation is achievable in
the
case. 
The interesting observation to note is demonstrated most clearly by both the
solid and the dashed green
lines with triangles in the left panel, which correspond to $k_m=3$ for
$\Omega = 0.1$ and $1$, respectively. The flows do not synchronise in these
two cases while they do in all other cases depicted in the panel. The
common feature of the spectra in these two cases is that the spectra peak at 
wavenumbers corresponding to 
the slaved modes. On the other hand, the spectra for the other cases all peak
at $k_m+1$. The same trend can be observed in the right panel of the figure,
and for cases with $N=192$ shown in Fig. \ref{fig:err_Ek_lya_f2f3_192}.
It appears that synchronisation can be achieved only when the
energy spectrum of the conditional Lyapunov vector does not have a local maximum
among the slave modes. 


The above results for the conditional Lyapunov vectors, though are of a
qualitative nature, also suggest that the 
threshold wavenumber $k_c\eta$ might be associated with the peak
of the spectrum of the Lyapunov vector. Given that our simulations cover
only a moderate range of Rossby numbers, with relatively low Reynolds
numbers, how this observation generalises to a wider range of parameter values
requires further investigation.

\subsection{The statistics of the local conditional Lyapunov exponents}

\bfig
\centering
\ig[width=0.5\lnw]{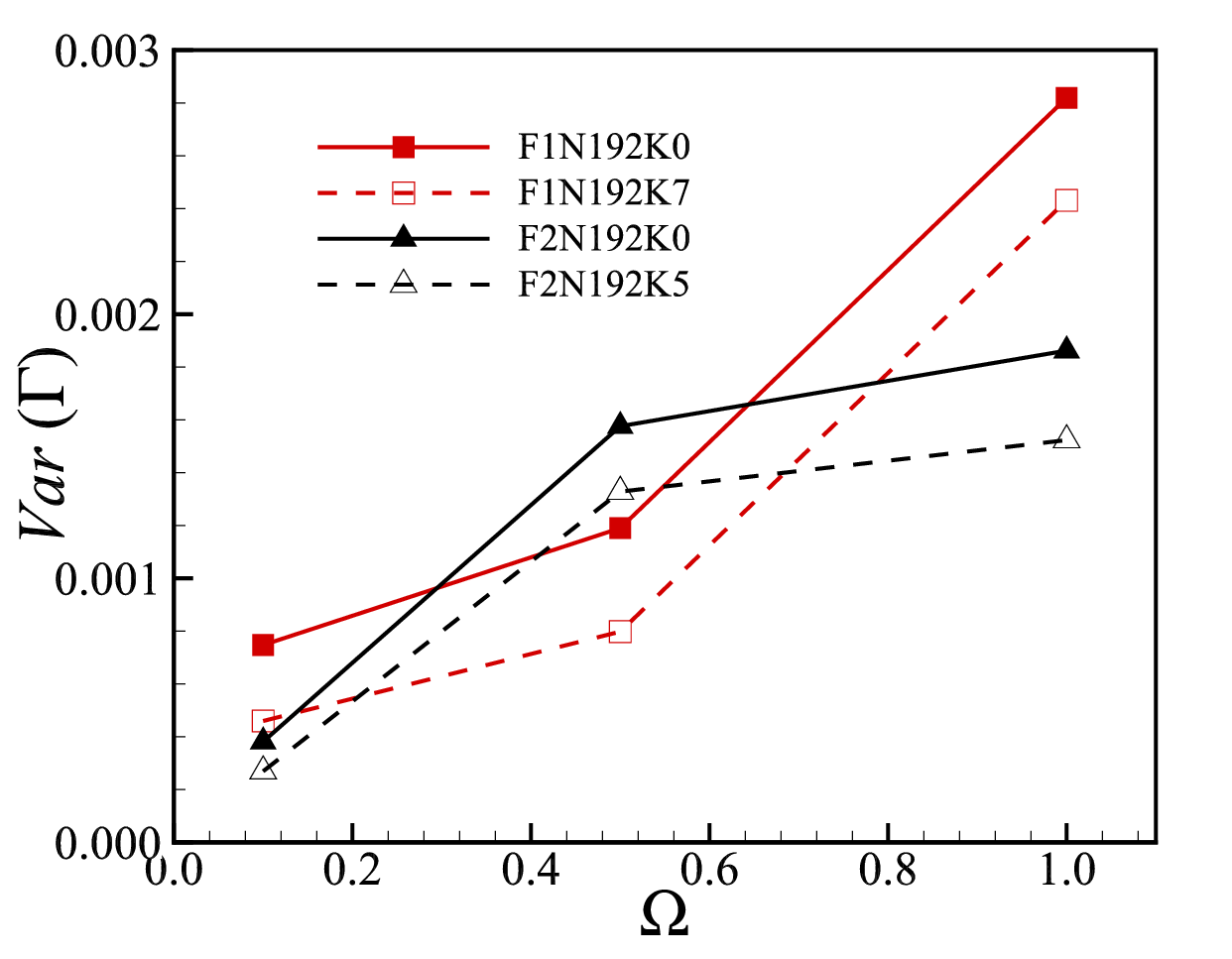}
\caption{\label{fig:lya_rot_var} The variance of the normalised local
Lyapunov exponent $\Gamma$ for selected cases.}
\efig

\bfig
\centering
\ig[width=0.48\lnw]{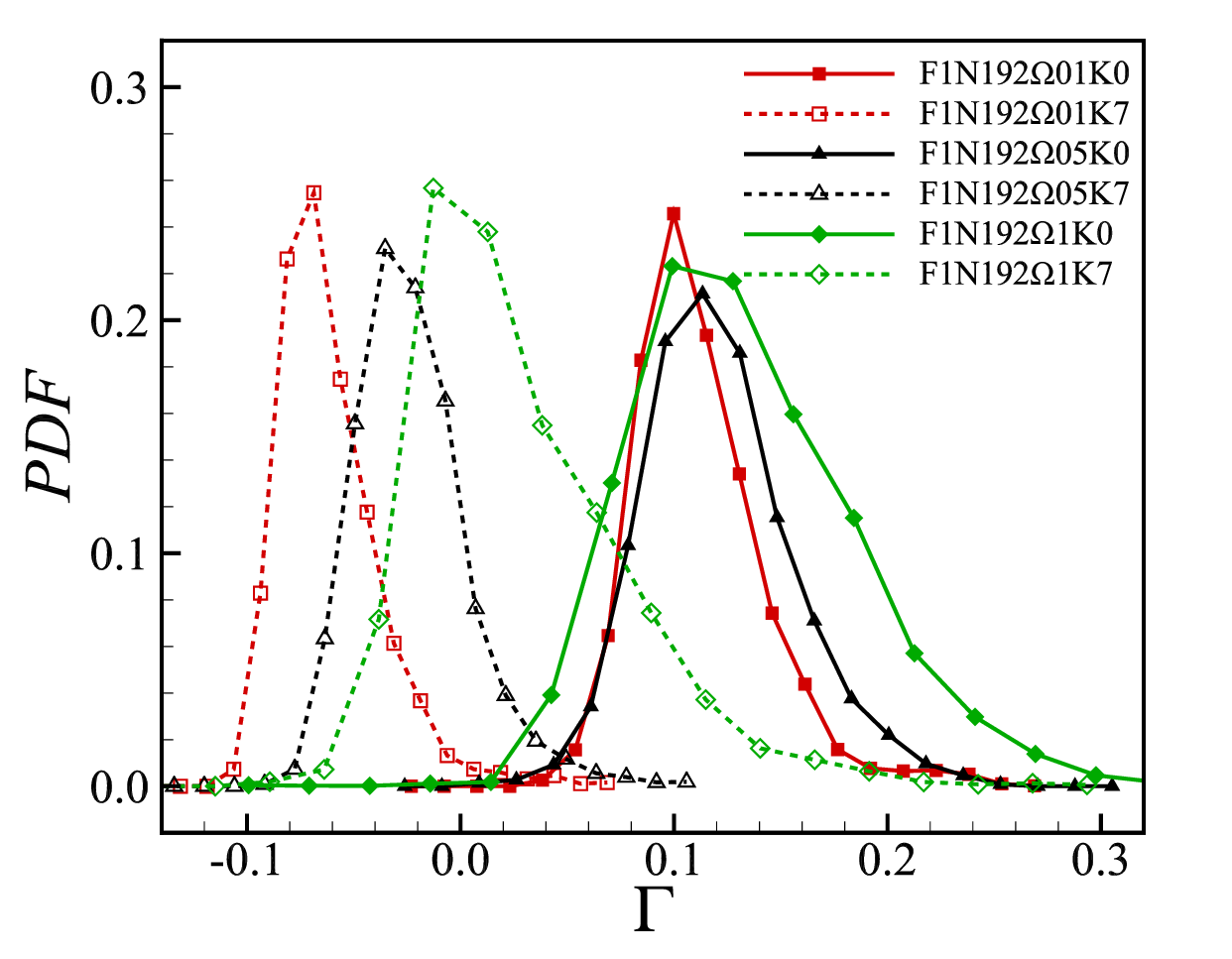} %
\ig[width=0.48\lnw]{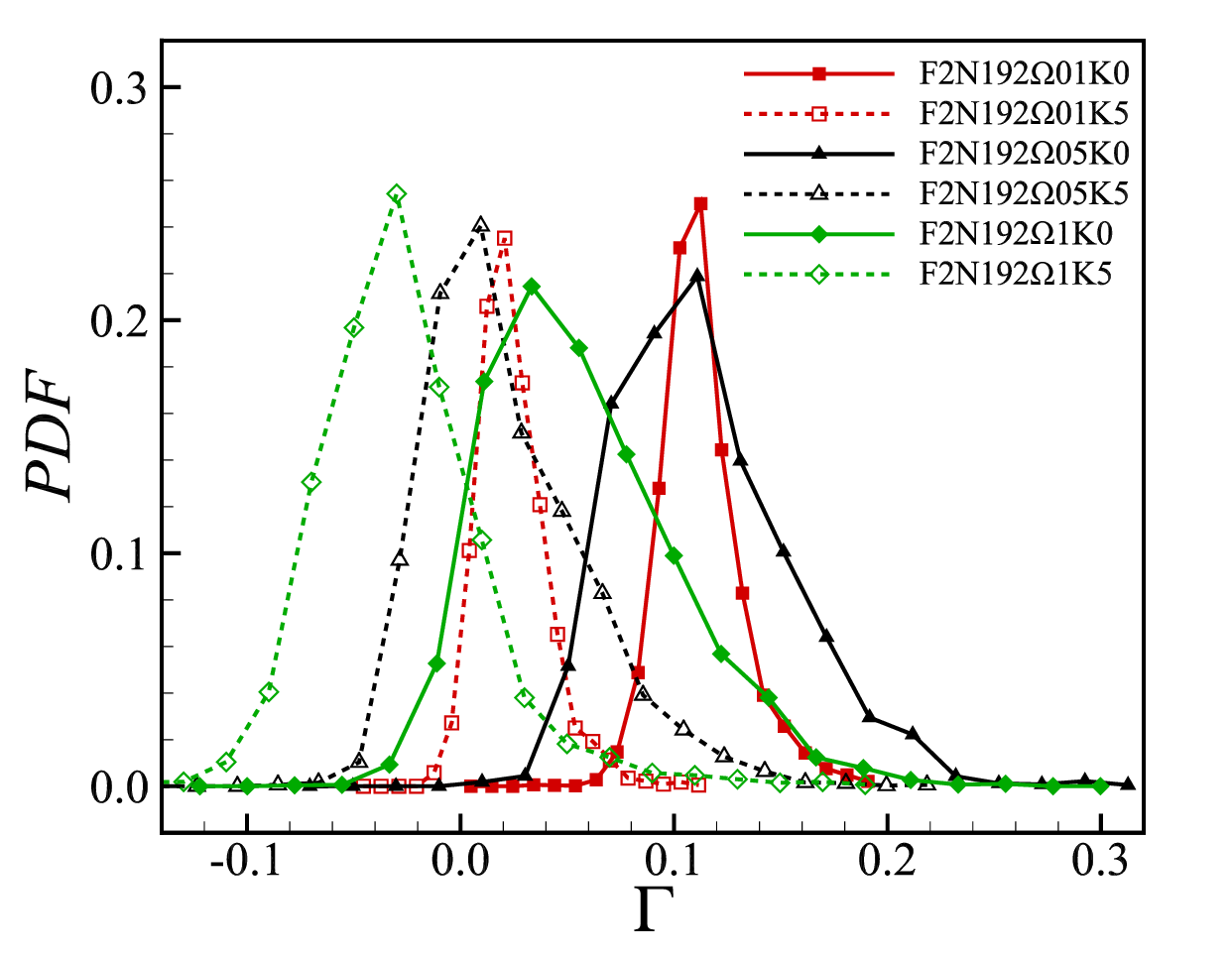}
\caption{\label{fig:lya_rot_pdf} The PDFs of the local
Lyapunov exponent $\Gamma$ for selected cases. Left: cases with Kolmogorov
forcing.
Right: cases with constant power forcing. Note the PDFs are not
normalised.}
\efig

As previously commented, the local CLEs $\Gamma$ display significant
fluctuations. In this subsection, we present statistics of $\Gamma$ for a few selected cases to
highlight the qualitative trends that are shared by the other cases. The
statistics in this subsection are all calculated by averaging over 
time as well as five independent
realisations. The average is denoted by $\lal ~ \ral$.  

The variance of $\Gamma$,
$
Var(\Gamma) = \lal (\Gamma - \lal \Gamma \ral)^2\ral,
$
is shown in Fig. \ref{fig:lya_rot_var} for the cases indicated in the figure.
The variance clearly increases with rotation in all cases for a given $k_m$,
and for a given $\Omega$, it is smaller for
larger $k_m$. The behaviours at high rotation rates are different for the two
forcing terms. The variance for constant power forcing seems to increase
slower at higher
rotation rates. 

The PDFs of $\Gamma$, shown in Fig. \ref{fig:lya_rot_pdf} for the same
selected cases, largely exhibit the same
behaviours already shown by the mean CLEs and the variances. 
A common feature is that the width of the PDF increases with the rotation
rate. Note that the PDFs are not normalised. The increase in the width is thus
a manifestation of increased variance. 

For Kolmogorov forcing (left panel), the PDFs of the 
unconditional $\Gamma$ (with $k_m=0$) do not move significantly with the rotation rate. 
For $k_m = 7$, the PDF moves towards the positive values as rotation is
strengthened, indicating increased mean CLE. 
These behaviours are 
consistent with Fig. \ref{fig:lya_rot_f2f3}. The behaviours of the PDFs for
constant power forcing (right panel) are also consistent with Fig.
\ref{fig:lya_rot_f2f3}. One notable difference with the results in the left
panel is the PDFs for the
unconditional Lyapunov exponents are affected more strongly by rotation in
this case. For example the PDF for $\Omega = 1$ is moved to the left
significantly, while the same is not observed for the corresponding case in
the left panel.   

At higher rotation rates, the PDFs often have
significant probabilities to take both positive and negative values, e.g.,
those for F1N192$\Omega$1K7, F1N192$\Omega$05K7 and those with constant power
forcing and $k_m=5$. Therefore, for these cases, even if synchronisation
is achieved in the long term,
the synchronisation error $\Delta(t)$ may increase temporarily when the local CLE is
positive. This behaviour is observed in Figs. \ref{fig:decay_f2_tauk} and
\ref{fig:decay_f3_tauk} for some $k_m$ values near the threshold
wavenumber.

\subsection{Statistics of energy production and dissipation}

\bfig
\centering
\ig[width=0.48\lnw]{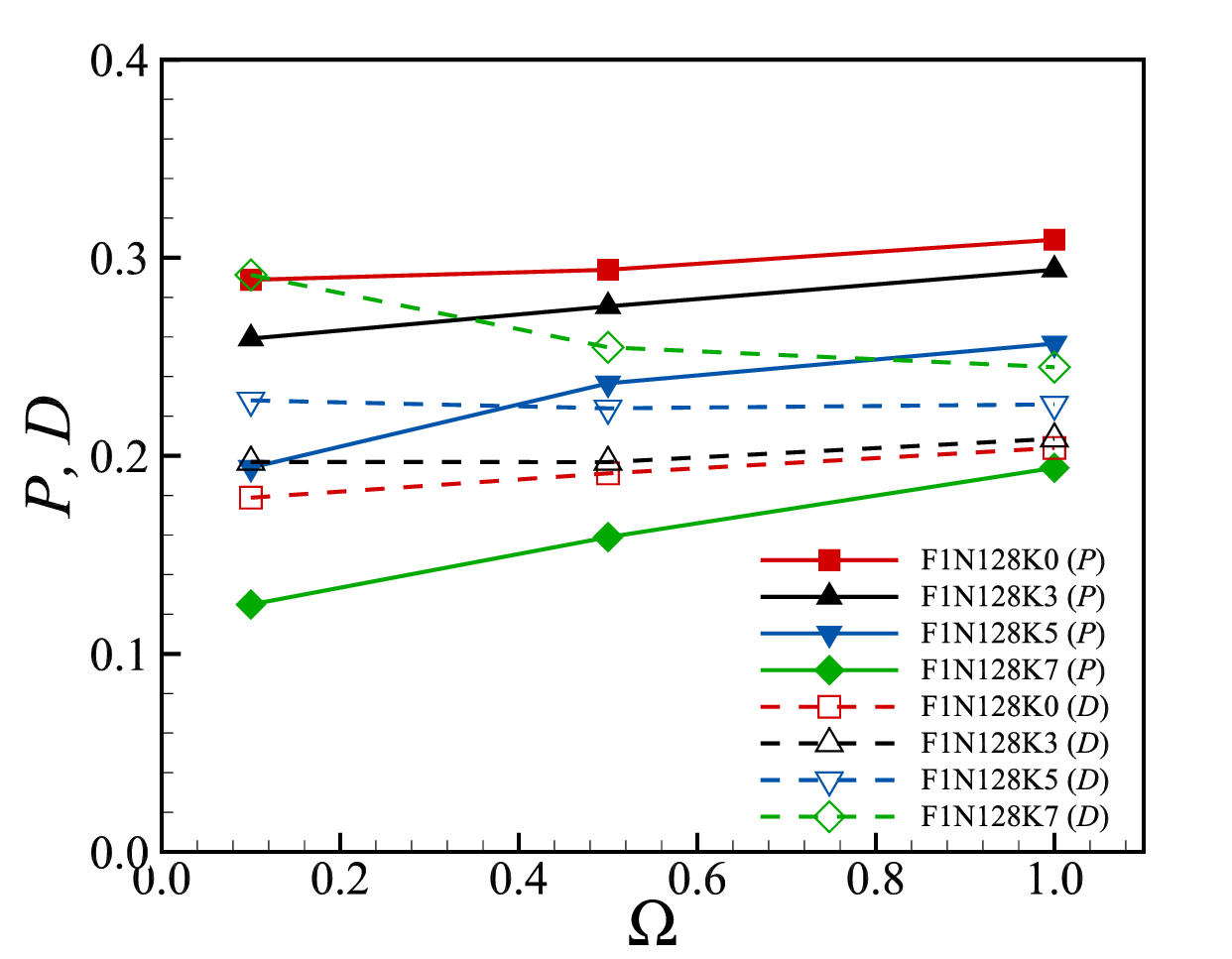} %
\ig[width=0.48\lnw]{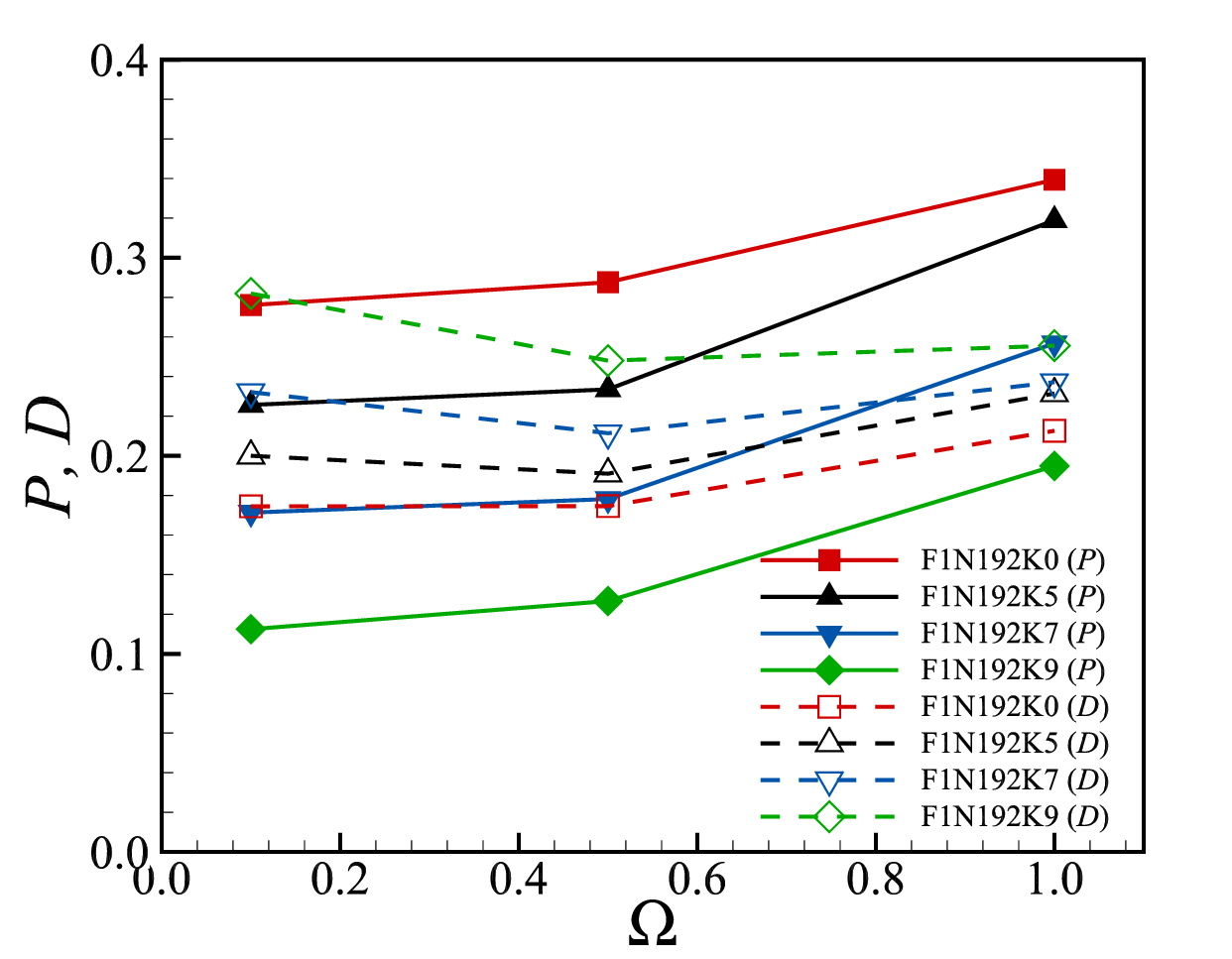}
\caption{\label{fig:err_pdds_mean_f2_pd_dsp} The production term $P$
and energy dissipation $D$ for cases with Kolmogorov forcing. Left:
$N=128$. Right: $N=192$. }
\efig


\bfig
\centering
\ig[width=0.48\lnw]{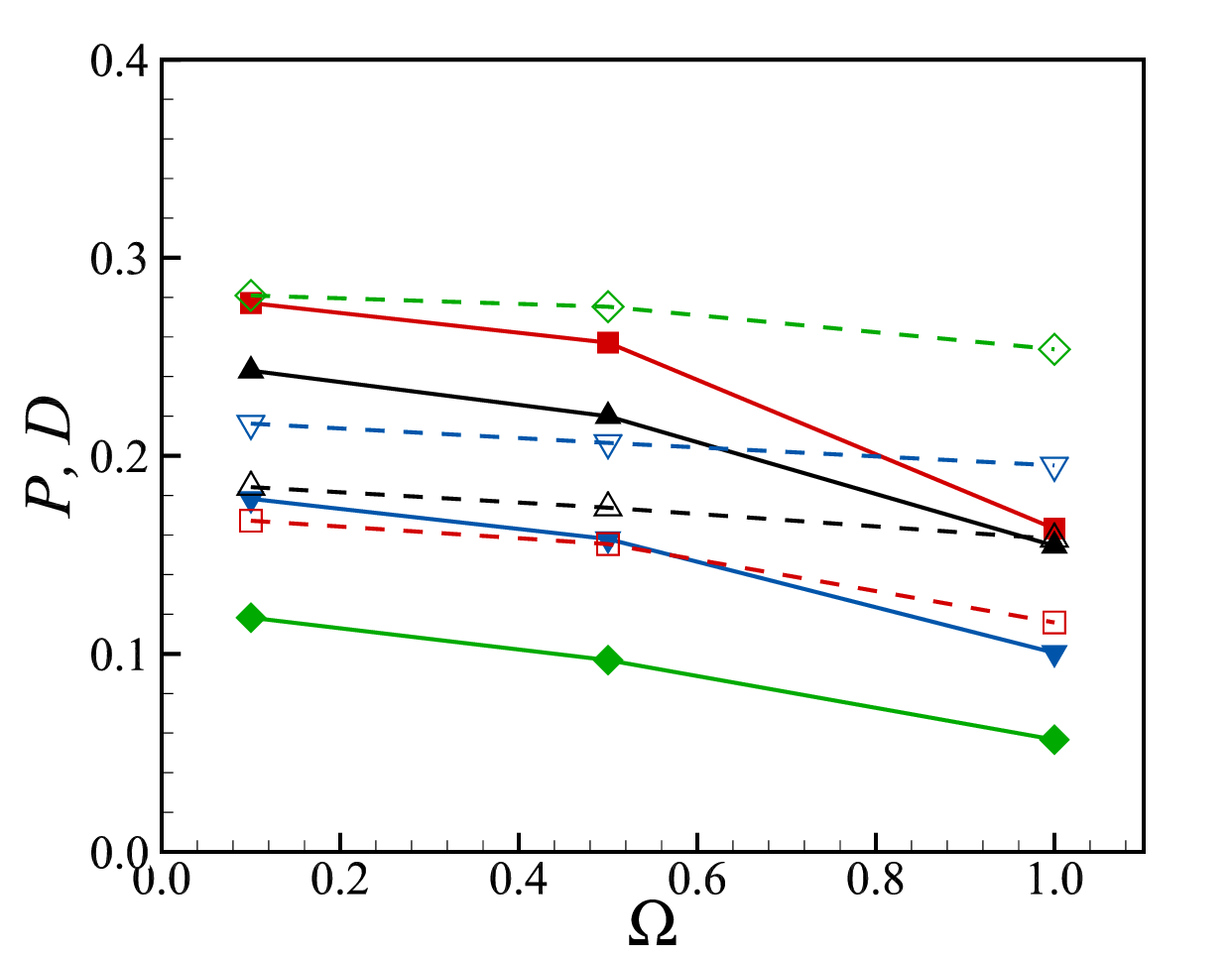} %
\ig[width=0.48\lnw]{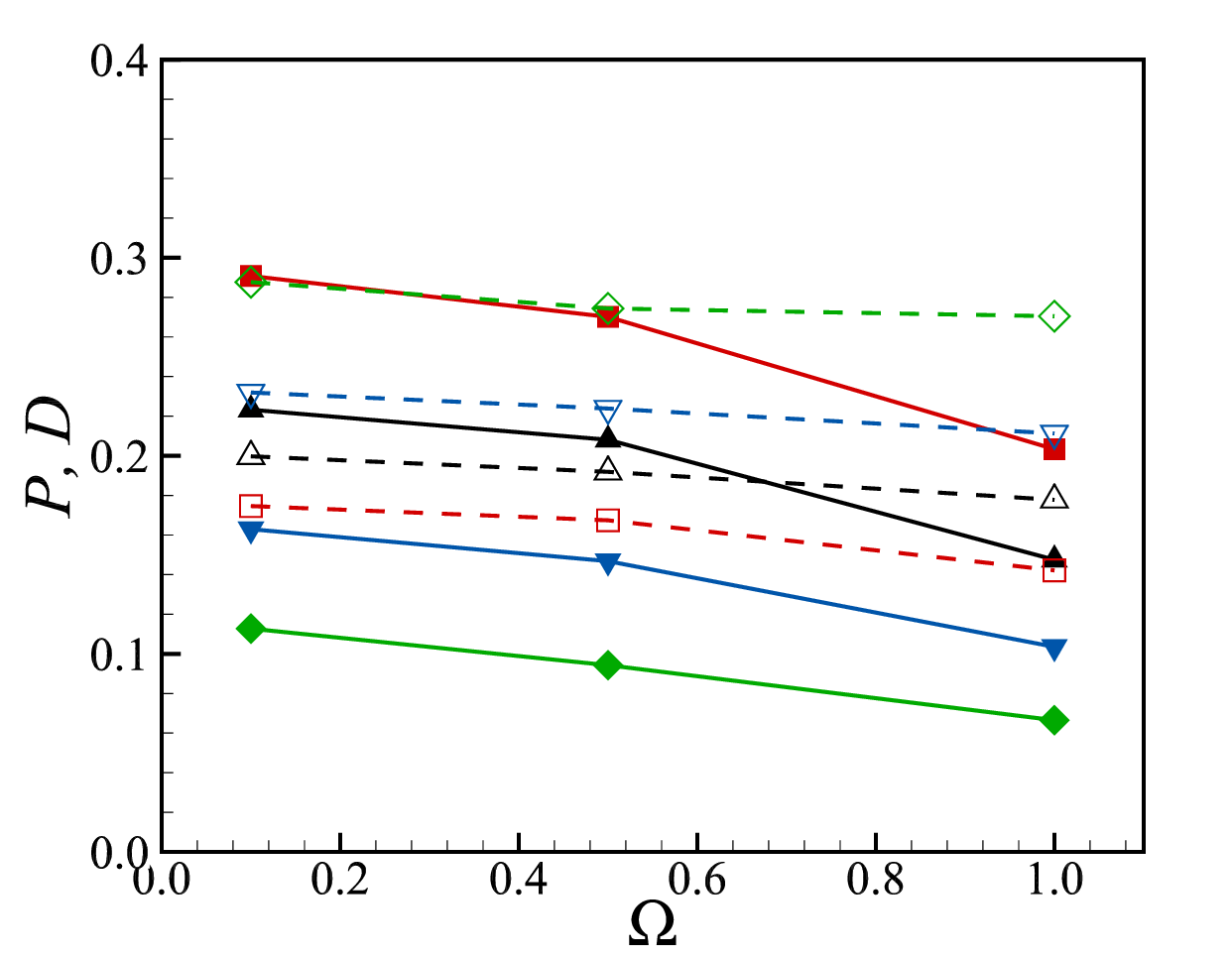}
\caption{\label{fig:err_pdds_mean_f3_pd_dsp} Same as Fig.
\ref{fig:err_pdds_mean_f2_pd_dsp} but for cases with constant power forcing.
}
\efig

Some understanding of $\Gamma$ and $\Lambda$ can be gained from
Eq. (\ref{eq:gamma}). Our calculation shows that
the contribution of the forcing term $\FF$ is always negligible.
In what follows we only present the results related to the
production term $\PP$ and the dissipation term $\DD$. We use 
\be
P \equiv \lla \frac{
  \tau_k\PP}{\Vert \bu^\delta \Vert^2} \rra , ~~ D \equiv
\lla \frac{\tau_k\DD }{\Vert
\bu^\delta \Vert^2} \rra
\ee
to denote the averaged non-dimensionalised production and dissipation terms,
respectively. 

The values of $P$ and $D$ are shown in Figs.
\ref{fig:err_pdds_mean_f2_pd_dsp}--\ref{fig:err_pdds_mean_f3_pd_dsp}. 
For completeness, results for $N=128$ and $192$ have both been included,
but we will mainly comment on those for $N=128$ as the trends are the same
for $N=192$. 
Fig. \ref{fig:err_pdds_mean_f2_pd_dsp} shows
the results for the cases with Kolmogorov forcing. We can see that both
$P$ and $D$ depend strongly on $k_m$, but are less sensitive to the value of $\Omega$. 
The production term $P$
decreases as $k_m$ increases, while the dissipation $D$ increases in the mean
time. Thus both contribute to the decrease in $\Gamma$, hence $\Lambda$, as
$k_m$ increases. 
For a given $k_m$, $P$ increases slightly with rotation rate $\Omega$. 
On the other hand, $D$ increases
slightly with $\Omega$ for smaller $k_m$, but decreases with $\Omega$ for
larger $k_m$. 
Overall, $P$ and $D$ change only slightly with $\Omega$. 


For the cases with constant power forcing, Fig.
\ref{fig:err_pdds_mean_f3_pd_dsp} shows that the main impact of
rotation is on the production term $P$. However, in this case $P$ decreases as
rotation is increased, i.e., the trend is opposite to what is shown in Fig.
\ref{fig:err_pdds_mean_f2_pd_dsp}. This trend is consistent with the
previous observation that in this case synchronisation is easier when $\Omega$
is larger. Dissipation $D$ does not
strongly depend on $\Omega$. 

\bfig
\centering
\ig[width = 0.48\lnw]{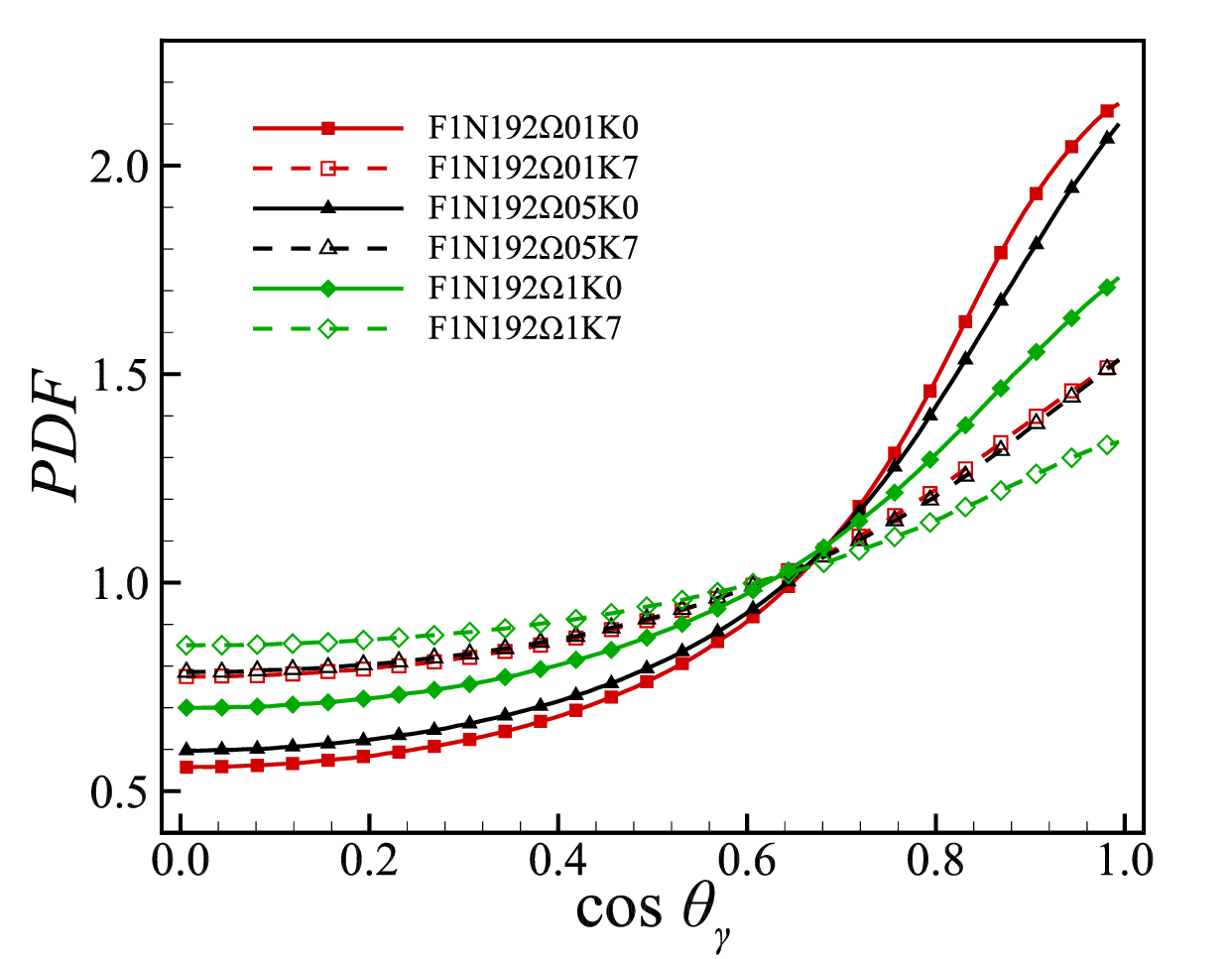}
\ig[width = 0.48\lnw]{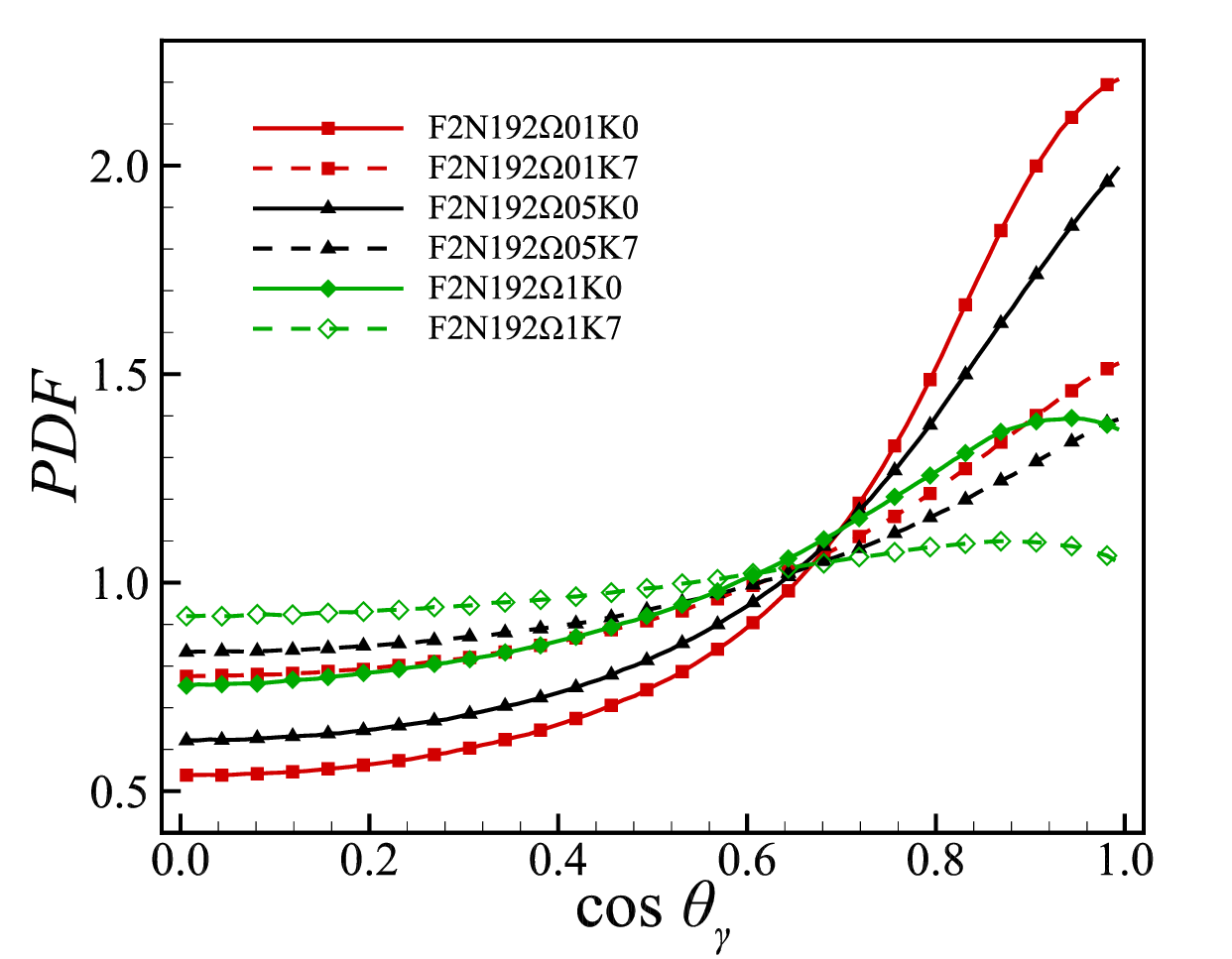}
\caption{\label{fig:alignment} The PDFs of $\cos \theta_\gamma$. Left: cases with
Kolmogorov forcing. Right: cases with constant power forcing.}
\efig
\bfig
\centering
\ig[width = 0.5\lnw]{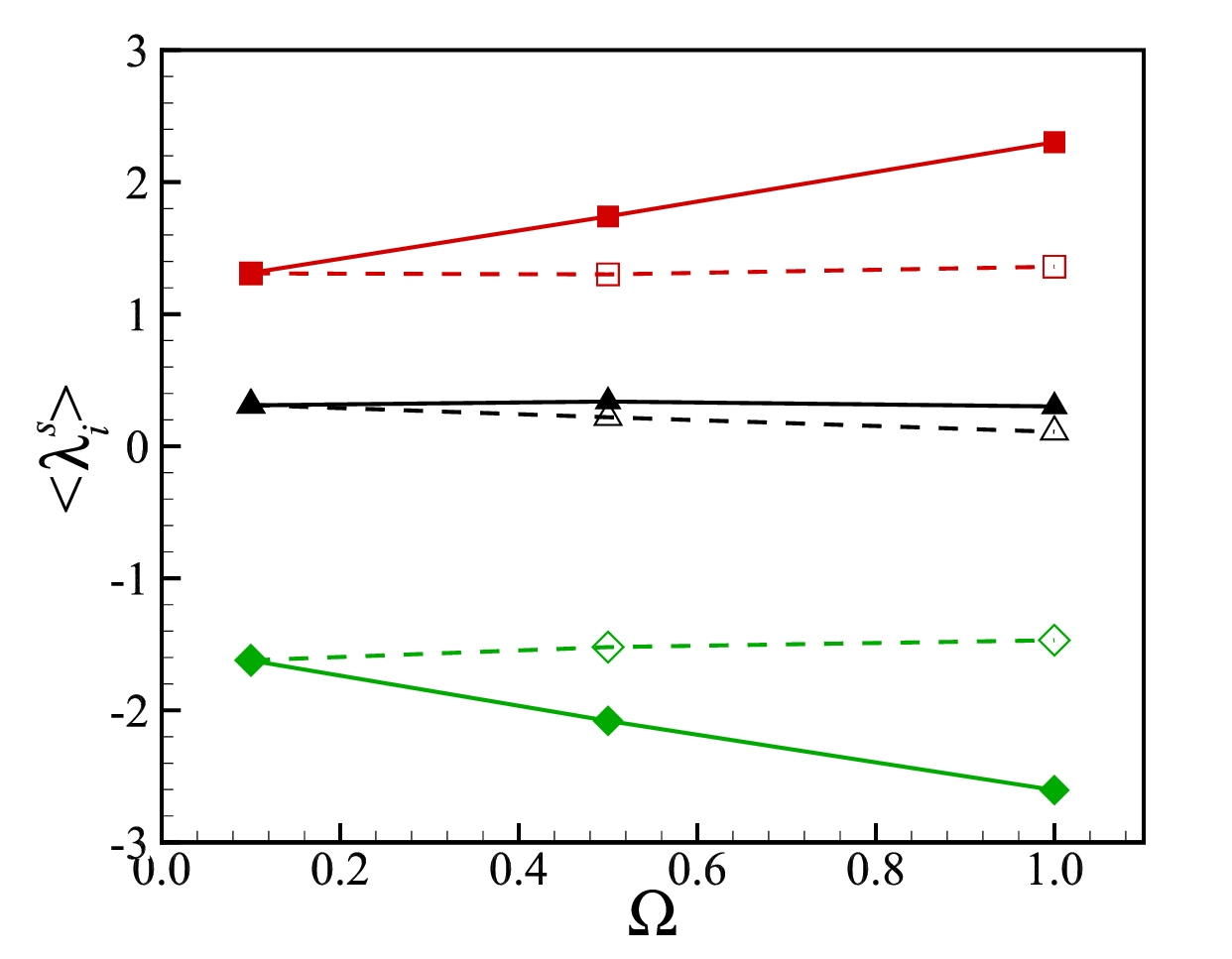}
\caption{\label{fig:lambda} The mean values of the eigenvalues of
the dimensionless strain rate tensor $ s^+_{ij}$. Solid lines: cases with
Kolmogorov
forcing. Dashed lines: cases with constant power forcing. Squares:
$\lal \lambda^s_\alpha\ral$. Triangles: $\lal \lambda^s_\beta\ral$. Diamonds:
$\lal \lambda^s_\gamma \ral$. }
\efig

Further insights into $P$ can be obtained by looking into the
alignment between $\bu^\delta$ and $s_{ij}$. 
Let $s^+_{ij} \equiv \tau_k s_{ij}$ be the dimensionless strain
rate tensor. 
Let $\bv = \bu^\delta/\Vert \bu^\delta \Vert$, and 
$\lambda^s_\al\ge \lambda^s_\beta\ge \lambda^s_\gamma$ be the 
eigenvalues of $s^+_{ij}$,
with corresponding eigenvectors $\e_i$ $(i = \al,
\beta, \gamma)$. Due to incompressibility, we have $\lambda^s_\alpha +
\lambda^s_\beta + \lambda^s_\gamma = 0$
with $\lambda^s_\alpha \ge 0$ and $\lambda^s_\gamma \le 0$. In isotropic turbulence, it is
well-known that $\lambda^s_\beta$ is more likely to take positive values so that
the magnitude of $\lambda^s_\gamma$ tends to be the largest among the three. 

Letting the angle between $\e_i$ and $\bv$ be $\theta_i$, we
may write
\be
P  = P_\alpha + P_\beta +
P_\gamma,
\ee
where
\be
P_\alpha = - \lla \ol{\lambda^s_\al \vert
\bv \vert^2 \cos^2\theta_\al}\rra  , 
P_\beta = - \lla \ol{\lambda^s_\beta  \vert
\bv \vert^2 \cos^2\theta_\beta} \rra , 
P_\gamma = 
- \lla \ol{\lambda^s_\gamma\vert
\bv \vert^2 \cos^2\theta_\gamma }\rra , 
\ee
with $P_\alpha \ge 0$ and $P_\gamma \le 0$. 
The above expressions show that $P$ is closely related to the alignment between $\bu^\delta$ and
the eigenvectors of $s^+_{ij}$, the magnitudes of the eigenvalues, and the
correlations between them. As $\bv$ is normalised, it
is reasonable to expect that its magnitude is insensitive to rotation, and 
that rotation will mainly affect $P$ through the eigenvalues and $\cos \theta_i$. 
Since $P$ is
always positive in our simulations (c.f., Figs.
\ref{fig:err_pdds_mean_f2_pd_dsp} and  
\ref{fig:err_pdds_mean_f3_pd_dsp}), 
$P_\gamma$ is the dominant term in $P$. As a result, we will only consider the
statistics of $\cos \theta_\gamma$ and the eigenvalues. 

The PDFs of $\cos\theta_\gamma$ are
given in Fig. \ref{fig:alignment} for selected cases, with 
the left panel showing the results for Kolmogorov
forcing and the right panel showing those for constant power forcing. 
It is evident that there is a preferable alignment between
$\e_\gamma$ and $\bu^\delta$ when rotation is weak, since the PDFs peak at $\cos
\theta_\gamma = 1$. 
Interestingly, the alignment is weaker for larger $k_m$, and is also weakened by rotation. 
These trends are consistently observed for both forcing terms. 
The mean values of the eigenvalues are given in Fig. \ref{fig:lambda}. Here,
the results for the two forcing terms display different trends. For constant
power
forcing, the mean eigenvalues are almost independent of the
rotation rate. For Kolmogorov forcing, the magnitude of the averaged $\lambda^s_\alpha$ and
$\lambda^s_\gamma$ both increase significantly with rotation. 

Putting the results in Figs.
\ref{fig:alignment} and \ref{fig:lambda} together, the physical picture for
flows with constant power forcing appears to be simple. The
production term $P$ decreases with rotation in this case, because the
preferable 
alignment between $\e_\gamma$ and $\bu^\delta$ is reduced by rotation. 
For the flows driven by Kolmogorov forcing, the preferable alignment 
is reduced by rotation, which tends to reduce $P$. However, this trend is
opposed by the trend where the eigenvalues of $s^+_{ij}$ increase with 
rotation. The
overall effect is that $P$ increases only slightly with rotation. 

\subsection{Discussion}

The different consequences of the two forcing terms have been made quite obvious from our
analysis so far. 
However, the cause of the difference is not yet elucidated. 
The Kolmogorov forcing term is inhomogeneous whereas the constant power forcing is
isotropic. However, if rotation is absent, this difference alone does not lead to
significant
differences in the statistics we have examined, notwithstanding the
fact that
the Reynolds numbers of the flows are not large. This assertion is supported by
the various statistics obtained for $\Omega = 0.1$, i.e., for 
weakest rotation. For example, Fig. \ref{fig:alignment} shows that the
alignment is roughly the same for the two forces when $\Omega = 0.1$. Fig.
\ref{fig:lambda} shows that the mean eigenvalues are also almost the same for
the two flows when $\Omega = 0.1$. The same can be said about the (conditional) Lyapunov exponents as
well (see, e.g., Figs.
\ref{fig:lya_rot_f2f3} and
\ref{fig:lya_keta_f2f3}). Therefore, if there is no rotation, the results
would be more or less independent of the forcing mechanism even if the Reynolds
number is
moderate, i.e. even if there is only moderate scale separation between the forced large scales
and the small scales. 

One may thus conclude that the drastic impacts of the forcing terms originate
from the interaction between forcing and 
rotation. The interaction 
seems to alter the spectral dynamics profoundly, eluding simple
phenomenological explanations (see, e.g., \citet{DallasTobias16} and references therein). 
A corollary is that it is also unlikely to obtain simple mechanical explanations for
the difference in the behaviours of the production term, hence those of the
Lyapunov exponents, shown in
previous subsections. 
Nevertheless, to shed some light on the physics behind the observations, we look into 
how different scales of the flow contribute to the production term, and how
these contributions depend on the rotation rate. 

\bfig
\centering
\ig[width = 0.48\lnw]{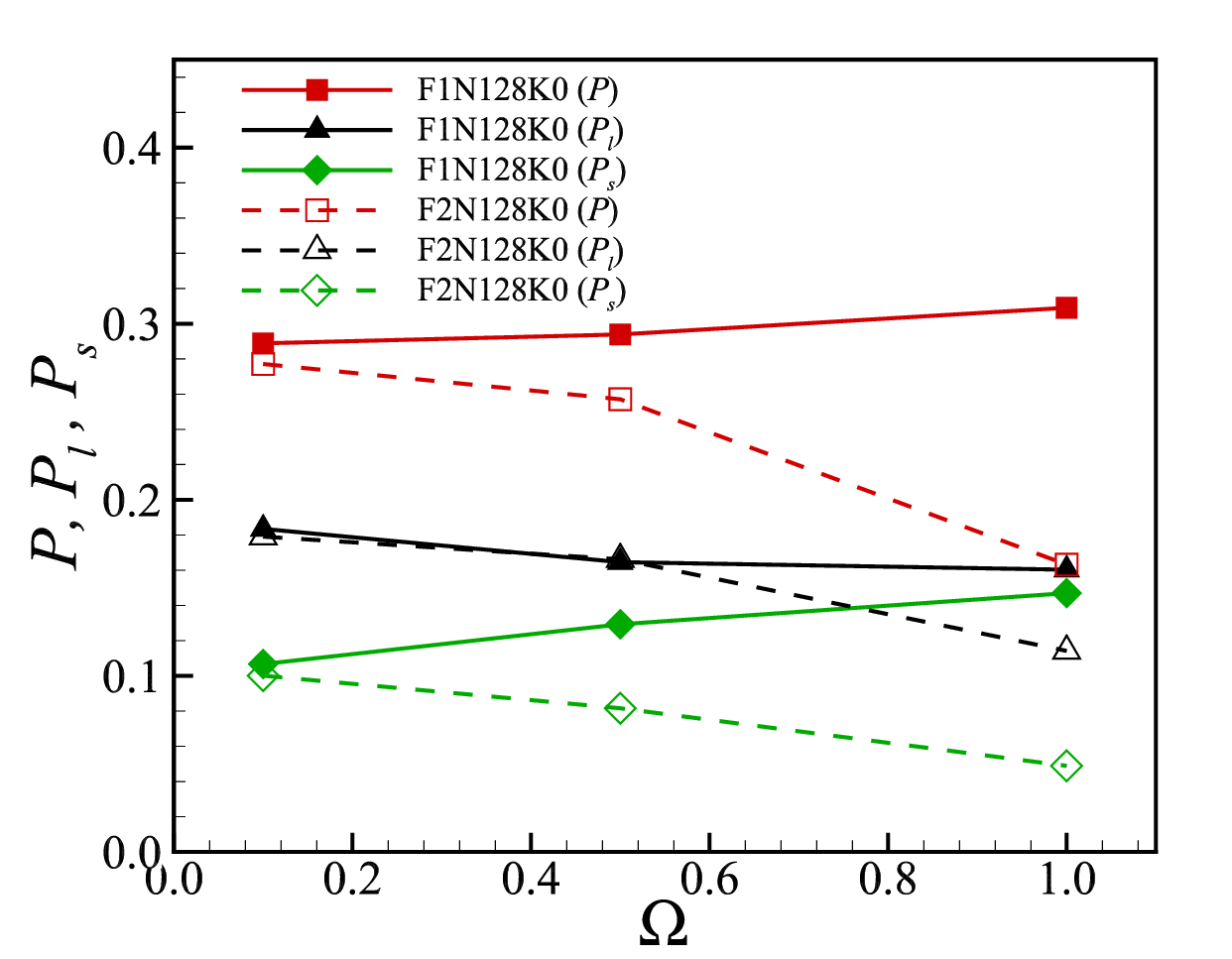}
\ig[width = 0.48\lnw]{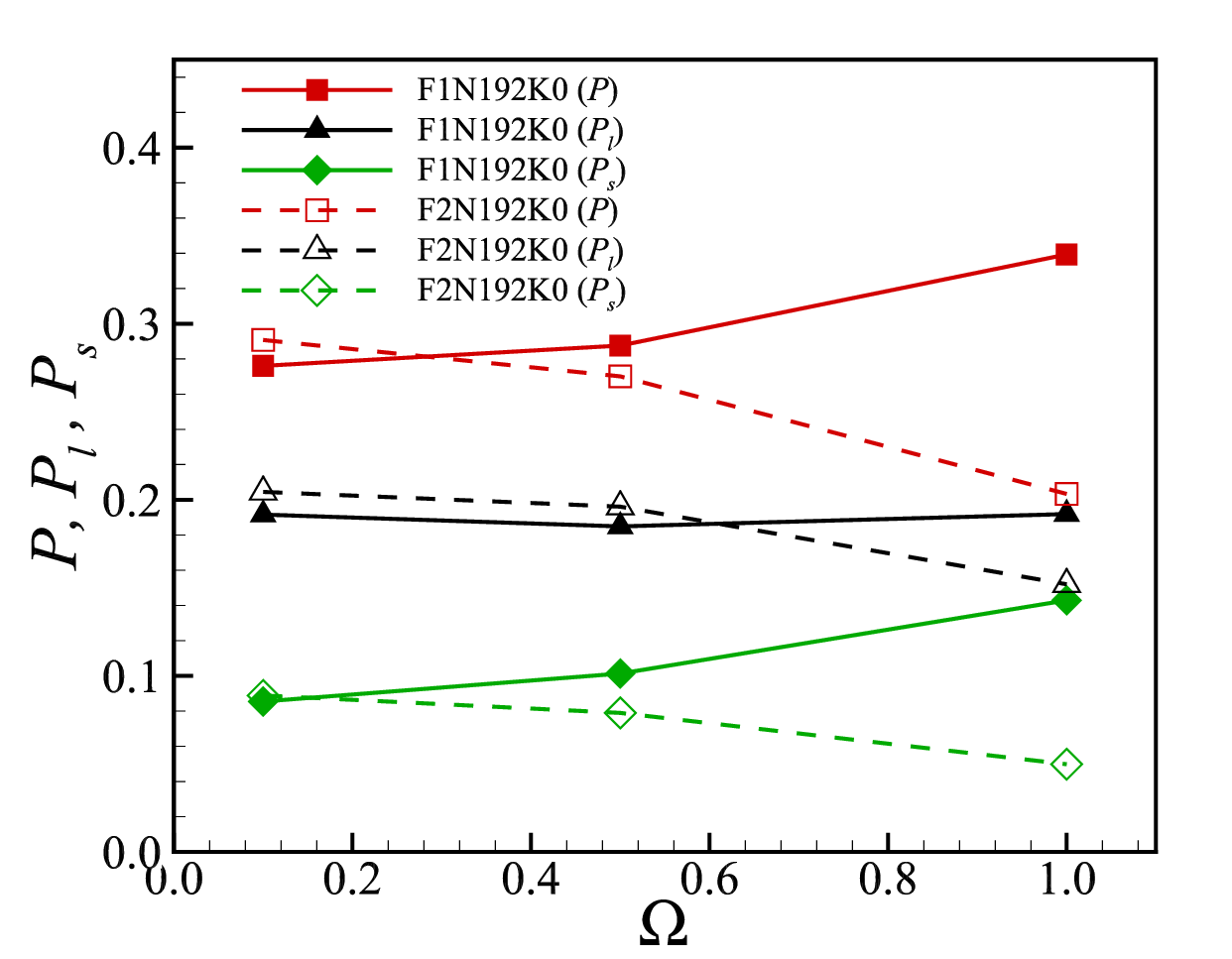}
\caption{\label{fig:pdls} The large scale ($P_l$) and small scale ($P_s$) contributions to the
production term ($P$). Left: $N=128$.
Right: $N=192$.}
\efig

To do so, we decompose the normalised strain rate tensor
$s^+_{ij}$ into a large scale component $s^{+>}_{ij}$ and a small scale
component $s^{+<}_{ij} \equiv s^+_{ij} - s^{+>}_{ij}$. 
$s^{+>}_{ij}$ is obtained by applying a low-pass filter on $s^+_{ij}$ \citep{Pope00}.
The production term
calculated with $s^{+>}_{ij}$ and $\bv$ is denoted by $P_l$, and that with
$s^{+<}_{ij}$ and $\bv$ is denoted by $P_s$, where obviously $P_l + P_s = P$.
$P_l$ and $P_s$ represent the contributions from large and small scale
straining, respectively, to the total
production $P$. 

We use the Gaussian filter \citep{Pope00} to calculate $s^{+>}_{ij}$. The filter length scale $\Delta$ 
is chosen to be eight times of the
grid size, with the corresponding filter wavenumber $k_\Delta \equiv \pi/\Delta$
being $12$ for $N=192$, and $8$ for $N=128$. 
The values for $P$, $P_l$ and $P_s$ at different rotation rates are plotted in
Fig. \ref{fig:pdls} for $k_m=0$. 
Several observations are evident. Firstly, $P_l$ is larger than $P_s$ for
both forcing terms and both Reynolds numbers. That
is, large scale straining makes larger contribution to the total
production term, which is consistent
with the fact that the spectrum of $\bu^\delta$ peaks at intermediate to low wavenumbers
(c.f. Fig. \ref{fig:lya_vector_errEk_f2f3}).   
Secondly, for constant power forcing,  
both $P_l$ and $P_s$ decrease as $\Omega$ increases. Both contribute roughly
equally to the
decrease of the total production. Thirdly, for Kolmogorov forcing, the picture
again is quite different. Interestingly, $P_l$ barely increases (or 
decreases slightly) as $\Omega$ increases, whereas $P_s$ 
increases with $\Omega$ at both Reynolds numbers. Even though $P_l$ makes
bigger contribution to $P$, the change in $P$ with $\Omega$ comes mainly from
$P_s$. 

The results in Fig. \ref{fig:pdls} are again non-trivial to interpret fully.
If the impacts of forcing are confined in
large scales, 
small scale contribution $P_s$ should behave 
in similar ways in flows driven by different forcing terms, but this is not supported
by Fig. \ref{fig:pdls}. If the effects of Kolmogorov forcing
(coupled with rotation) on smaller scales
decrease with increasing scale separation according to
naive Kolmogorov phenomenology, then one should reasonably expect $P_l$
depends more strongly on $\Omega$ compared with $P_s$, which again is not supported by
Fig. \ref{fig:pdls}.  
Overall, like previous research
\citep{DallasTobias16}, these observations suggest that large scale forcing affects
the spectral dynamics of rotating turbulence in highly non-trivial ways, 
which is the root cause of the different synchronisability of the flows. 
 
\section{Conclusions \label{sect:conclusions}}

We investigated the synchronisation of rotating turbulence numerically, with a
focus on the effects of the rotation rates and the forcing mechanism. The
phenomenon is analysed through the decay rate of the synchronisation error,
the threshold value of the coupling wavenumber, the conditional Lyapunov
exponents, the conditional Lyapunov vector and the
dynamical equation for the velocity perturbations. 

One main finding is that
the ability to synchronise rotating turbulence varies significantly with the
forcing mechanism. 
For Kolmogorov flows, which are driven by a constant
sinusoidal forcing term, the conditional Lyapunov exponent for a given
coupling wavenumber increases with
rotation, which means the flows are more difficult to synchronise with a given
coupling wavenumber. However, the dimensionless threshold value for the coupling
wavenumber is essentially independent of rotation within the range of rotation
rates we have investigated, and is unchanged from the value
found in isotropic turbulence even though the energy spectrum of the flow
clearly is
steeper (consistent with the $k^{-2}$ power law). 

For a different forcing scheme which is characterised by a prescribed constant energy
injection rate, the conditional Lyapunov exponent decreases as rotation is
strengthened so synchronisation is easier to achieve. The dimensionless
threshold coupling wavenumber can be significantly smaller when rotation is
strong and the slope of the energy spectrum approaching $-3$. 

We find that the energy spectra of the Lyapunov vector as well as the
conditional Lyapunov vectors have a close relationship 
with the threshold coupling wavenumber for both forcing schemes. 
The threshold coupling wavenumber and 
the wavenumber where the 
energy spectrum of the Lyapunov vector peaks appear to show same dependence on
rotation. Meanwhile, we find that, for both forcing terms, the flows
do not synchronise when the energy spectrum of the conditional Lyapunov vector
has a peak in the wavenumber range for the slaved Fourier modes. 

Rotation is also shown to increase the fluctuation in the local conditional
Lyapunov exponent. In some cases, though the long time conditional Lyapunov
exponent is negative, the fluctuating local conditional Lyapunov exponent can 
frequently become positive. This behaviour explains why for some numerical
experiments, the synchronisation error can oscillate about an overall
exponential decay curve, and shows that rotation can reduce the stability of
the synchronised state. 

An analysis of the production term in the dynamical equation for velocity
perturbations shows that rotation tends to reduce the preferential alignment
between the perturbation velocity and the eigenvectors of the strain rate
tensor. This behaviour tends to reduce the conditional Lyapunov exponent,
which is the reason why the flows driven by the second forcing term is easier
to synchronise. However, this effect is counter-balanced in the Kolmogorov
flows by increased eigenvalues of the strain rate tensor. 

A limitation of current investigation is that the Reynolds number is
relatively small. Our results indicate that the threshold coupling wavenumber 
depends on the slope of the energy spectrum of the flow to some extent. To
ascertain the relation, simulations with an extended inertial range are
needed, which can be achieved only when the Reynolds number of the flow is
much higher. The relation between the threshold wavenumber and the energy
spectrum of the Lyapunov vector also requires further scrutiny at higher Reynolds
numbers. Another limitation is that the anisotropy of rotating turbulence has
not yet been
accounted for. Due to the formation of columnar vortices along the direction
of the rotation axis, the threshold wavenumber can be
different in the axial and the transversal directions. A two component threshold
wavenumber is likely to provide a more precise description. 

The drastically different physical pictures yielded by the two forcing naturally
suggest that we might obtain different results again for yet another forcing
mechanism (e.g., when the Kolmogorov forcing introduces shearing along the
rotating axis). A more extensive investigation is warranted. 

Though rotation has profound effects on turbulence, the rotation rate does
not appear explicitly in the
energy budget of the flow. However, it does directly enter the spectral dynamics
and the equations for
higher order statistics. It would be interesting to investigate the behaviours
of higher order
statistics such as the generalised
Lyapunov exponents \citep{Fujisaka83,Cencinietal10} or the generalised
conditional Lyapunov exponents. 
They are the natural measures for the
strong fluctuations in finite time amplification of synchronisation errors.
Such an investigation would lead to more refined characterisation of the
synchronisation process.

\backsection[Acknowledgments]
{The authors gratefully acknowledge the anonymous referees for their
insightful comments which have helped to improve the
manuscript. We are particularly grateful for their comments related to generalised Lyapunov
exponents and the anisotropic threshold wavenumber.   
Huda Khaleel Mohammed acknowledges 
the School of Mathematics and Statistics, University of Sheffield,
Sheffield, UK for hosting her academic visit.
}

\backsection[Funding]
{Jian Li acknowledges the support of the National Natural Science Foundation of China (No.
12102391).
Huda Khaleel Mohammed acknowledges the Iraqi Ministry of Higher Education and
Scientific Research for the research leave based on the ministerial directive
No. 29743 on November 11, 2021, and Ninevah University for the research
scholarship based on the university directive No. 05/06/3595 on December 8,
2021.
}

\backsection[Data availability statement]{The data that support the findings of this study are available 
from the corresponding author upon reasonable request.}

\backsection[Declaration of Interests]{The authors report no conflict of
interest.}

\bibliographystyle{jfm}
\bibliography{/home/yili/GoogleWork/turbref}

\begin{thebibliography}{32}
\expandafter\ifx\csname natexlab\endcsname\relax\def\natexlab#1{#1}\fi
\def\au#1{#1} \def\ed#1{#1} \def\yr#1{#1}\def\at#1{#1}\def\jt#1{\textit{#1}}
  \def\bt#1{#1}\def\bvol#1{\textbf{#1}} \def\vol#1{#1} \def\pg#1{#1}
  \def\publ#1{#1}\def\arxiv#1{#1}\def\org#1{#1}\def\st#1{\textit{#1}}

\bibitem[Alexakis(2015)]{Alexakis15}
{\sc \au{Alexakis, A.}} \yr{2015}  \at{Rotating taylor-green flows}.  \jt{J.
  Fluid Mech.}  \bvol{769},  \pg{46--78}.

\bibitem[Bartello {\em et~al.\/}(1994)Bartello, M{\`e}tais \&
  Lesieur]{Bartelloetal94}
{\sc \au{Bartello, P.}, \au{M{\`e}tais, O.} \& \au{Lesieur, M.}} \yr{1994}
  \at{Coherent structures in rotating three-dimensional turbulence}.  \jt{J.
  Fluid Mech.}  \bvol{273},  \pg{1--29}.

\bibitem[Boccaletti {\em et~al.\/}(2002)Boccaletti, Kurths, Osipov, Valladares
  \& Zhou]{Boccalettietal02}
{\sc \au{Boccaletti, S.}, \au{Kurths, J.}, \au{Osipov, G.}, \au{Valladares,
  D.~L.} \& \au{Zhou, C.~S.}} \yr{2002}  \at{The synchronization of chaotic
  systems}.  \jt{Phys. Rep.}  \bvol{366},  \pg{1--101}.

\bibitem[Boffetta \& Musacchio(2017)]{BoffettaMusacchio17}
{\sc \au{Boffetta, G.} \& \au{Musacchio, S.}} \yr{2017}  \at{Chaos and
  predictability of homogeneous-isotropic turbulence}.  \jt{Phys. Rev. Lett.}
  \bvol{119},  \pg{054102}.

\bibitem[Bohr {\em et~al.\/}(1998)Bohr, Jensen, Paladin \&
  Vulpiani]{Bohretal98}
{\sc \au{Bohr, T.}, \au{Jensen, M.~H.}, \au{Paladin, G.} \& \au{Vulpiani, A.}}
  \yr{1998} {\em Dynamical Systems Approach to Turbulence\/}.  \publ{Cambridge
  University Press}.

\bibitem[Borue \& Orszag(1996)]{BorueOrszag96}
{\sc \au{Borue, V.} \& \au{Orszag, S.~A.}} \yr{1996}  \at{Numerical study of
  three-dimensional kolmogorov flow at high reynolds numbers}.  \jt{J. Fluid
  Mech.}  \bvol{306},  \pg{293--323}.

\bibitem[Buzzicotti \& Leoni(2020)]{BuzzicottiLeoni20}
{\sc \au{Buzzicotti, Michele} \& \au{Leoni, Patricio Clark~De}} \yr{2020}
  \at{Synchronizing subgrid scale models of turbulence to data}.  \jt{Phys.
  Fluids}  \bvol{32},  \pg{125116}.

\bibitem[Cencini {\em et~al.\/}(2010)Cencini, Cecconi \&
  Vulpiani]{Cencinietal10}
{\sc \au{Cencini, M.}, \au{Cecconi, F.} \& \au{Vulpiani, A.}} \yr{2010} {\em
  Chaos: from simple models to complex systems\/}.  \publ{World Scientific,
  Singapore}.

\bibitem[Dallas \& Tobias(2016)]{DallasTobias16}
{\sc \au{Dallas, Vassilios} \& \au{Tobias, Steven~M.}} \yr{2016}
  \at{Forcing-dependent dynamics and emergence of helicity in rotating
  turbulence}.  \jt{J. Fluid Mech.}  \bvol{798}.

\bibitem[Eroglu {\em et~al.\/}(2017)Eroglu, Lamb \& Pereira]{Erogluetal17}
{\sc \au{Eroglu, D.}, \au{Lamb, J. S.~W.} \& \au{Pereira, T.}} \yr{2017}
  \at{Synchronisation of chaos and its applications}.  \jt{Contemporary
  Physics}  \bvol{58},  \pg{207}.

\bibitem[Fujisaka(1983)]{Fujisaka83}
{\sc \au{Fujisaka, H.}} \yr{1983}  \at{Statistical dynamics generated by
  fluctuations of local lyapunov exponents}.  \jt{Prog. Theor. Phys.}
  \bvol{70},  \pg{1264--1275}.

\bibitem[Fujisaka \& Yamada(1983)]{FujisakaYamada83}
{\sc \au{Fujisaka, H.} \& \au{Yamada, T.}} \yr{1983}  \at{Stability theory of
  synchronized motion in coupled-oscillator systems}.  \jt{Prog. Theor. Phys.}
  \bvol{69},  \pg{32}.

\bibitem[Godeferd \& Moisy(2015)]{GodeferdMoisy15}
{\sc \au{Godeferd, F.~S.} \& \au{Moisy, F.}} \yr{2015}  \at{Structure and
  dynamics of rotating turbulence: A review of recent experimental and
  numerical results}.  \jt{Applied Mechanics Review}  \bvol{67},  \pg{030802}.

\bibitem[Greenspan(1969)]{greenspan69}
{\sc \au{Greenspan, H.~P.}} \yr{1969}  \at{On the nonlinear interaction of
  inertial waves}.  \jt{J. Fluid Mech.}  \bvol{36},  \pg{257--286}.

\bibitem[Henshaw {\em et~al.\/}(2003)Henshaw, Kreiss \&
  Ystr\'{o}m]{Henshawetal03}
{\sc \au{Henshaw, W.D.}, \au{Kreiss, H.-O.} \& \au{Ystr\'{o}m, J.}} \yr{2003}
  \at{Numerical experiments on the interaction between the large- and
  small-scale motions of the navier–stokes equations}.  \jt{Multiscale Model.
  Simul.}  \bvol{1},  \pg{119–149}.

\bibitem[Lalescu {\em et~al.\/}(2013)Lalescu, Meneveau \& Eyink]{Lalescuetal13}
{\sc \au{Lalescu, C.~C.}, \au{Meneveau, C.} \& \au{Eyink, G.~L.}} \yr{2013}
  \at{Synchronization of chaos in fully developed turbulence}.  \jt{Phys. Rev.
  Lett.}  \bvol{110},  \pg{084102}.

\bibitem[Leoni {\em et~al.\/}(2018)Leoni, Mazzino \& Biferale]{Leonietal18}
{\sc \au{Leoni, P. C.~Di}, \au{Mazzino, A.} \& \au{Biferale, L.}} \yr{2018}
  \at{Inferring flow parameters and turbulent configuration with
  physics-informed data assimilation and spectral nudging}.  \jt{Phys. Rev.
  Fluids}  \bvol{3},  \pg{104604}.

\bibitem[Leoni {\em et~al.\/}(2020)Leoni, Mazzino \& Biferale]{Leonietal20}
{\sc \au{Leoni, P. C.~Di}, \au{Mazzino, A.} \& \au{Biferale, L.}} \yr{2020}
  \at{Synchronization to big data: Nudging the navier-stokes equations for data
  assimilation of turbulent flows}.  \jt{Phys. Rev. X}  \bvol{10},
  \pg{011023}.

\bibitem[Li {\em et~al.\/}(2022)Li, Tian \& Li]{Lietal22}
{\sc \au{Li, J.}, \au{Tian, M.} \& \au{Li, Y.}} \yr{2022}  \at{Synchronizing
  large eddy simulations with direct numerical simulations via data
  assimilation}.  \jt{Phys. Fluids}  \bvol{34},  \pg{065108}.

\bibitem[Li {\em et~al.\/}(2020)Li, Zhang, Dong \& Abdullah]{Lietal20}
{\sc \au{Li, Y.}, \au{Zhang, J.}, \au{Dong, G.} \& \au{Abdullah, N.~S.}}
  \yr{2020}  \at{Small-scale reconstruction in three-dimensional kolmogorov
  flows using four-dimensional variational data assimilation}.  \jt{J. Fluid
  Mech.}  \bvol{885},  \pg{A9}.

\bibitem[Morize {\em et~al.\/}(2005)Morize, Moisy \& Rabaud]{Morizeetal05}
{\sc \au{Morize, C.}, \au{Moisy, F.} \& \au{Rabaud, M.}} \yr{2005}
  \at{Decaying grid-generated turbulence in a rotating tank}.  \jt{Phys.
  Fluids}  \bvol{17},  \pg{095105}.

\bibitem[Nikolaidis \& Ioannou(2022)]{NikolaidisIoannou22}
{\sc \au{Nikolaidis, M.-A.} \& \au{Ioannou, P.~J.}} \yr{2022}
  \at{Synchronization of low reynolds number plane couette turbulence}.  \jt{J.
  Fluid Mech.}  \bvol{933}.

\bibitem[Ohkitani \& Yamada(1989)]{OhkitaniYamada89}
{\sc \au{Ohkitani, K.} \& \au{Yamada, M.}} \yr{1989}  \at{Temporal
  intermittency in the energy cascade process and local lyapunov analysis in
  fully developed model turbulence}.  \jt{Prog. Theor. Phys.}  \bvol{81},
  \pg{329--341}.

\bibitem[Pecora \& Carroll(1990)]{PecoraCarroll90}
{\sc \au{Pecora, L.~M.} \& \au{Carroll, T.~L.}} \yr{1990}  \at{Synchronization
  in chaotic systems}.  \jt{Phys. Rev. Lett.}  \bvol{64},  \pg{821--824}.

\bibitem[Pecora \& Carroll(2015)]{PecoraCarroll15}
{\sc \au{Pecora, L.~M.} \& \au{Carroll, T.~L.}} \yr{2015}  \at{Synchronization
  of chaotic systems}.  \jt{Chaos}  \bvol{25},  \pg{097611}.

\bibitem[Pope(2000)]{Pope00}
{\sc \au{Pope, S.~B.}} \yr{2000} {\em Turbulent flows\/}.  \publ{Cambridge
  University Press, Cambridge}.

\bibitem[Sagaut \& Cambon(2008)]{SagautCambon08}
{\sc \au{Sagaut, P.} \& \au{Cambon, C.}} \yr{2008} {\em Homogeneous Turbulence
  Dynamics\/}.  \publ{CUP}.

\bibitem[Vela-Martin(2021)]{VelaMartin21}
{\sc \au{Vela-Martin, A.}} \yr{2021}  \at{The synchronisation of intense
  vorticity in isotropic turbulence}.  \jt{J. Fluid Mech.}  \bvol{913},
  \pg{R8}.

\bibitem[Wang \& Zaki(2022)]{WangZaki21}
{\sc \au{Wang, M.} \& \au{Zaki, T.~A.}} \yr{2022}  \at{Synchronization of
  turbulence in channel flow}.  \jt{J. Fluid Mech.}  \bvol{943},  \pg{A4}.

\bibitem[Wolf {\em et~al.\/}(1985)Wolf, Swift, Swinney \& Vastano]{Wolfetal85}
{\sc \au{Wolf, A.}, \au{Swift, J.~B.}, \au{Swinney, H.~L.} \& \au{Vastano,
  J.~A.}} \yr{1985}  \at{Determining lyapunov exponents from a time series}.
  \jt{Physica D}  \bvol{16}.

\bibitem[Yeung \& Zhou(1998)]{YeungZhou98}
{\sc \au{Yeung, P.~K.} \& \au{Zhou, Ye}} \yr{1998}  \at{Numerical study of
  rotating turbulence with external forcing}.  \jt{Phys. Fluids}  \bvol{10},
  \pg{2895}.

\bibitem[Yoshida {\em et~al.\/}(2005)Yoshida, Yamaguchi \&
  Kaneda]{Yoshidaetal05}
{\sc \au{Yoshida, K.}, \au{Yamaguchi, J.} \& \au{Kaneda, Y.}} \yr{2005}
  \at{Regeneration of small eddies by data assimilation in turbulence}.
  \jt{Phys. Rev. Lett.}  \bvol{94},  \pg{014501}.

\end{thebibliography}

\end{document}